\documentclass
   [
      aps,
      prc,
      preprint,
      tightenlines,
      showpacs,
      floatfix
   ]
   {revtex4-1}

\usepackage{graphicx,amsmath}
\usepackage{dcolumn}             %= align table columns on decimal point
\usepackage{rccol}               %= as dcolumn + rounding
\usepackage{multirow}            %= merging rows in tables
\usepackage{soul}                %= underline text
\usepackage{url}

\newcommand{\nuc}[2] {$^{#1}$#2}

\newcommand{\overlap}[2] {\mbox{$\langle$#1$\,|\,$#2$\rangle$}}

\newcommand{\figref}[1] {Fig.~\ref{#1}}
\newcommand{\secref}[1] {Sec.~\ref{#1}}
\newcommand{\tabref}[1] {Table~\ref{#1}}
\newcommand{\myeqref}[1] {Eq.~\eqref{#1}}

\def\beq{\begin{equation}}
\def\eeq{\end{equation}}
\def\beqa{\begin{eqnarray}}
\def\eeqa{\end{eqnarray}}
\def\bfig{\begin{figure}}
\def\efig{\end{figure}}

\newcommand{\ud}{\,\mathrm{d}}
\begin{document}

\title{Quantum Monte Carlo calculations of spectroscopic overlaps in $A \leq 7$ 
       nuclei}

\author{I. Brida}
\email{brida@anl.gov}
\author{Steven C. Pieper}
\email{spieper@anl.gov}
\author{R. B. Wiringa}
\email{wiringa@anl.gov}
\affiliation{Physics Division, Argonne National Laboratory,
Argonne, IL 60439, USA}

\date{\today}

\begin{abstract}
We present Green's function Monte Carlo calculations of spectroscopic overlaps for $A \leq 7$ nuclei. The realistic Argonne $v_{18}$ two-nucleon and Illinois-7 three-nucleon interactions are used to generate the nuclear states. The overlap matrix elements are extrapolated from mixed estimates between variational Monte Carlo and Green's function Monte Carlo wave functions. The overlap functions are used to obtain spectroscopic factors and asymptotic normalization coefficients, and they can serve as an input for low-energy reaction calculations. 
% In most cases, our spectroscopic factors and asymptotic normalization coeffiicients are in a very good agreement with values deduced from experiments.
\end{abstract}


\maketitle

%==================================== intro ====================================
\section{Introduction}
\label{sec introduction}

One-nucleon spectroscopic overlaps, defined as the expectation value of the nucleon removal operator between 
% projections onto one another of 
states of nuclei differing by one particle, have been used in the treatment of processes in which one particle is added to or removed from the system~\cite{Cl73, BG85}. In particular, these functions provide the interface between nuclear structure and direct reaction theories~\cite{MF60,Sa83, HT03,TN09}. Traditionally, direct reaction theories of processes such as one-nucleon transfer, capture, knockout, or electron-induced proton knockout assume that the state of motion of only one particle is changed as a result of the projectile-target interaction, while the other particles are left essentially undisturbed. Under such a drastic approximation, the structure input needed to analyze experimental reaction data consists of single-particle-like functions of the involved nucleon inside the interacting nuclei, the one-nucleon overlaps. 

For decades, direct reactions have been the main tool for extracting spectroscopic factors (SFs) defined as norms of spectroscopic overlaps. Given the single-particle-like nature of these overlaps, SFs are often associated with single-particle-level occupancies and compared mainly with shell-model values. In practical applications, ``experimental" SFs are extracted from experimental cross sections in a model-dependent way with structure and reaction uncertainties entangled. This dependence was illustrated, for example, in~\cite{PS97,KB01}, where SFs from transfer and electron-scattering data were shown to differ unless extracted in a consistent way. Strictly speaking, however, the overlaps and the SFs are merely theoretical concepts that cannot be measured The overlaps should not be interpreted as the measurable probability amplitudes of finding a nucleon at some distance from the rest of the nucleus~\cite{BD87}. 
% Moreover, several authors recently have reiterated that neither SFs nor the underlying overlaps are observables since they depend on short-range effects and as such would change upon a unitary transformation of the Hamiltonian~\cite{FH02,FS10,MK10}. 
Nevertheless, until more advanced reaction models become available, spectroscopic overlaps and factors need to be calculated theoretically for a chosen Hamiltonian to enable the analysis of direct reaction data.
% facilitate the comparison of experiment with theory.
%``we should be satisfied with extracting limited information in the form imposed upon us by the theoretical framework of the analysis"~\cite{BD87} 
These concerns are less severe for the long-range scaling factors of overlaps, the asymptotic normalization coefficients (ANCs), which enter the cross section of peripheral direct reactions, as these coefficients remain invariant under finite-range unitary transformations of the Hamiltonian~\cite{MK10}. The spectroscopic overlaps and factors are still useful to assess the relative importance of clustering patterns in a given nucleus.

Typically, spectroscopic overlaps are approximated by shell-model or mean-field single-particle functions, or they are taken as solutions of single-particle (Woods-Saxon (WS)) potential wells with some commonly accepted but otherwise arbitrary potential parameters~\cite{Sa83,BG85,LT06}. Of a more recent vintage are some calculations attempting to account for contributions from missing model space, for example results obtained within the correlated basis function theory~\cite{BS07} or overlaps generated as solutions of an inhomogeneous equation with a shell-model source term~\cite{Ti10}. Realistic calculations of one-nucleon overlaps based on modern nuclear interactions are scarce and limited to light nuclei as they are complicated even in the simplest cases: in the $s$-shell, they were computed in the hyperspherical harmonics (HH) method~\cite{KR97,VK05}; some overlaps and SFs for $p$-shell nuclei were calculated within the no-core shell model~\cite{Na04,NB07} and the coupled-cluster method~\cite{JH10}. Of particular relevance for the present work are previous variational Monte Carlo (VMC) overlap calculations for nuclei up to $A=10$ (unpublished, available from~\cite{www-overlaps}) and a more recent VMC-based calculation of ANCs in light nuclei~\cite{NW11}. VMC overlaps have been used in some analyses of hadronic~\cite{WR05,WR05a,WS08,KA08,GB11} and leptonic~\cite{LW99} experiments.

This work provides a systematic study of one-nucleon spectroscopic overlaps, SFs, and ANCs in light nuclei calculated within the Green's function Monte Carlo (GFMC) method based on realistic two- and three-nucleon interactions. The GFMC method is designed to project the exact solutions out of trial wave functions by propagating them in imaginary time. The GFMC propagations are initiated by VMC wave functions. Over the years, the VMC/GFMC method has been found to accurately describe the structure and some reaction aspects of light  nuclei~\cite{PW01,PV02,PW04,Pi05,NP07,PP07,MP08}. Given our experience, GFMC is expected to improve VMC overlaps, especially in the $p$-shell. In this paper, we present GFMC overlap results for nuclei up to $A=7$; heavier nuclei up to $A \lesssim 10$ will be the subject of a forthcoming paper.

This paper is organized as follows. In \secref{sec QMC}, the theoretical framework is established, the VMC and GFMC methods are briefly reviewed, and all overlap-related quantities are defined. \secref{sec inner workings} contains technical and computational details and an error analysis of the GFMC overlap calculations. Results are presented and discussed in \secref{sec results}; final remarks and conclusions are given in \secref{sec conclusions}. 

%===================================== QMC =====================================
\section{Quantum Monte Carlo}
\label{sec QMC}

To calculate spectroscopic overlaps, we first construct wave functions $\Psi(J^\pi,T,T_z)$ for nuclei of interest as solutions of the nonrelativistic many-body Schr\"{o}dinger equation:
\beq
\label{eq schrodinger}
   H \Psi(J^\pi,T,T_z) = E \Psi(J^\pi,T,T_z),
\eeq
with $J^\pi$, $T$, and $T_z$ denoting the total spin-parity, isospin, and isospin projection, respectively. The $A$-body Hamiltonians used in this work have the form
\beq
\label{eq H}
   H = \sum_{i}^{A} K_i + \sum_{i<j}^{A} v_{ij} + \sum_{i<j<k}^A V_{ijk},
\eeq
where $K_i$ is the nonrelativistic kinetic energy and $v_{ij}$ and $V_{ijk}$ are two-nucleon ($2N$) and three-nucleon ($3N$) potentials, respectively. Most of the results presented in this paper have been obtained for a combination of Argonne $v_{18}$ (AV18)~\cite{WS95} $2N$ and Illinois-7 (IL7)~\cite{Pi08} $3N$ realistic interactions. For testing and benchmarking purposes, some calculations involved AV18 supplemented by the older Urbana IX (UIX)~\cite{PP95} $3N$ force. The kinetic energy of the center of mass is zero because the wave functions used in this work are translationally invariant.
% or a simpler, but still fairly realistic, Argonne $v^{\prime}_8$ (AV8$^\prime$)~\cite{PP97} two-body force alone.

%===================================== VMC =====================================
\subsection{Variational Monte Carlo}
\label{sec VMC}

The wave functions are constructed in two steps. First, a trial VMC approximation of $\Psi$, $\Psi_V$, is written and optimized by minimizing the energy expectation value as computed by Metropolis Monte Carlo integration~\cite{MR53}. Our latest variational functions have the form~\cite[Eq.~3.13]{WP00}
\beq
\label{eq psi VMC}
   |\Psi_V\rangle = 
      {\cal S} \prod_{i<j}^A
      \left[1 + U_{ij} + \sum_{k\neq i,j}^{A}\tilde{U}^{3N}_{ijk} \right]
      |\Psi_J\rangle.
\eeq
The Jastrow wave function, $\Psi_J$, is a fully antisymmetric 
% single-particle-like 
state having the desired $J^\pi,T,T_z$ quantum numbers of the state of interest; the $U_{ij}$ and $\tilde{U}^{3N}_{ijk}$ are two- and three-nucleon short-range correlation operators induced by dominant parts of $2N$ and $3N$ forces, respectively, while $\cal S$ is a symmetrization operator that restores the antisymmetry violated by the noncommuting character of $U_{ij}$ and $\tilde{U}^{3N}_{ijk}$. At long range, appropriate boundary conditions are imposed on $\Psi_V$~\cite{Wi91}.

The Jastrow wave function for $s$-shell nuclei consists of a simple product of central pair and triplet correlation functions and an antisymmetrized spin-isospin state. For $p$-shell nuclei, we use two types of $\Psi_J$: a shell-model-like trial function, which we call Type~I, and a clusterized version denoted Type~II. These two kinds of wave functions differ primarily in their treatment of correlations between $p$-shell particles and the long-range asymptotics. 

In Type~I trial functions, the single-particle radial functions of $p$-shell particles are exponentially decaying solutions of WS potential wells centered at the $s$-shell $\alpha$ core with potential parameters subject to a variational search. Pair correlations between s- and $p$-shell particles and among $p$-shell particles themselves are chosen to allow clusterization in the $p$-shell. 
% Details of Type~I functions for nuclei with $A=6,7$ are given in~\cite{PP97}.
Being less restrictive on long-range cluster decomposition compared with Type~II functions, Type~I functions are more widely applicable. Details of Type~I functions for nuclei with $A=6,7$ can be found in~\cite{PP97}.

Type~II functions that may be used for nuclei with low-lying cluster break-up thresholds impose a strict cluster-cluster asymptotic decomposition. Examples are the ground states of \nuc{6}{Li} and \nuc{7}{Li} that may asymptotically decouple into $\alpha+d$ and $\alpha+t$, respectively. In Type~II trial functions, the $p$-shell single-particle radial functions are again solutions of $p$-wave differential equations with potentials containing WS and Coulomb terms, but with an added Lagrange multiplier that turns on at long range. The role of the Lagrange multiplier is to impose the cluster boundary condition:
% on the product of the single-particle functions:
%
\beq
\label{eq Type II asymptotics}
   \Psi_V(r \rightarrow \infty)
   \propto
   \psi_{\alpha} \psi_x W_{-\eta, l+1/2}(2k r) / r,
\eeq
where $x=\,d$ or $t$ for \nuc{6}{Li} and \nuc{7}{Li}, respectively, and $r$ and $l$ are the relative $\alpha$-$x$ distance and orbital angular momentum. The Sommerfeld parameter $\eta$ and the wave number $k$ are related to the $\alpha$-$x$ separation energy in a given nucleus, which is set to its experimental value. Here, $W$ is the Whittaker function defined in Eq.~(13.1.33) of~\cite{AS72}. The correlations between $p$-shell particles in \nuc{6}{Li} and \nuc{7}{Li} are adopted from exact deuteron and variational triton wave functions, respectively. More details on Type~II trial functions are given in~\cite{NW01, No01, PP07}. 

For either type of trial function, a diagonalization is carried out to find the optimal values of mixing parameters for states of different spatial symmetries among $p$-shell particles~\cite{PW01}.

%==================================== GFMC =====================================
\subsection{Green's function Monte Carlo}
\label{sec GFMC}

Being variational solutions of \myeqref{eq schrodinger}, the VMC wave functions $\Psi_V$ can be improved further by action of the operator $\lim_{\tau \rightarrow \infty} \exp\left[-\left(H^{\prime} - E_{0}\right)\tau\right]$, which, for a given set of quantum numbers, projects out the exact lowest-energy state $\Psi_0$ of a possibly simplified version $H^{\prime}$ of the desired Hamiltonian $H$. In practice, the operator is applied in small increments of the imaginary time $\tau$ up to some finite value to yield a propagated wave function $\Psi(\tau)$:
\beq
\label{eq GFMC propagation}
   \Psi(\tau) = e^{-\left(H^{\prime} - E_{0}\right)\tau} \Psi_{V}.
\eeq
Obviously, $\Psi(\tau=0)=\Psi_V$ and $\Psi(\tau \rightarrow \infty)=\Psi_0$. In practice, \myeqref{eq GFMC propagation} is turned into an integral equation involving Green's functions with integrations performed by Monte Carlo methods, hence the name Green's function Monte Carlo. The energy $E_0$ is an approximate guess for the true eigenenergy corresponding to $\Psi_0$.

For a given state of the nucleus, quantities of interest are evaluated in terms of a ``mixed'' estimate between $\Psi_V$ and $\Psi(\tau)$:
\beq
\label{eq mixed estimate}
   \langle O(\tau) \rangle_M 
   =
   \frac{ \langle \Psi(\tau) | O |\Psi_V \rangle }
        { \langle \Psi(\tau) |    \Psi_V \rangle }.
\eeq
The desired expectation values would have $\Psi(\tau)$ on both
sides; by writing $\Psi(\tau) = \Psi_{T} + \delta\Psi(\tau)$ and neglecting
terms of order $[\delta\Psi(\tau)]^2$, we obtain the approximate expression
\beq
\label{eq gfmc mean value}
   \langle O (\tau)\rangle
   =
   \frac{ \langle \Psi(\tau)| O |\Psi(\tau) \rangle }
        { \langle \Psi(\tau)    |\Psi(\tau) \rangle }
   \approx
   \langle O (\tau)\rangle_M 
   + 
   [\langle O (\tau)\rangle_M - \langle O \rangle_V],
\eeq
where $\langle O \rangle_V$ is the variational expectation value. The expectation value of $H'$ is an exception to \myeqref{eq gfmc mean value} since for this operator the mixed estimate already gives the correct value~\cite{PW01}. For off-diagonal matrix elements between two different $A$-body nuclear states, \myeqref{eq gfmc mean value} can be generalized to a form involving two mixed estimates~\cite{PP07,MP08}. 

We note that the GFMC propagator involves a simplified Hamiltonian $H'$ = AV8$'$ +  $H'_{3N}$ based on a reprojection AV8$'$ of the full AV18 two-body potential~\cite{PP97}. Therefore, the GFMC wave functions are really eigenstates of $H'$ rather than $H$, which, in general, should not be a problem given that $H'$ is a good approximation of $H$. In this work, the three-nucleon part $H'_{3N}$ is either UIX or IL7, with the strength of their central repulsive parts adjusted so that $\langle H' \rangle \approx \langle H \rangle$.
% or the $3N$ part is omitted completely in pure AV8$'$ calculations. 
Although energies can be corrected perturbatively by adding $\langle H - H' \rangle$ to $\langle H' \rangle$, this kind of correction is not possible for other observables. In what follows, we will be referring to the desired Hamiltonian $H$ remembering that, in fact, we use $H'$ in GFMC propagations.

For a detailed description of the nuclear GFMC method, see the review~\cite{PW01} and references therein.

%======================= overlaps, spec. factors and ANCs ======================
\subsection{Spectroscopic overlaps and factors, and asymptotic normalization coefficients}
\label{sec overlaps theory}

Let us consider a decomposition of an $A$-body nucleus, the parent, into the core $C$ and a valence particle $v$ (a proton or a neutron), namely $A \rightarrow C + v$. The terms ``core" and ``valence" are used in a figurative sense considering that the parent wave functions are explicitly antisymmetrized states of indistinguishable particles. Then, a one-nucleon spectroscopic overlap is defined as a projection of a parent state onto an antisymmetrized core-valence product form:
\beqa
\label{eq overlap antisym}
   R(\alpha,\gamma,\nu;r)
   & = &
   C_{T_C T_{z,C} tt_z}^{T_A T_{z,A}} O(\alpha,\gamma,\nu;r)
   \\
   & \equiv &
   \left\langle
      \mathcal{A}_{Cv}
      \left[
         \left[
            \Psi(\gamma;\xi_C) 
            \otimes 
            \mathcal{Y}(\nu;\hat{r}_{Cv},\sigma_v,\tau_v)
         \right]_{J_A}
         \frac{ \delta( r - r_{Cv} ) }
              { r_{Cv}^2 }
      \right]
      \bigg|
      \Psi(\alpha;\xi_A)
   \right\rangle,
   \nonumber
\eeqa
where $C_{T_C T_{z,C} tt_z}^{T_A T_{z,A}}$ is an isospin Clebsch-Gordan coefficient; $\alpha \equiv \{A,J_A^{\pi},T_A,T_{z,A} \}$, $\gamma \equiv \{C,J_C^{\pi},T_C,T_{z,C} \}$, and $\nu \equiv \{v,l,s, j,t,t_z\}$ are composite indexes for the parent, the core, and the valence particle, respectively; $\xi$ comprise all degrees of freedom in a given nucleus; $\vec{r}_{Cv}$ is the vector extending from the core's center of mass to the valence particle; and $\mathcal{A}_{Cv}$ is the core-valence antisymmetrizer: $\mathcal{A}_{Cv} = (1/\sqrt{A}) \sum_{1}^{A-1} (-1)^p P_{Cv}$ with $P_{Cv}$ being the permutation operators exchanging the valence particle with those inside the core and $p$ being the permutation parity $\pm1$.
% An overlap channel is specified by an $\{\alpha,\gamma,\nu\}$ set. 
The symbols $A$ and $C=A-1$ also denote the nuclear mass numbers, and the  subscripts $A$ and $C$ may be dropped when convenient. In \myeqref{eq overlap antisym}, the parent and core states are assumed to be normalized to unity, and the integral is done over all $3A$ dimensions. The valence angle-spin-isospin function is defined as
\beq
\label{eq valence part}
   \mathcal{Y}(\nu;\hat{r}_{Cv},\sigma_v,\tau_v)
   \equiv
   \left[ 
      Y_l(\hat{r}_{Cv})
      \otimes
      \chi_s (\sigma_v)
   \right]_j
   \chi_{t,t_z}(\tau_v),
\eeq
where $Y_l$ is a spherical harmonic, $\chi_{s}(\sigma_v)$ and 
$\chi_{t,t_z}(\tau_v)$ denote the valence spin and isospin states with $s=t=1/2$, and $t_z$ gives the valence isospin projection $\pm 1/2$. Taking into account the antisymmetry of the parent state, one can turn the integral in \myeqref{eq overlap antisym} into a simpler form:
\beq
\label{eq overlap direct}
   R(\alpha,\gamma,\nu;r)
   =
   \sqrt{A} 
   \left\langle
      \left[
         \Psi(\gamma) \otimes \mathcal{Y}(\nu)
      \right]_{J_A}
      \left|
      \frac{ \delta( r - r_{Cv}) }
           { r_{Cv}^2 }
      \right|
      \Psi(\alpha)
   \right\rangle.
\eeq
By definition \eqref{eq overlap antisym} and by using translationally invariant wave functions $\Psi$, the overlap functions $R$ are translationally invariant. For a given parent-core combination, different angular momentum channels will be denoted $l_j$.

The theoretical SF is then defined as the norm of the overlap:
\beq
\label{eq SF}
   S(\alpha,\gamma,\nu)
   =
   \int \left| R(\alpha,\gamma,\nu;r) \right|^2 r^2 \ud r.
\eeq
Among other sum rules, all possible (proton, neutron) SFs for a given state of the parent nucleus add up to the parent's number of nucleons (protons, neutrons)~\cite{Cl73}. Our definition of
%= I decided to omit "overlaps" because to make our overlaps fully consistent
%= with \cite{Ti10} requires an extra angular-momentum Clebsch-Gordan due to
%= different coupling orders: we use core * valence, \cite{Ti10} uses 
%= valence * core
% overlaps and 
SFs is consistent with those of some other works, for example~\cite{Ti10}
%= there seems to be a typo in Ken's paper in Eq.1 where his coupling order
%= is core * (spin * Y). I think it should have been core * (Y * spin) because
%= he was using the same VMC code to generate the valence part as I did.
% \cite{NW11}
but differs from others~\cite{CK67, Sa83} by inclusion of the isospin Clebsch-Gordan coefficient in \myeqref{eq overlap antisym} into overlaps and consequently into SFs (these works define SFs as norms of $O(r)$ in \myeqref{eq overlap antisym}).

To calculate the overlap functions $R$ within GFMC, one can, neglecting terms of order $\left[ \Psi(\alpha;\tau) - \Psi_V(\alpha) \right]^2$ and $\left[ \Psi(\gamma;\tau) - \Psi_V(\gamma) \right]^2$, derive the following expression similar to Eq.~(19) in~\cite{PP07}:
\beqa
\label{eq overlap GFMC}
   R(\alpha,\gamma,\nu;r;\tau)
   & = &
   \sqrt{A} \,
   \frac{ \left\langle
             \left[
                \Psi(\gamma;\tau) \otimes \mathcal{Y}(\nu)
             \right]_{J_A}
             \left|
             \frac{ \delta( r - r_{Cv}) }
                  { r_{Cv}^2 }
             \right|
             \Psi(\alpha;\tau)
          \right\rangle
        }
        { \sqrt{ \left\langle \left| \Psi(\gamma;\tau) \right|^2 \right\rangle}
          \,
          \sqrt{ \left\langle \left| \Psi(\alpha;\tau) \right|^2 \right\rangle}
        }
   \nonumber \\
   & \approx &
   \langle R(\alpha,\gamma,\nu;r;\tau) \rangle_{M_A} +
   \langle R(\alpha,\gamma,\nu;r;\tau) \rangle_{M_C} -
   \langle R(\alpha,\gamma,\nu;r) \rangle_V,
\eeqa
where
\beqa
   \label{eq overlap mixed A}
   \langle R(\alpha,\gamma,\nu;r;\tau) \rangle_{M_A}
   & = &
   \sqrt{A} \,
   \frac{ \left\langle
             \left[
                \Psi_V(\gamma) \otimes \mathcal{Y}(\nu)
             \right]_{J_A}
             \left|
             \frac{ \delta( r - r_{Cv}) }
                  { r_{Cv}^2 }
             \right|
             \Psi(\alpha;\tau)
          \right\rangle
        }
        { \left\langle 
             \Psi_V(\alpha) | \Psi(\alpha;\tau)
          \right\rangle 
        }
   \sqrt{ \mathcal{N} },
   \\
   \label{eq overlap mixed A-1}
   \langle R(\alpha,\gamma,\nu;r;\tau) \rangle_{M_C}
   & = &
   \sqrt{A} \,
   \frac{ \left\langle
             \left[
                \Psi(\gamma;\tau) \otimes \mathcal{Y}(\nu)
             \right]_{J_A}
             \left|
             \frac{ \delta( r - r_{Cv}) }
                  { r_{Cv}^2 }
             \right|
             \Psi_V(\alpha)
          \right\rangle
        }
        { \left\langle 
             \Psi(\gamma;\tau) | \Psi_V(\gamma)
          \right\rangle 
        }
   \sqrt{ \frac{ 1           }
               { \mathcal{N} } },
   \\
   \label{eq overlap VMC}
   \langle R(\alpha,\gamma,\nu;r) \rangle_V
   & = &
   \sqrt{A} \,
   \frac{ \left\langle
             \left[
                \Psi_V(\gamma) \otimes \mathcal{Y}(\nu)
             \right]_{J_A}
             \left|
             \frac{ \delta( r - r_{Cv}) }
                  { r_{Cv}^2 }
             \right|
             \Psi_V(\alpha)
          \right\rangle
        }
        { \left\langle 
             \Psi_V(\alpha) | \Psi_V(\alpha)
          \right\rangle 
        }
   \sqrt{ \mathcal{N} }
\eeqa
with a self-normalization factor:
\beq
\label{eq self-norm}
   \mathcal{N}
   = 
   \frac{ \left\langle \Psi_V(\alpha) | \Psi_V(\alpha) \right\rangle }
        { \left\langle \Psi_V(\gamma) | \Psi_V(\gamma) \right\rangle }.
%= just a different notation
%   \frac{ \left\langle \left| \Psi_V(\alpha) \right|^2 \right\rangle }
%        { \left\langle \left| \Psi_V(\gamma) \right|^2 \right\rangle }.
\eeq
The overlap function $\langle R \rangle_V$ in \myeqref{eq overlap VMC} is a pure variational estimate, to be called a VMC overlap, whereas the other two estimates, $\langle R \rangle_{M_A}$ in \myeqref{eq overlap mixed A} and $\langle R \rangle_{M_C}$ in \myeqref{eq overlap mixed A-1}, involve combinations of GFMC-propagated and VMC wave functions, and as such will be called $A$- and ($A$-1)-body mixed overlaps and referred to by the corresponding GFMC-propagated nucleus. For each estimate Eqs.~\eqref{eq overlap mixed A}-\eqref{eq overlap VMC}, one can define a SF similar to \myeqref{eq SF}. The GFMC-extrapolated overlap $R$ in \myeqref{eq overlap GFMC} will be called a GFMC overlap. Obviously, at $\tau=0$, $R = \langle R \rangle_V = \langle R \rangle_{M_A} = \langle R \rangle_{M_C}$.

Overlap functions $R$ satisfy a one-body Schr\"{o}dinger equation with the  appropriate source terms~\cite{BG85}. Asymptotically, for $r \rightarrow \infty$, these source terms contain core-valence Coulomb interaction at most, and hence the long-range part of overlap functions for parent states below core-valence separation thresholds is proportional to a Whittaker function $W_{-\eta,l+1/2}$:
\beq
\label{eq ANC}
   R(\alpha,\gamma,\nu;r)
   \xrightarrow{r \rightarrow \infty}
   C(\alpha,\gamma,\nu)
   \frac{ W_{-\eta,l+1/2} (2kr) }{r},
\eeq
where $\eta = Z_C Z_v (e^2/\hbar c) \sqrt{ \mu c^2 / 2B}$ depends on proton numbers $Z_C$ and $Z_v$, and the core-valence reduced mass $\mu$ and separation energy $B$ (positive for parent states below core-valence separation thresholds). The wave number $k$ is defined as $\sqrt{2\mu B}/\hbar$, and $l$ is the orbital angular momentum from \myeqref{eq valence part}. Note that $W$, $\mu$, $B$, and $k$ implicitly carry channel labels $\alpha$, $\gamma$, and $\nu$. The Whittaker function is defined in Eq.~(13.1.33) of~\cite{AS72}, and it has an (approximately) exponentially decaying tail. The proportionality constant $C(\alpha,\gamma,\nu)$ in \myeqref{eq ANC} is the ANC.

To provide a convenient parametrization of overlaps suitable for reaction calculations and to extract ANCs, we perform $\chi^2$-fits of the overlaps by the eigenstates of a single-particle-like Hamiltonian $-[\hbar^{2}/2\mu]\triangle + V(r)$ containing a WS plus  spin-orbit (so) plus Coulomb (Coul) potential
\beqa
\label{eq fitting potential}
   V(r)
   & = &
   V_{WS} 
   \left[ \frac{1}{1+\exp((r-R_{WS})/a_{WS})} - \beta\exp(-(r/\rho)^2) \right]
   +
   \nonumber \\
   &&
   \left( 4\vec{l} \cdot \vec{s} \right) 
   \frac{ V_{so} }{ r }
   \frac{\ud}{\ud r} \left[ \frac{1}{1+\exp((r-R_{so})/a_{so})} \right]
   + V_{Coul}
\eeqa
with
\beq
\label{eq fitting potential Coulomb part}
   V_{Coul} 
   =
   \left\{
      \begin{array}{ll}
         Z_C Z_v e^2 / r 
         &
         \qquad : r \geq R_{Coul}
         \\
         Z_C Z_v e^2 \left[ 3 - (r/R_{Coul})^2 \right] / [2R_{Coul}]
         &
         \qquad : r < R_{Coul}
      \end{array}
   \right.
\eeq
%
% In addition to $V(r)$, there is also the standard core-valence centrifugal barrier. 
The central WS part of the potential includes a Gaussian ``wine-bottle" term to provide an additional flexibility at short range. The depths $V_{WS}$ and $V_{so}$ are in the same units, the factor 4 in the spin-orbit part is approximately twice the square of the pion Compton wave length $\hbar/(m_{\pi}c)$ in fm, and $\vec{l}$ and $\vec{s}$ are operators of the core-valence orbital angular momentum and of the valence spin, respectively, both in units of $\hbar$. The Coulomb radius is $R_{Coul}=2$~fm.
% and \ul{$e^2$=? Steve?? we also should give our values for hbar/m and fine-structure constant or some other combination of these}. 
The potential parameters are varied freely to provide the best fit under the constraint of the eigenenergy being equal to the desired value $-B$. This fitting procedure provides good overlap fits at short and medium distances; at large distances, the fits have the desired form shown in \myeqref{eq ANC}. 
% overlap tail should fall off as prescribed by the desired core-valence separation energy $B$. 
These overlap parametrizations can be used in reaction codes such as \textsc{Ptolemy}~\cite{MP78} or \textsc{Fresco}~\cite{Th88}, and they will be refered to as WS fits throughout the paper.

\section{Inner workings of overlap calculations and error analysis}
\label{sec inner workings}

In this section, we elaborate on some technical aspects of the overlap calculations, show several detailed examples, and assess systematic errors of quantities being computed. The discussion of results is given in \secref{sec results}.

The VMC overlap $\langle R \rangle_V$ in \myeqref{eq overlap VMC} is calculated on a random walk guided by the parent's $|\Psi_V(\alpha)|^2$. On the same walk, we also evaluate the self-normalization factor from \myeqref{eq self-norm}; to extend the denominator of $\mathcal{N}$ into the $3A$-dimensional space being sampled,  $\Psi_V(\gamma)$ is replaced by $\Psi_V(\gamma)\times p(\vec{r}_{Cv})$, where $p$, normalized to unity as $\int p^2(\vec{r}_{Cv}) \ud \vec{r}_{Cv}$=1, acts as a single-particle function of a virtual valence particle taken with respect to the core's center of mass. In principle, any function can be used for $p$ as long as the product form
$|\Psi_V(\gamma)\times p|^2$ is a reasonably good approximation of the $A$-body sampling probability density in order to yield reasonably small statistical errors on the denominator of $\mathcal{N}$.
% in other words fluctuations of local values of $|\Psi_{\gamma,V}\times p_v|^2 / |\Psi_{\alpha,V}|^2$ around their mean value should be reasonably small. 
In practice, we find it sufficient to use purely radial ($s$-wave) functions $p(r_{Cv})$ of either a Gaussian shape or one generated by a single-particle-like Hamiltonian containing a WS potential well with parameters adjusted to minimize the error on the self-normalization factor. The WS-generated single-particle functions have the advantage of better approximating the $A$-body sampling density at large core-valence distances, especially when the potential depth is set to approximately reproduce the (experimental or VMC) core-valence separation energy. In the present work, the self-normalization factors $\mathcal{N}$ were calculated (after accounting for auto-correlations between local samples)  with an accuracy of the order of $0.1\%$ or better. 
%The self-normalization factor is calculated only once to be used in all Eqs.~\eqref{eq overlap VMC}-\eqref{eq overlap mixed A-1}.

The two mixed overlaps in \myeqref{eq overlap mixed A} and \myeqref{eq overlap mixed A-1} are calculated on random walks guided by GFMC propagations for the corresponding nucleus. For the ($A$-1)-body mixed overlap, the GFMC sampling density only spans a 3($A$-1)-dimensional subspace of the full $3A$-dimensional space of the parent nucleus; we draw the position of the valence nucleon from the same single-particle density $p^2(\vec{r}_{Cv})$ used in the computation of the variational self-normalization factor $\mathcal{N}$.

Generating new VMC or GFMC samples is computationally more expensive than evaluating a sample's contribution to Eqs.~\eqref{eq overlap mixed A}-\eqref{eq overlap VMC}. Therefore, in order to improve the computational efficiency, each VMC or GFMC sample is used several times: in \myeqref{eq overlap mixed A} and \myeqref{eq overlap VMC}, we consider all $A$ cyclic particle permutations to rotate the valence particle over all possible positions within the parent nucleus, while in \myeqref{eq overlap mixed A-1} the position of the valence particle is drawn several ($A$) times from its single-particle distribution $p^2$ described above for each core's GFMC sample.

We illustrate the method with the \overlap{\nuc{3}{H}}{\nuc{4}{He}} overlap for AV18+IL7 taken as a typical example of overlaps in the $s$-shell. First, the VMC overlap is computed by \myeqref{eq overlap VMC}. Then, the wave functions of the parent and the core are independently propagated and the two mixed and the GFMC overlaps are computed in each radial bin following Eqs.~\eqref{eq overlap mixed A}, \eqref{eq overlap mixed A-1}, and \eqref{eq overlap GFMC}. The corresponding SFs are plotted in \figref{fig H3p IL7 SFs}. In all figures, the quantities 
corresponding to the variational, the ($A$-1)- and the $A$-body mixed, and to the GFMC overlaps are plotted as black, red, blue, and green, respectively.

\bfig[t]
   \includegraphics[width = 0.48\textwidth]
                   {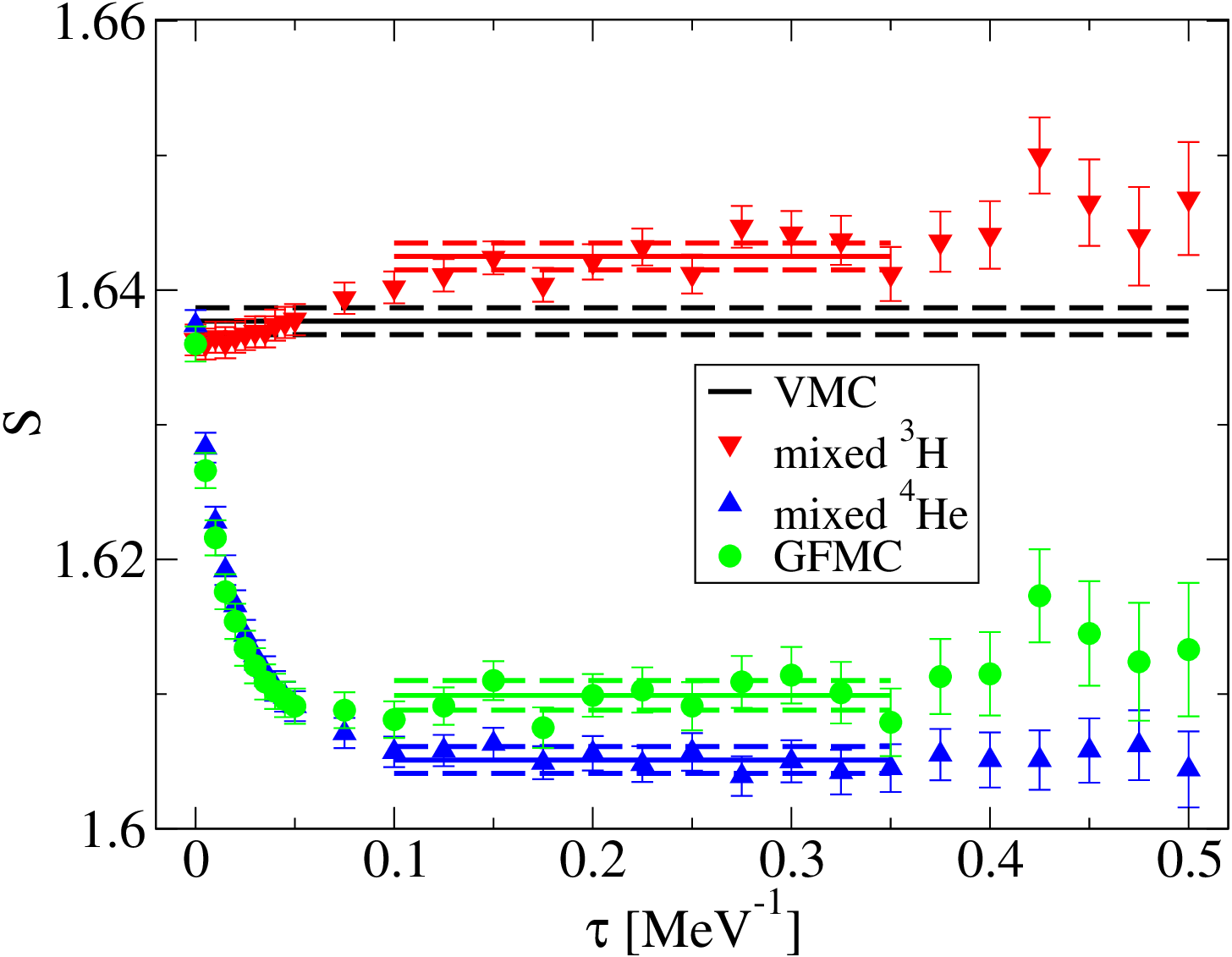}
   \caption{(Color online) Imaginary time evolution of SFs in the $s_{1/2}$ 
            channel of the \overlap{\nuc{3}{H}}{\nuc{4}{He}} overlap obtained 
            for the AV18+IL7 potential. Only statistical errors are shown. 
            Horizontal lines are SFs for VMC and time-averaged mixed and GFMC 
            overlaps, as described in the text.}
   \label{fig H3p IL7 SFs}
\efig

For $s$-shell nuclei, we do unconstrained propagations~\cite{PP97} and thus obtain essentially exact solutions for a given Hamiltonian $H'$ in \myeqref{eq GFMC propagation}. 
% The exponentially growing statistical noise due to the fermion sign problem is kept under control by propagating large sets ($\sim 12$ million) of Monte Carlo samples. 
Because of the fermion sign problem, the statistical noise grows with time, as can be inferred from \figref{fig H3p IL7 SFs}. Fortunately, 
% as for expectation values of energy and other observables~\cite{PW01}, 
GFMC quickly eliminates excited-state impurities from VMC wave functions resulting in a rapid convergence of mixed and GFMC overlaps and SFs.
% in \figref{fig H3p IL7 SFs}. 
For $s$-shell nuclei, GFMC propagations are fully converged long before $\tau = 0.5$~MeV$^{-1}$, at which point they are terminated to avoid the increasing statistical noise.
% due to the fermion sign problem creeping up.

In order to improve the statistical accuracy further, the ($A$-1)- and $A$-body mixed overlaps are separately averaged over time. These mixed time-averages are then combined with the VMC overlap to obtain the final time-averaged GFMC overlap from \myeqref{eq overlap GFMC}. In \figref{fig H3p IL7 SFs} and similar figures, the variational estimate, although computed at $\tau=0$~MeV$^{-1}$, is plotted across the full range of times, while the other solid lines give SFs for time-averaged mixed and GFMC overlaps with averaging done over the range of times indicated by the horizontal extent of these lines; the dashed lines show statistical errors. We have checked that our time-averaged results have little dependence on the exact time interval being averaged over as long as this interval is safely within the GFMC converged region and does not include large $\tau$'s for unconstrained propagations.

\bfig[t]
   \includegraphics[width = 0.48\textwidth]
                   {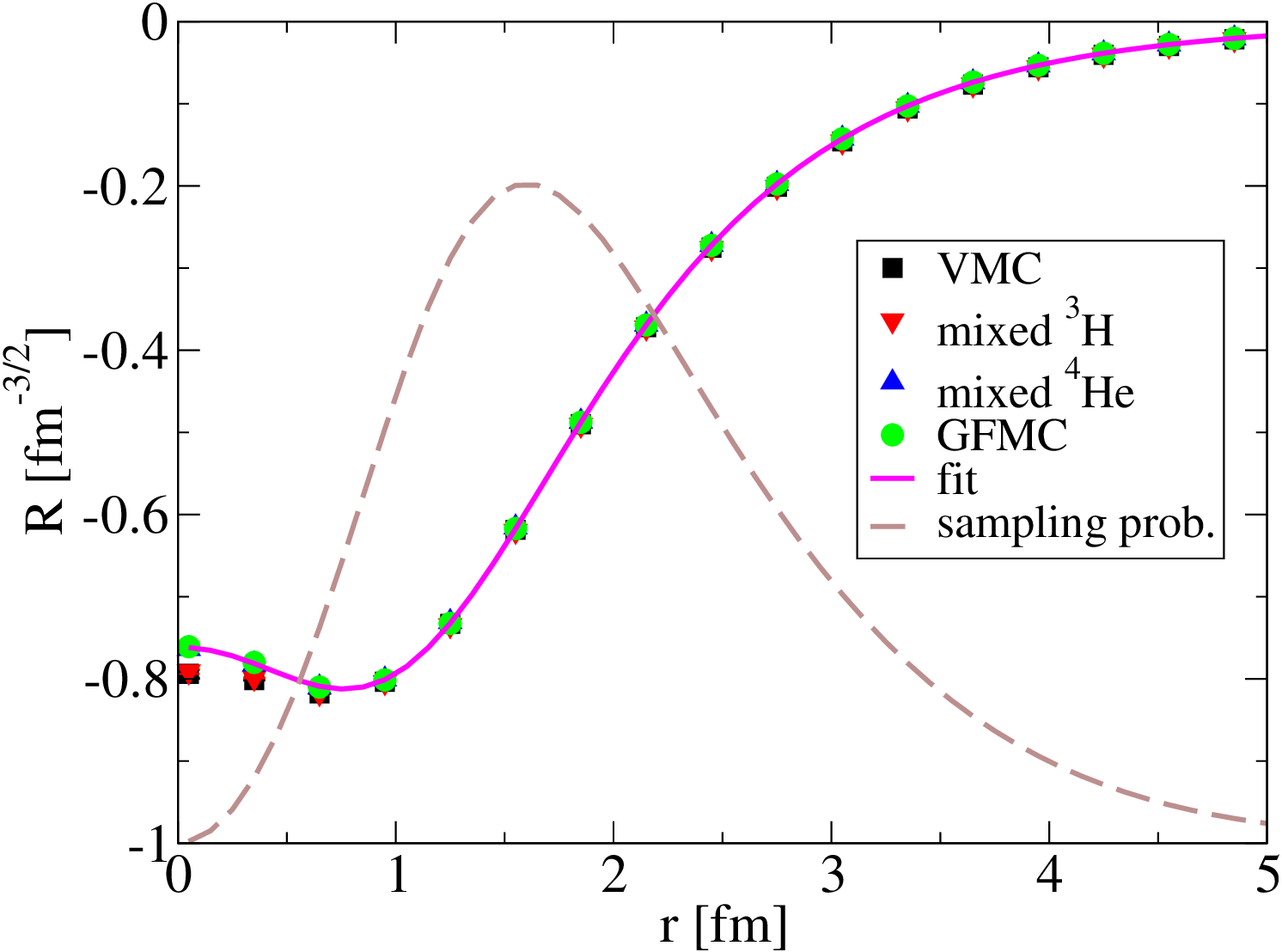}
   \hspace{0.02\textwidth}
   \includegraphics[width = 0.48\textwidth]
                   {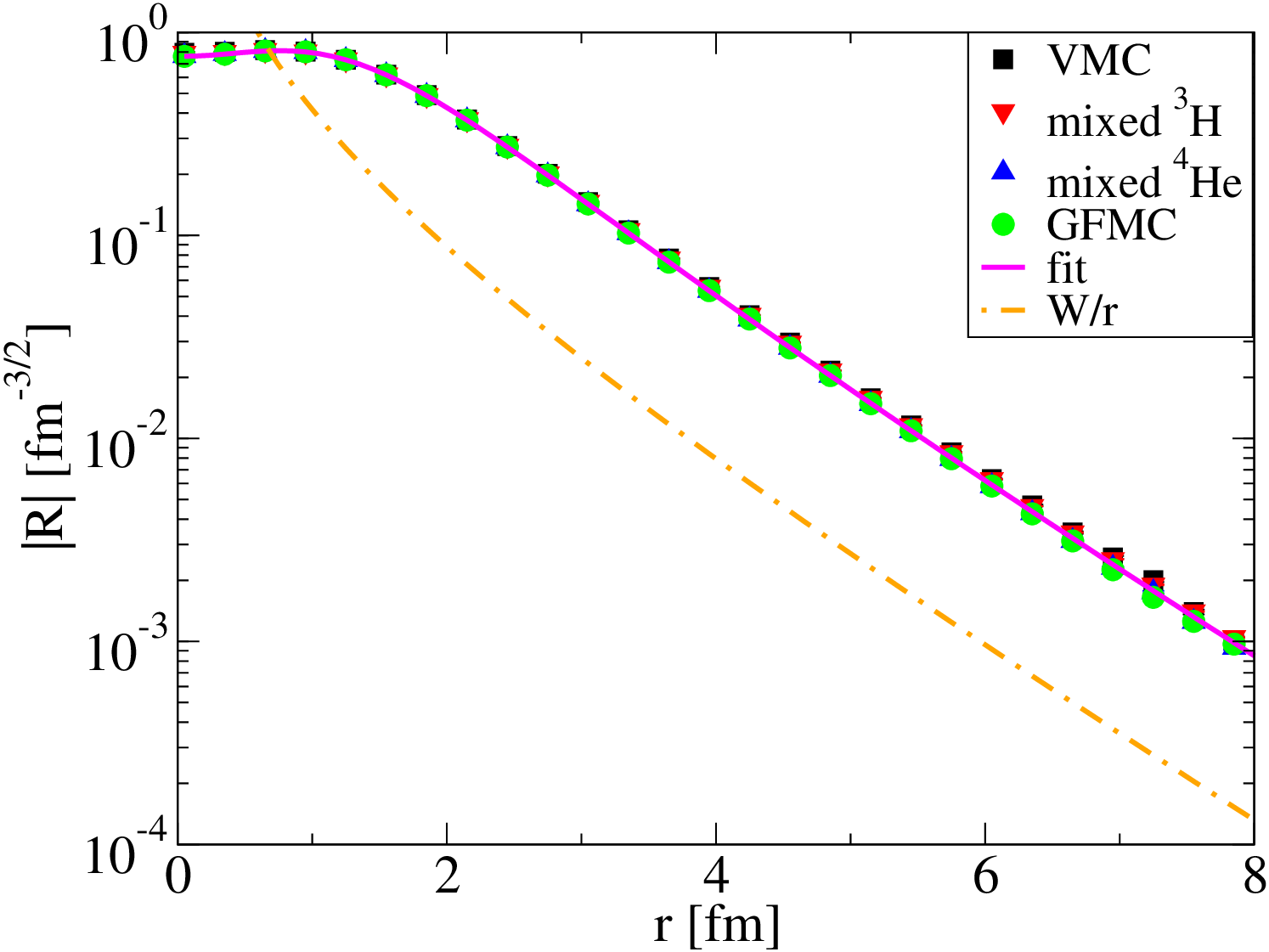}
   \caption{(Color online) Linear (left) and logarithmic (right) plots of VMC 
            and time-averaged mixed and GFMC $s_{1/2}$ 
            \overlap{\nuc{3}{H}}{\nuc{4}{He}} overlaps obtained for the AV18+IL7
            potential. Only statistical errors are shown. Also shown are a 
            WS fit to the GFMC overlap and the asymptotic Whittaker function 
            corresponding to the experimental separation energy in the right 
            panel. Superimposed in the linear plot is the sampling probability 
            (arbitrary scale).}
   \label{fig H3p IL7 overlaps}
\efig

In \figref{fig H3p IL7 overlaps}, the VMC and time-averaged mixed and GFMC \overlap{\nuc{3}{H}}{\nuc{4}{He}} overlaps are shown along with the asymptotic Whittaker function corresponding to the experimental separation energy (discussed below), a WS fit to the GFMC overlap, and the radial sampling probability. In general, the differences between $s$-shell VMC and time-averaged mixed and GFMC overlaps in a given overlap channel are very small because the starting VMC wave functions are already very good approximate solutions of \myeqref{eq schrodinger}; these small differences are reflected by small ($\sim$2\%) differences between VMC and GFMC SFs as illustrated in \figref{fig H3p IL7 SFs}. Statistical errors on SFs are small because the dominant contribution to \myeqref{eq SF} comes from the volume region well covered by the Monte Carlo sampling probability, also shown in \figref{fig H3p IL7 overlaps}. 
% Note that the sign of the overlaps depends on the detailed construction of our wave functions and on the isospin Clebsch-Gordan coefficient in \myeqref{eq overlap antisym}. For these reasons absolute overlap phases are somewhat arbitrary; however, relative phases of overlaps in different angular-momentum channels for a given parent-core combination are meaningful since they are important for interference effects as in the case of the $s_{1/2}$ and $d_{3/2}$ overlaps  in $A=3$ nuclei discussed in \secref{sec results}.

\bfig[b]
   \includegraphics[width = 0.48\textwidth]
                   {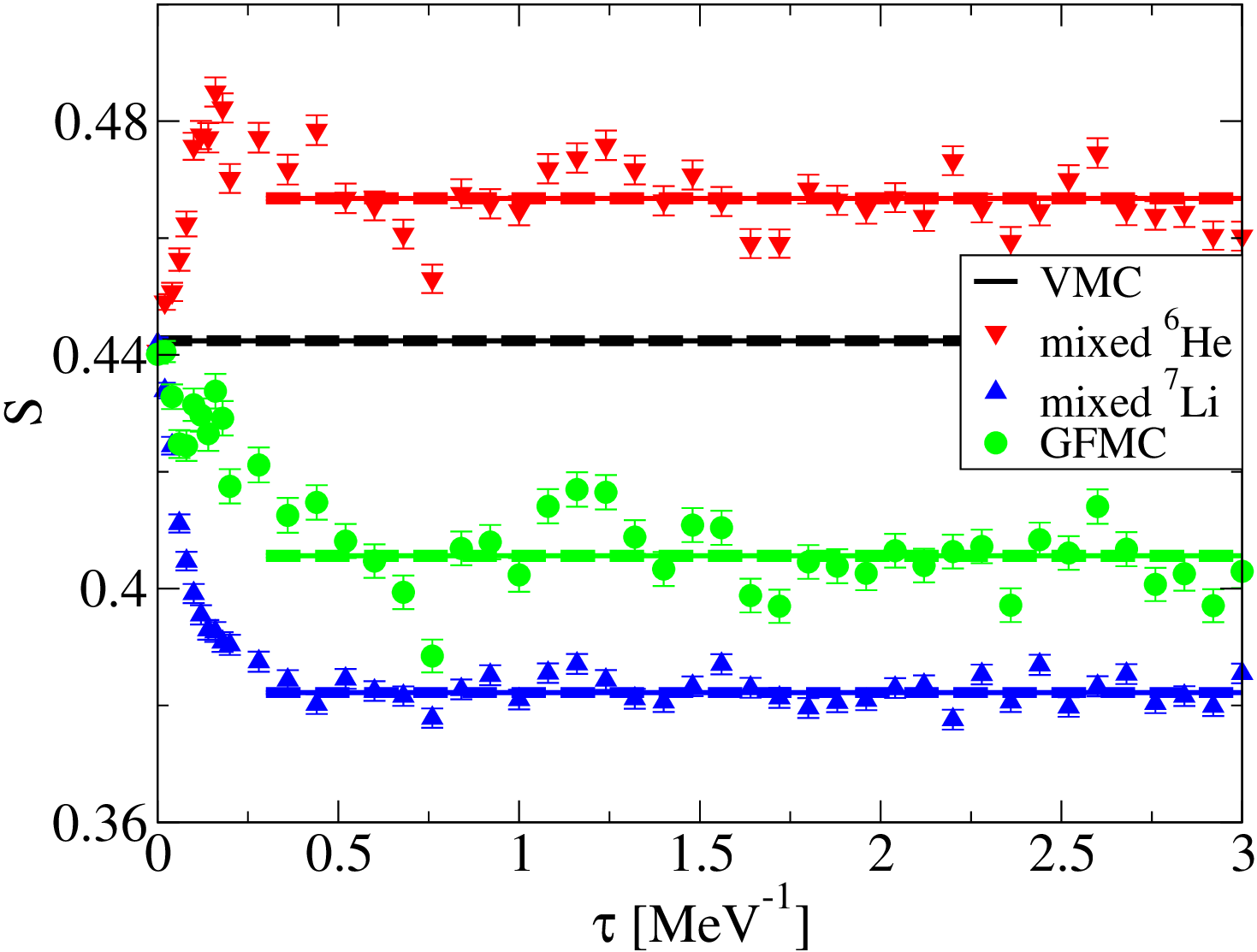}
   \caption{(Color online) Imaginary time evolution of SFs in the $p_{3/2}$ 
            channel of the \overlap{\nuc{6}{He}$(0^+)$}{\nuc{7}{Li}$(3/2^-)$} 
            overlap obtained for the AV18+IL7 potential. The GFMC wave functions
            originated from Type~I VMC wave functions. Only statistical errors 
            are shown. Horizontal lines are SFs for VMC and time-averaged mixed 
            and GFMC overlaps, as described in the text.}
   \label{fig he6p IL7 SFs}
\efig

Overlaps between $p$-shell nuclei are computed by using the same algorithm with some technical modifications. Compared with $s$-shell nuclei, the fermion sign problem becomes more severe in the $p$-shell, and we retreat to constrained path GFMC sampling~\cite{WP00}. Consequently, statistical errors are well under control, as illustrated in \figref{fig he6p IL7 SFs} for SFs in the $p_{3/2}$ channel of the \overlap{\nuc{6}{He}$(0^+)$}{\nuc{7}{Li}$(3/2^-)$} overlap. Given our experience indicating that mixed estimates of many observables tend to fluctuate more in the $p$-shell than in the $s$-shell, we chose to carry out propagations for longer times up to $\tau=3$~MeV$^{-1}$ to ensure full convergence in the $p$-shell. VMC, and time-averaged mixed and GFMC overlaps corresponding to \figref{fig he6p IL7 SFs} are shown in \figref{fig he6p IL7 overlaps}. The difference of about 8\% between the VMC and GFMC SFs reflects the change in the shape of the underlying overlaps. 

\bfig[t]
   \includegraphics[width = 0.48\textwidth]
                {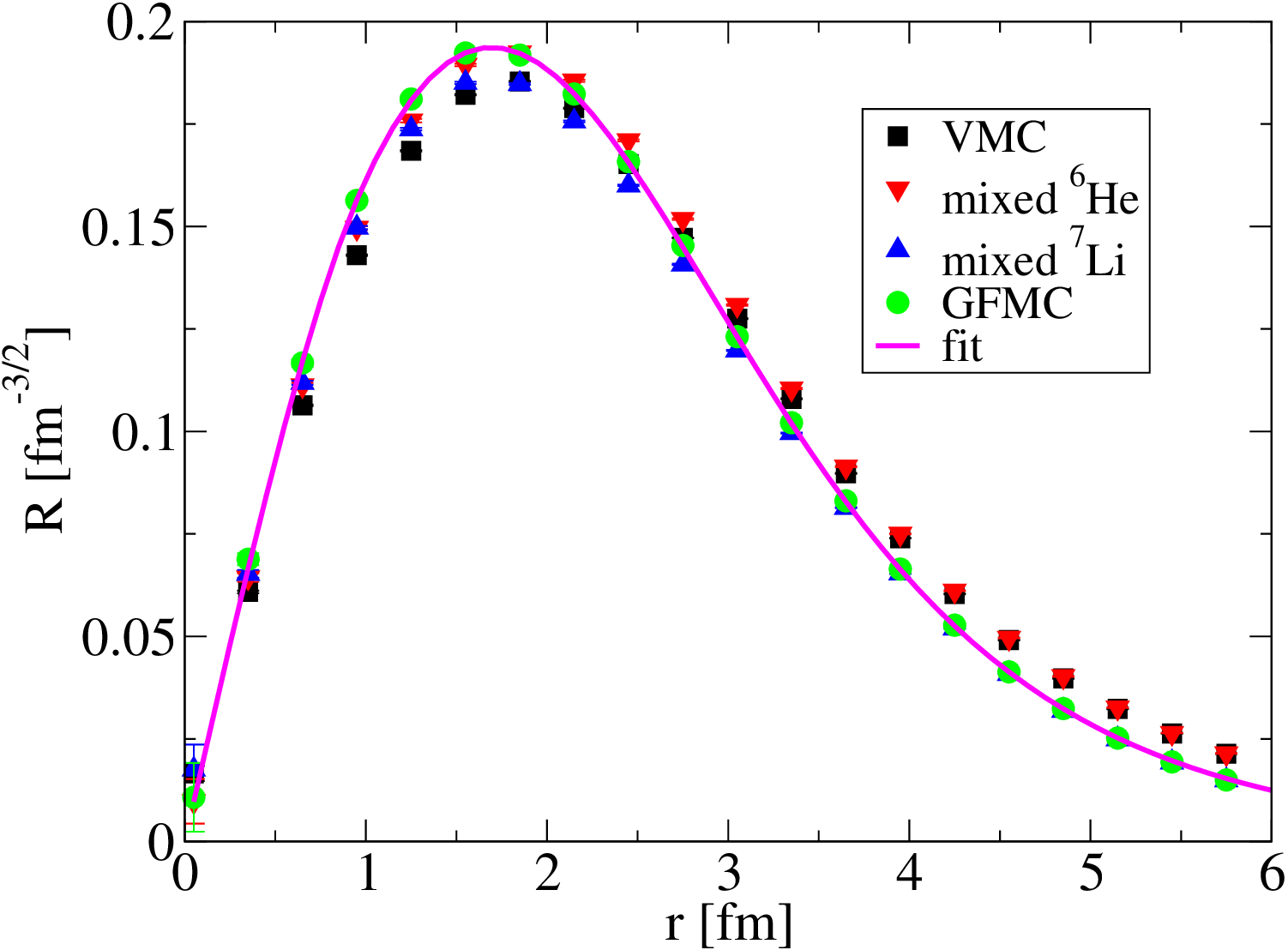}
   \hspace{0.02\textwidth}
   \includegraphics[width = 0.48\textwidth]
                {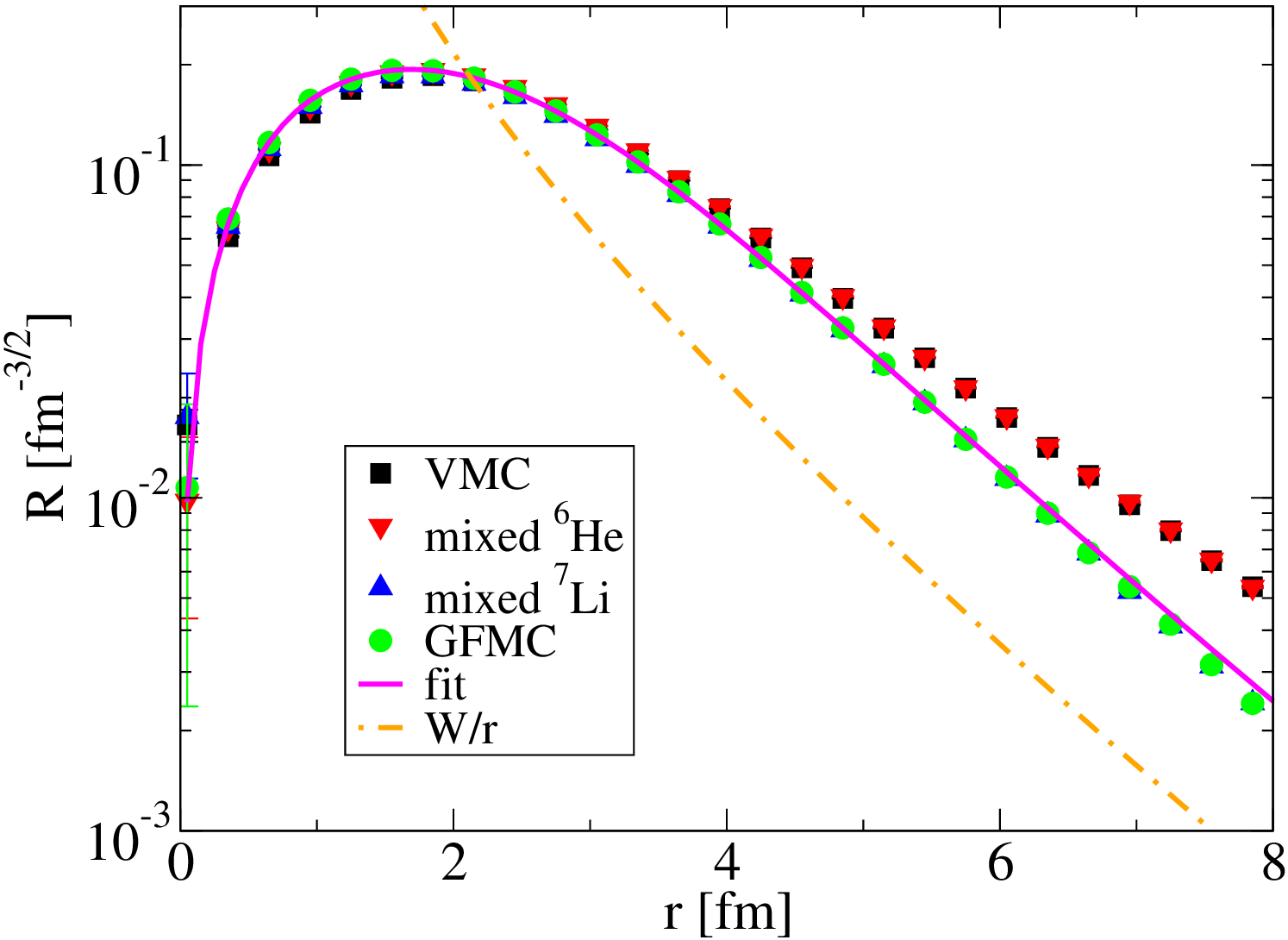}
   \caption{(Color online) Linear (left) and logarithmic (right) plots of VMC 
            and time-averaged mixed and GFMC $p_{3/2}$
            \overlap{\nuc{6}{He}$(0^+)$}{\nuc{7}{Li}$(3/2^-)$} overlaps obtained
            for the AV18+IL7 potential. The GFMC wave functions
            originated from Type~I VMC wave functions. Only statistical errors 
            are shown. Also shown are a WS fit to the GFMC overlap, and the 
            asymptotic Whittaker function corresponding to the experimental 
            separation energy in the right panel.}
   \label{fig he6p IL7 overlaps}
\efig

At short range, the GFMC overlap is a result of fine cancellations between VMC and mixed overlaps in \myeqref{eq overlap GFMC}. At large core-valence distances, the GFMC overlaps tend to follow the $A$-body mixed overlaps, while the ($A$-1)-body mixed overlaps usually stay close to the VMC ones, in agreement with our expectation that it is the long-range fall-off of the parent's wave function that primarily sets the tail of the overlap whereas the detailed structure of the core plays a less important role in this region although the core still needs to be described reasonably well. 
% The improvement of the asymptotic shape provided by GFMC compared to VMC in \figref{fig he6p IL7 overlaps} is one of the most dramatic ones in our calculations. 
From this point on, mixed overlaps and estimates will not be shown, and the time-averaged GFMC overlap will be called a GFMC overlap.

Because of the vanishing sampling probabilities at large distances (an example is shown in \figref{fig H3p IL7 overlaps}), reliable sampling of overlap tails requires very large Monte Carlo sets; we propagated about 12 million samples in the $s$-shell and about 250,000 for $p$-shell nuclei. As a consequence, statistical errors are small. It is therefore important to assess systematic errors in addition to statistical errors. The most notable sources of systematic errors are: the difference between the Hamiltonian $H'$ in the GFMC propagator in \myeqref{eq GFMC propagation} and the desired $H$ in \myeqref{eq schrodinger}, possible errors due to the constrained path sampling in the $p$-shell, and a residual dependence of GFMC results on starting VMC wave functions. In addition, ANCs depend on the separation energy used in their determination. We now attempt to place limits on such systematic errors.

For $s$-shell nuclei, we can compare our results with those obtained by the hyperspherical harmonics (HH) method~\cite{KR08,KR97,VK05,VK11}.
% , in particular, to updated and improved AV18+UIX HH results previously published in~\cite{KR97, VK05} and to those for Av8' alone~\cite{VK11}. 
For AV18+UIX, the agreement between HH and GFMC is very good, as can be seen from \figref{fig h2n and h3p AV18+UIX} for overlaps and from \tabref{tab SFs s-shell} for SFs in \secref{sec results}. Testing the bias due to $H'$ for $p$-shell overlaps is not possible since, to our knowledge, no $p$-shell overlaps have been published previously for the realistic interactions employed here. To test the dependence of GFMC SFs on VMC wave functions, we include in \tabref{tab SFs p-shell} results involving the ground state of \nuc{7}{Li}  obtained for several combinations of Type~I and II (see \secref{sec VMC}) trial wave functions. Although VMC SFs for different $\Psi_V$ may differ by as much as 20\%,
% depending on VMC wave functions,
% as in the case of $p_{3/2}$ overlaps between the ground states of \nuc{7}{Li} and \nuc{6}{Li}, 
GFMC reduces the spread to no more than 3\%. Hence, we estimate systematic errors on GFMC SFs to be no more than 2-3\%.

Typically, extracting ANCs directly from overlaps in \myeqref{eq overlap direct} is problematic for most many-body methods. First, these methods may not yield the correct exponential asymptotic form given in \myeqref{eq ANC}, as is the case of methods employing harmonic oscillator bases~\cite{Na04, NB07}. In variational methods, including VMC, it may be hard to impose the correct asymptotics while preserving short-range properties. This problem is less prevalent in integral methods of computing ANCs in which the desired asymptotics is imposed by definition regardless of the actual asymptotic shape of the wave functions involved~\cite{Ti10,NW11}. Second, ANCs extracted from \myeqref{eq ANC} depend on the separation energy $B$. To determine ANCs from overlaps in a fully consistent manner, one should use the separation energy $B_H$ obtained for a given Hamiltonian; on the other hand, for ANCs to be practically usable for reaction calculations, experimental values $B_{exp}$ should be used. Ideally, $B_H \approx B_{exp}$, a condition that is often violated in literature and therefore relaxed without error analysis for ANCs~\cite{NB07,Ti10,NW11}. We now discuss these aspects applied to GFMC.

\bfig[t]
   \includegraphics[width = 0.485\textwidth]
                   {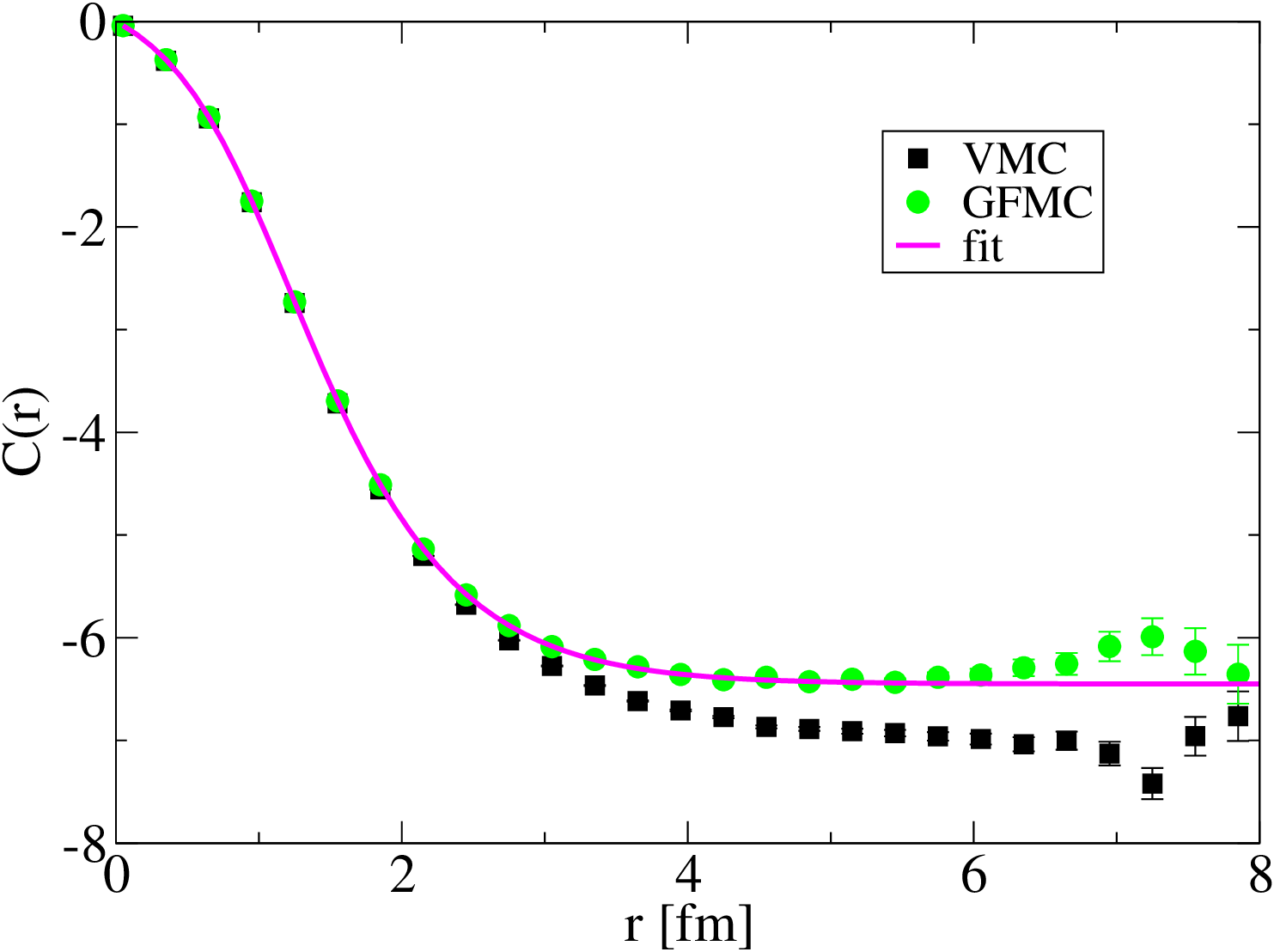}
   \hspace{0.02\textwidth}
   \includegraphics[width = 0.478\textwidth]
                   {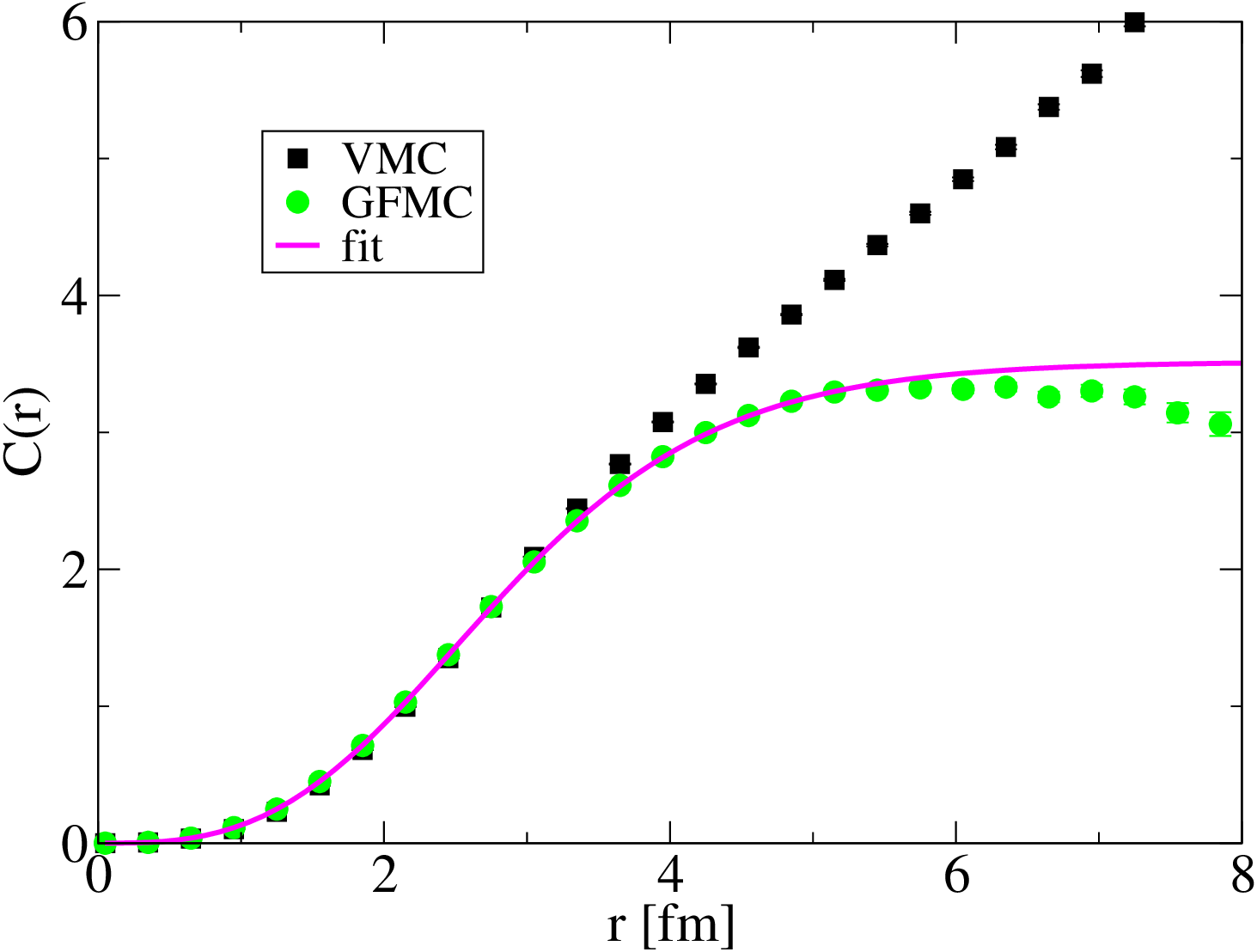}
   \caption{(Color online) The ratio $C(r)$ of VMC, GFMC, and fitted GFMC 
            overlaps to asymptotic Whittaker functions corresponding to 
            experimental core-valence separation energies. Left panel: 
            \overlap{\nuc{3}{H}}{\nuc{4}{He}} overlaps from \figref{fig H3p IL7 
            overlaps}; right panel:    
            \overlap{\nuc{6}{He}$(0^+)$}{\nuc{7}{Li}$(3/2^-)$} overlaps from
            \figref{fig he6p IL7 overlaps}.}
   \label{fig H3p he6p IL7 ANCs zoom-out}
\efig

It is gratifying to observe that the GFMC overlaps in \figref{fig H3p IL7 overlaps} and \figref{fig he6p IL7 overlaps} transition nicely from the interior into the asymptotic region and that in the computationally safe region ($r\lesssim 8$~fm) they follow closely the right asymptotic trend represented by the Whittaker functions. To illustrate this in more detail for overlaps from Figs.~\ref{fig H3p IL7 overlaps} and \ref{fig he6p IL7 overlaps}, we show in \figref{fig H3p he6p IL7 ANCs zoom-out} the ratios
\beq
\label{eq radial ANC}
   C(r)
   \equiv
   R(r) / \left[ W_{-\eta,l+1/2}(2kr) / r \right].
\eeq
%
%with channel indexes $\{\alpha,\gamma,\nu\}$ suppressed. 
Ideally,  $C(r\rightarrow \infty)$ should approach $C$ from \myeqref{eq ANC}, the ANC. As can be seen in the figure, the GFMC curves may not flatten out perfectly at larger distances where the method being driven by $H'$ in \myeqref{eq GFMC propagation} may break down because of insufficient contributions of the asymptotic part of the configuration space to the energy. To correct this imperfection, the GFMC overlaps are extended to larger distances by means of WS fits (\myeqref{eq fitting potential}). For $s$-shell overlaps, $C(r)$ flatten out at $r$$\sim$2.5--3~fm where $C(r)/C$$\sim$0.9, and they are almost fully converged at $r$$\sim$5~fm; for $p$-shell overlaps, $C(r)/C$$\sim$0.9 is realized at $r$$\sim$5~fm. Therefore, the overlaps are (almost) asymptotic at about 5~fm, which is the upper radial limit used in our fitting procedure. ANCs are then determined by applying \myeqref{eq radial ANC} to GFMC overlap fits at large distances. We have checked that including the region beyond 5~fm in the fitting procedure does not result in significant changes of ANCs.

In GFMC, the binding energy is thought to be computed with systematic errors of about 1--2\%~\cite{PP97, WP00, PW01}. Most GFMC overlaps in this work were obtained for the AV18+IL7 potentials. For this potential, GFMC energies in \tabref{tab energies} differ by less than 1\% from experimental values, in most cases the difference is less than 0.5\%. These deviations, though being small fractions of total binding energies, may translate into bigger (fractional) deviations on core-valence separation energies, as can be seen from Tables~\ref{tab separation energies s-shell} and \ref{tab separation energies p-shell}. Even then, however, the difference between experimental and GFMC separation energes is $\lesssim 3\%$, with a noticeable exception being the \overlap{\nuc{6}{He}$(0^+)$}{\nuc{7}{He}$(3/2^-)$} overlap involving a particle-unbound \nuc{7}{He} (for this overlap, the ANC is not a meaningful quantity). Since GFMC wave functions are really eigenstates of $H^\prime$ rather than $H$, it is equally important to observe in Tables~\ref{tab separation energies s-shell} and \ref{tab separation energies p-shell} that separation energies for these Hamiltonians are about the same ($H'$ was adjusted to get $\langle H' \rangle \approx \langle H \rangle$ as mentioned in \secref{sec GFMC}). To estimate the sensitivity of ANCs on separation energies, we refitted the GFMC overlaps from Figs.~\ref{fig H3p IL7 overlaps} and \ref{fig he6p IL7 overlaps} with solutions of \myeqref{eq fitting potential} imposing separation energies corresponding to $\pm 3\%$ deviations from $B_{exp}$. Using $B_{exp}$ from Tables~\ref{tab separation energies s-shell} and \ref{tab separation energies p-shell}, we obtain the following ANCs for $B=0.97B_{exp}$, $B=B_{exp}$, and $B=1.03B_{exp}$: -6.13, -6.45, -6.78 for the \overlap{\nuc{3}{H}}{\nuc{4}{He}} overlap, and 3.33, 3.52, 3.71 for the \overlap{\nuc{6}{He}}{\nuc{7}{Li}} overlap. Thus, any uncertainty of about $3\%$ or less on the separation energy translates into an ANC uncertainty of about 5--6\% or less. Furthermore, the $s$-shell GFMC ANCs in \tabref{tab ANCs s-shell} agree well with those obtained within the HH method except for the weak $d$-waves in $A=3$ nuclei for which the two methods differ for unknown reasons (the HH ANCs assume the actual separation energies for a given Hamiltonian that are within a fraction of percent of the experimental values). To test the dependence of GFMC ANCs in the $p$-shell on starting VMC wave functions, we include in \tabref{tab ANCs p-shell} results involving the ground state of \nuc{7}{Li} obtained for several combinations of Type~I/II (see \secref{sec VMC}) trial wave functions; depending on the starting VMC wave function, GFMC ANCs differ by at most 7\% in a given overlap channel.

Based on these arguments, we chose to determine the GFMC ANCs assuming experimental separation energies. We estimate the systematic errors on our ANCs to be $\lesssim 5\%$.

%=================================== results ===================================
\section{Results}
\label{sec results}

In this section we present GFMC spectroscopic overlaps, SFs, and  ANCs and compare them with those obtained by some other methods and with  experimentally deduced values. A convenient parametrization of the overlaps is also provided.

Table \ref{tab energies} shows computed and experimental binding energies for $A\leq7$ nuclear states relevant for this work. The errors shown in parentheses are only the Monte Carlo statistical errors; in addition there may be systematic errors from the GFMC algorithm of the order of 1--2\%~\cite{PP97, WP00, PW01}. Compared with previous GFMC works, the statistical accuracy on binding energies is much better thanks to the high sample counts needed for a  statistically reliable sampling of overlap tails as mentioned in \secref{sec inner workings}. Most GFMC results were obtained for the AV18+IL7 potential. For this potential, GFMC energies in \tabref{tab energies} differ by less than 1\% from experimental values; in most cases the difference is less than 0.5\%. 
% This excellent agreement is due to \ul{a sentence on what makes IL7 so damn good compared to UIX or IL2. Steve?} 
For the ground states of \nuc{6}{Li} and \nuc{7}{Li}, both Type~I and II VMC wave functions defined in \secref{sec VMC} were used. For comparison of GFMC overlaps with those obtained by other methods, we also constructed wave functions of $s$-shell nuclei bound by AV18+UIX whose energies are also shown in \tabref{tab energies}.

\begin{table}
\caption{GFMC binding energies $E$ for $A\leq7$ nuclei for several $2N$+$3N$ 
         potentials $H$. Only statistical errors are shown on GFMC values. For 
         $p$-shell nuclei, the VMC wave function type is indicated. The energy 
         shown for \nuc{2}{H} was obtained in a stochastic sampling of an 
         explicit solution of \myeqref{eq schrodinger}. Experimental values are 
         also shown.}
\label{tab energies}
\begin{ruledtabular}
\begin{tabular}{cccD{.}{.}{3.7}D{.}{.}{3.4}}
   $^AZ(J^\pi,T)$                &
   Type                          &
   $H$                           &
   \multicolumn{2}{c}{$E$ [MeV]}
   \\ \cline{4-5}
                            &
                            &
                            &
   \multicolumn{1}{c}{GFMC} &
   \multicolumn{1}{c}{exp.}
   \\ \hline
   \nuc{2}{H}$(1^{+},0)$                    &    & AV18     & -2.2247(3)&-2.2246
   \\
   \multirow{2}{*}{\nuc{3}{H}$(\frac{1}{2}^+,\frac{1}{2})$}  
                                            &    & AV18+IL7 & -8.47(0)  & -8.48 
   \\
                                            &    & AV18+UIX & -8.47(0)  & -8.48 
   \\
   \multirow{2}{*}{\nuc{3}{He}$(\frac{1}{2}^+,\frac{1}{2})$}
                                            &    & AV18+IL7 & -7.72(0)  & -7.72 
   \\
                                            &    & AV18+UIX & -7.72(0)  & -7.72 
   \\
   \multirow{2}{*}{\nuc{4}{He}$(0^+,0)$}
                                            &    & AV18+IL7 & -28.43(0) & -28.30    \\
                                            &    & AV18+UIX & -28.34(1) & -28.30    \\ \hline
%= p-shell
   \nuc{6}{He}$(0^+,1)$                     & I  & AV18+IL7 & -29.20(3) & -29.27
   \\
%   \nuc{6}{He}$(2^+,1)$                     & I  & AV18+IL7 & xxx.xxx   & -27.47
%   \\
   \multirow{2}{*}{\nuc{6}{Li}$(1^+,0)$}
                                            & I  & AV18+IL7 & -31.93(3) & -31.99
   \\
                                            & II & AV18+IL7 & -31.88(3) & -31.99
   \\
   \nuc{6}{Li}$(3^+,0)$                     & I  & AV18+IL7 & -29.59(3) & -29.80
   \\
   \nuc{6}{Li}$(0^+,1)$                     & I  & AV18+IL7 & -28.43(3) & -28.43
   \\
   \nuc{7}{He}$(\frac{3}{2}^-,\frac{3}{2})$ & I  & AV18+IL7 & -28.66(3) & -28.83
   \\
   \multirow{2}{*}{\nuc{7}{Li}$(\frac{3}{2}^-,\frac{1}{2})$}
                                            & I  & AV18+IL7 & -39.08(3) & -39.24
   \\
                                            & II & AV18+IL7 & -39.00(3) & -39.24
   \\
   \nuc{7}{Li}$(\frac{1}{2}^-,\frac{1}{2})$ & I  & AV18+IL7 & -38.88(3) & -38.76
   \\
   \nuc{7}{Be}$(\frac{3}{2}^-,\frac{1}{2})$ & I  & AV18+IL7 & -37.61(4) & -37.60
\end{tabular}
\end{ruledtabular}
\end{table}

Besides getting the absolute binding energies right, it is important for consistent overlap calculations to work with nuclear states having the correct one-nucleon separation energies.
% , a condition often violated and therefore relaxed in literature without a detailed analysis of consequent uncertainties~\cite{NB07,Ti10,NW11}. 
% This aspect of the calculation applied to GFMC was discussed in \secref{sec inner workings}. 
In Tables~\ref{tab separation energies s-shell} and \ref{tab separation energies p-shell}, we show GFMC one-nucleon separation energies computed for the desired Hamiltonian $H$ from \myeqref{eq schrodinger} and for the simplified $H'$ from \myeqref{eq GFMC propagation} along with experimental values. Given the agreement between GFMC and the experiment, and the related discussion in \secref{sec inner workings}, we assume the experimental separation energies
% experimental separation energies 
when determining ANCs from GFMC overlaps and account for possible systematic errors inflicted by this choice.
% Considering that GFMC wave functions resulting from \myeqref{eq GFMC propagation} are really eigenstates of 
% a somewhat simplified Hamiltonian $H^\prime$ 
% described in \secref{sec GFMC} 
% and that it is this $H^\prime$ that determines the asymptotic decay of overlaps, Tables~\ref{tab separation energies s-shell} and \ref{tab separation energies p-shell} also lists separation energies for $H'$. Because $H'$ was adjusted to give $\langle H' \rangle \approx \langle H \rangle$, separation energies for $H$ and $H^\prime$ agree very well and, more importantly, the difference from experiment is $\lesssim 3\%$ with a noticeable exception being the \overlap{\nuc{6}{He}}{\nuc{7}{He}} overlap involving a particle-unbound \nuc{7}{He} (for this overlap, the ANC is not a meaningful quantity). Given this agreement, we assume experimental separation energies when determining ANCs from GFMC overlaps and account for possible systematic errors inflicted by this choice as discussed in \secref{sec inner workings}.

\begin{table}
\caption{GFMC core-valence separation energies in the $s$-shell for several 
         $2N$+$3N$ potentials $H$ and for the corresponding $H'$ from 
         \myeqref{eq GFMC propagation}. The combined statistical errors on the 
         core and the parent energies from \tabref{tab energies} are all $\leq 
         0.01$~MeV. Experimental values are also shown.}
\label{tab separation energies s-shell}
\begin{ruledtabular}
\begin{tabular}{cccD{.}{.}{3.2}D{.}{.}{3.2}D{.}{.}{3.2}}
   \multicolumn{1}{c}{parent}    &
   \multicolumn{1}{c}{core}      &
   $H$                           &
   \multicolumn{3}{c}{$B$ [MeV]}
   \\ \cline{1-1} \cline{2-2} \cline{4-6}
   $^{A}Z(J^\pi,T)$               &
   $^{A-1}Z(J^\pi,T)$             &
                                  &
   \multicolumn{1}{c}{GFMC, $H$}  & 
   \multicolumn{1}{c}{GFMC, $H'$} & 
   \multicolumn{1}{c}{exp.}
   \\ \hline
   \multirow{2}{*}{\nuc{3}{H}$(\frac{1}{2}^+,\frac{1}{2})$}  &
   \multirow{2}{*}{\nuc{2}{H}$(1^+,0)$}                      &
   AV18+IL7                                                  & 
   6.24                                                      &
   6.24                                                      &
   6.26
   \\
                                                             &
                                                             &
   AV18+UIX                                                  &
   6.24                                                      &
   6.24                                                      &
   6.26
   \\ \hline
   \multirow{2}{*}{\nuc{3}{He}$(\frac{1}{2}^+,\frac{1}{2})$} &
   \multirow{2}{*}{\nuc{2}{H}$(1^+,0)$}                      &
   AV18+IL7                                                  & 
   5.49                                                      &
   5.49                                                      &
   5.49
   \\
                                                             &
                                                             &
   AV18+UIX                                                  & 
   5.50                                                      &
   5.50                                                      &
   5.49
   \\ \hline
   \multirow{4}{*}{\nuc{4}{He}$(0^+,0)$}                     &
   \multirow{2}{*}{\nuc{3}{H}$(\frac{1}{2}^+,\frac{1}{2})$}  &
   AV18+IL7                                                  & 
   19.96                                                     &
   19.94                                                     &
   19.81
   \\
                                                             &
                                                             &
   AV18+UIX                                                  &
   19.89                                                     &
   19.88                                                     &
   19.81
   \\ \cline{2-6}
                                                             &
   \multirow{2}{*}{\nuc{3}{He}$(\frac{1}{2}^+,\frac{1}{2})$} &
   AV18+IL7                                                  &
   20.71                                                     &
   20.69                                                     &
   20.58
   \\
                                                             &
                                                             &
   AV18+UIX                                                  &
   20.63                                                     &
   20.62                                                     &
   20.58
\end{tabular}
\end{ruledtabular}
\end{table}

\begin{table}[b]
\caption{Same as \tabref{tab separation energies s-shell} for $A \leq 7$ 
         $p$-shell nuclei bound by AV18+IL7. (Negative) Positive values mean 
         that the $A$-body nucleus is particle (un)bound relative to the core. 
         The VMC wave function type is indicated.}
\label{tab separation energies p-shell}
\begin{ruledtabular}
\begin{tabular}{ccccD{.}{.}{3.5}D{.}{.}{3.5}D{.}{.}{3.2}}
   \multicolumn{2}{c}{parent}    &
   \multicolumn{2}{c}{core}      &
   \multicolumn{3}{c}{$B$ [MeV]}
   \\ \cline{1-2} \cline{3-4} \cline{5-7}
   $^{A}Z(J^\pi,T)$               &
   Type                           &
   $^{A-1}Z(J^\pi,T)$             &
   Type                           &
   \multicolumn{1}{c}{GFMC, $H$}  & 
   \multicolumn{1}{c}{GFMC, $H'$} & 
   \multicolumn{1}{c}{exp.}
   \\ \hline
   \nuc{7}{He}$(\frac{3}{2}^-,\frac{3}{2})$                  &
   I                                                         &
   \nuc{6}{He}$(0^+,1)$                                      &
   I                                                         &
   -0.54(4)                                                  &
   -0.47(1)                                                  &
   -0.45
   \\ \hline
   \multirow{8}{*}{\nuc{7}{Li}$(\frac{3}{2}^-,\frac{1}{2})$} &
   I                                                         &
   \multirow{2}{*}{\nuc{6}{He}$(0^+,1)$}                     &
   \multirow{2}{*}{I}                                        &
   9.88(4)                                                   &
   9.82(1)                                                   &
   9.98
   \\
                                                             &
   II                                                        &
                                                             &
                                                             &
   9.81(4)                                                   &
   9.74(1)                                                   &
   9.98
   \\ \cline{2-7}
                                                             &
   I                                                         &
   \multirow{2}{*}{\nuc{6}{Li}$(1^+,0)$}                     &
   I                                                         &
   7.15(4)                                                   &
   7.16(1)                                                   &
   7.25
   \\
%   \nuc{7}{Li}$(\frac{3}{2}^-,\frac{1}{2})$ & I  &
%   \nuc{6}{Li}(1+,0)                        & II &
%   AV18+IL7                                 &
%   7.20(4)                                  &
%   7.14(1)                                  &
%   7.25
%   \\   
%   \nuc{7}{Li}$(\frac{3}{2}^-,\frac{1}{2})$ & II &
%   \nuc{6}{Li}(1+,0)                        & I  &
%   AV18+IL7                                 &
%   7.08(4)                                  &
%   7.08(1)                                  &
%   7.25
%   \\
                                                             &
   II                                                        &
                                                             &
   II                                                        &
   7.12(4)                                                   &
   7.06(1)                                                   &
   7.25
   \\ \cline{2-7}
                                                             &
   I                                                         &
   \multirow{2}{*}{\nuc{6}{Li}$(3^+,0)$}                     &
   \multirow{2}{*}{I}                                        &
   9.49(4)                                                   &
   9.45(1)                                                   &
   9.44
   \\
                                                             &
   II                                                        &
                                                             &
                                                             &
   9.42(4)                                                   &
   9.37(1)                                                   &
   9.44
   \\ \cline{2-7}
                                                             &
   I                                                         &
   \multirow{2}{*}{\nuc{6}{Li}$(0^+,1)$}                     &
   \multirow{2}{*}{I}                                        &
   10.65(4)                                                  &
   10.60(1)                                                  &
   10.81
   \\
                                                             &
   II                                                        &
                                                             &
                                                             &
   10.57(4)                                                  &
   10.52(1)                                                  &
   10.81
   \\ \hline
   \nuc{7}{Li}$(\frac{1}{2}^-,\frac{1}{2})$                  &
   I                                                         &
   \nuc{6}{Li}$(1^+,0)$                                      &
   I                                                         &
   6.95(4)                                                   &
   7.00(1)                                                   &
   6.77
   \\ \hline
   \multirow{3}{*}{\nuc{7}{Be}$(\frac{3}{2}^-,\frac{1}{2})$} &
   \multirow{3}{*}{I}                                        &
   \nuc{6}{Li}$(1^+,0)$                                      &
   I                                                         &
   5.69(4)                                                   &
   5.68(2)                                                   &
   5.61
   \\ \cline{3-7}
                                                             &
                                                             &
   \nuc{6}{Li}$(3^+,0)$                                      &
   I                                                         &
   8.03(5)                                                   &
   7.98(2)                                                   &
   7.79
   \\ \cline{3-7}
                                                             &
                                                             &
   \nuc{6}{Li}$(0^+,1)$                                      &
   I                                                         &
   9.18(4)                                                   &
   9.13(2)                                                   &
   9.17
\end{tabular}
\end{ruledtabular}
\end{table}

Tables~\ref{tab SFs s-shell} and \ref{tab SFs p-shell} summarize, respectively, spectroscopic factors for $s$- and $p$-shell nuclei with $A \leq 7$. In these tables, the errors on VMC and GFMC SFs are only statistical; possible systematic errors on GFMC values were estimated in \secref{sec inner workings} to be 2--3\% or less.

\begin{table}[t]
\caption{SFs for overlaps between $s$-shell nuclei for a given $2N$+$3N$ 
         Hamiltonian $H$ and angular momentum channel $l_j$. The statistical 
         errors on VMC and GFMC values are less than 1 in the last digit shown;
         systematic uncertainties in GFMC values are 2--3\% or less, as 
         discussed in \secref{sec inner workings}. Also shown are the results of
         the HH method
         % hyperspherical harmonics method (HH)~\cite{KR97, VK05, VK11} 
         and experimentally deduced values.}
\label{tab SFs s-shell}
\begin{ruledtabular}
\begin{tabular}{ccccD{.}{.}{1.3}D{.}{.}{1.4}D{.}{.}{1.4}c}
   parent                        &
   core                          &
   $l_j$                         &
   $H$                           &
   \multicolumn{4}{c}{$S$}
   \\ \cline{1-1} \cline{2-2} \cline{5-8}
   $^{A}Z(J^\pi,T)$              &
   $^{A-1}Z(J^\pi,T)$            &
                                 &
                                 &
   \multicolumn{1}{c}{HH}        & 
   \multicolumn{1}{c}{VMC}       & 
   \multicolumn{1}{c}{GFMC}      & 
   \multicolumn{1}{c}{exp.}
   \\ \hline
   \multirow{4}{*}{\nuc{3}{H}$(\frac{1}{2}^+,\frac{1}{2})$}  &
   \multirow{4}{*}{\nuc{2}{H}$(1^+,0)$}                      &
   \multirow{2}{*}{$s_{1/2}$}                                &
   AV18+IL7                                                  &
                                                             &
   1.32                                                      &
   1.30                                                      &
   
   \\
                                                             &
                                                             &
                                                             &
   AV18+UIX                                                  &
   1.30                                                      &
   1.32                                                      &
   1.30                                                      &
   
   \\ \cline{3-8}
                                                             &
                                                             &
   \multirow{2}{*}{$d_{3/2}$}                                &
   AV18+IL7                                                  &
                                                             &
   0.0194                                                    &
   0.0224                                                    &
   
   \\
                                                             &
                                                             &
                                                             &
   AV18+UIX                                                  &
   0.0225                                                    &
   0.0195                                                    &
   0.0223                                                    &
   
   \\ \hline
   \multirow{4}{*}{\nuc{3}{He}$(\frac{1}{2}^+,\frac{1}{2})$} &
   \multirow{4}{*}{\nuc{2}{H}$(1^+,0)$}                      &
   \multirow{2}{*}{$s_{1/2}$}                                &
   AV18+IL7                                                  &
                                                             &
   1.32                                                      &
   1.31                                                      &
   
   \\
                                                             &
                                                             &
                                                             &
   AV18+UIX                                                  & 
   1.31                                                      &
   1.32                                                      &
   1.31                                                      &
   
   \\  \cline{3-8}
                                                             &
                                                             &
   \multirow{2}{*}{$d_{3/2}$}                                &
   AV18+IL7                                                  & 
                                                             &
   0.0190                                                    &
   0.0221                                                    &
   
   \\
                                                             &
                                                             &
                                                             &
   AV18+UIX                                                  & 
   0.0222                                                    &
   0.0191                                                    &
   0.0221                                                    &
   
   \\ \hline
   \multirow{4}{*}{\nuc{4}{He}(0+,0)}                        &
   \multirow{2}{*}{\nuc{3}{H}$(\frac{1}{2}^+,\frac{1}{2})$}  &
   \multirow{2}{*}{$s_{1/2}$}                                &
   AV18+IL7                                                  & 
                                                             &
   1.64                                                      &
   1.61                                                      &
   \multirow{2}{*}{$\sim$1.4--1.6~\cite{PS97}}
   \\
                                                             &
                                                             &
                                                             &
   AV18+UIX                                                  & 
   1.60                                                      &
   1.63                                                      &
   1.61                                                      &
   
   \\ \cline{2-8}
                                                             &
   \multirow{2}{*}{\nuc{3}{He}$(\frac{1}{2}^+,\frac{1}{2})$} &
   \multirow{2}{*}{$s_{1/2}$}                                &
   AV18+IL7                                                  & 
                                                             &
   1.62                                                      &
   1.60                                                      &
   
   \\
                                                             &
                                                             &
                                                             &
   AV18+UIX                                                  & 
   1.58                                                      &
   1.62                                                      &
   1.60                                                      &
   
   \\
\end{tabular}
\end{ruledtabular}
\end{table}

In the $s$-shell, where VMC wave functions are already very good approximations of true eigenstates~\cite{PW01}, the $s_{1/2}$ spectroscopic factors in \tabref{tab SFs s-shell} change by less than 2\% between VMC and GFMC. VMC, however, seems to underestimate the small amount of $d$-waves in $A=3$ nuclei by about 15\%.
% as measured by SFs which is presumably corrected by GFMC. \ul{do we have any other indication for this defficiency of VMC?}. 
Not surprising, $s$-shell results for AV18+UIX and AV18+IL7 are similar because both $3N$ forces, IL7 and especially UIX, were tuned to reproduce the binding of $s$-shell nuclei.
% Despite many electron scattering measurements~\cite{xxx}, 
The experimental information on SFs in the $s$-shell is scarce; for the \overlap{\nuc{3}{H}}{\nuc{4}{He}} overlap, the GFMC value of 1.61 is in good agreement with an experimental value $\sim$1.4--1.6 deduced from electron scattering~\cite{PS97}.

The $A=3$ nuclei contain 1.5 $S,T=1,0$ nucleon pairs~\cite{FP96,Wi06}.
% To the extent that $A=3$ nuclei are pure $T=1/2$ states \ul{is this OK or should it be to the extend that they are 2S[3] states?}, they contain \ul{at most??} 1.5 $S,T=1,0$ nucleon pairs~\cite{Wi06}. 
The sums of $s$- and $d$-wave SFs for overlaps of \nuc{3}{H} and \nuc{3}{He} with a deuteron are about 1.32 and 1.33, respectively. If these values are interpreted as numbers of deuterons~\cite{SP86}, it appears that about 90\% of $T=0$ pairs in $A=3$ nuclei are in the deuteron state. We conjecture that the slightly bigger \overlap{\nuc{2}{H}}{\nuc{3}{He}} SF compared with that between \nuc{2}{H} and \nuc{3}{H} is due to the fact that it is easier to compact a spatially extended deuteron into \nuc{3}{He}, which is somewhat bigger than the triton. 

Using a similar probabilistic interpretation for \nuc{4}{He}, where all one-nucleon (neutron or proton) SFs add up to 2~\cite{Cl73} and those for overlaps with $A=3$ nuclei in \tabref{tab SFs s-shell} are about 1.6, we find that about 80\% of all nucleon triples in \nuc{4}{He} are in the triton or \nuc{3}{He} state. Because both $A=3$ nuclei are spatially more extended than \nuc{4}{He} and the triton is somewhat smaller than \nuc{3}{He}, we conjecture that it is more likely to find a triton than \nuc{3}{He} inside \nuc{4}{He}, which is reflected by a slight difference in the corresponding SFs in \tabref{tab SFs s-shell}. In the $s$-shell, these conjectures are plausible since the $s$-wave SFs in \tabref{tab SFs s-shell} account for such a big fraction of the total spectroscopic strength.

In \tabref{tab SFs s-shell}, the AV18+UIX values labeled HH are those of the hyperspherical harmonics method~\cite{KR08}; in particular, the overlaps for $A=3$ parent nuclei are from~\cite{KR97} and those for \nuc{4}{He} are improved and updated versions of previously published results~\cite{VK05, VK11}. The HH results were converted to our conventions for overlaps.
% The hyperspherical overlaps are used to benchmark our GFMC results, especially to check that using the simplified Hamiltonian $H'$ in \myeqref{eq GFMC propagation} can be justified. 
The agreement between GFMC and HH is very good, as can be seen from SFs in \tabref{tab SFs s-shell} and from \figref{fig h2n and h3p AV18+UIX}, where the actual \overlap{\nuc{2}{H}}{\nuc{3}{H}} and \overlap{\nuc{3}{H}}{\nuc{4}{He}} overlaps are shown. Small discrepancies exist between GFMC and HH overlaps at short distances, and the $d$-wave ANCs in $A=3$ nuclei differ; the sources of these discrepancies remain unknown. An equally good agreement is obtained for overlaps involving \nuc{3}{He}.

\bfig[b]
   \includegraphics[width = 0.48\textwidth]
                   {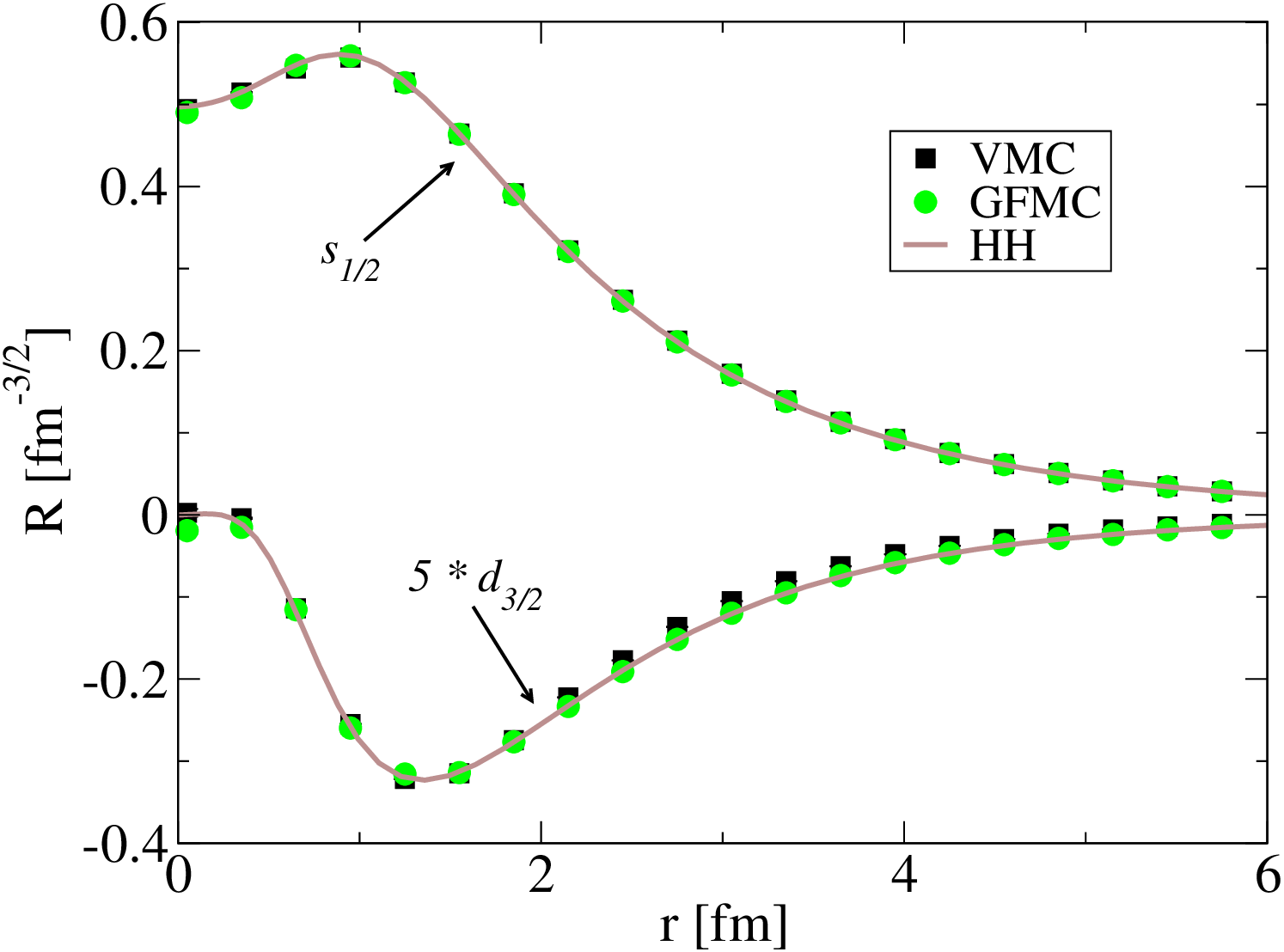}
   \hspace{0.02\textwidth}
   \includegraphics[width = 0.48\textwidth]
                   {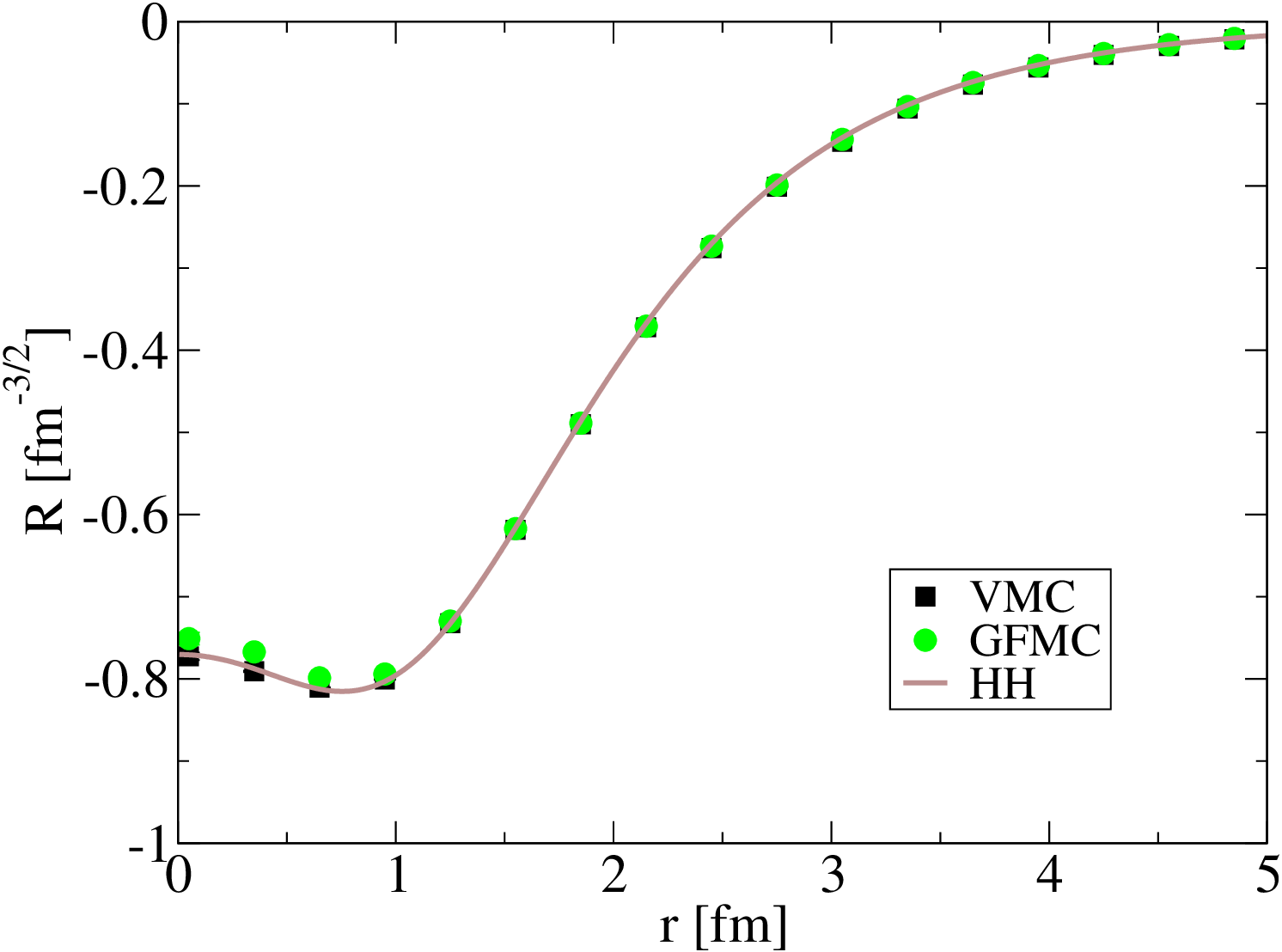}
   \caption{(Color online) Comparison of the $s_{1/2}$ and the $d_{3/2}$ 
            \overlap{\nuc{2}{H}}{\nuc{3}{H}} (left), and the 
            $s_{1/2}$ \overlap{\nuc{3}{H}}{\nuc{4}{He}} (right) overlaps
            obtained for the AV18+UIX potential within VMC, GFMC, and HH. Only
            statistical errors are shown on Monte Carlo overlaps.}
   \label{fig h2n and h3p AV18+UIX}
\efig

As examples of our $s$-shell overlaps obtained for the AV18+IL7 potential, we show in \figref{fig h2n and h3p AV18+IL7} the \overlap{\nuc{2}{H}}{\nuc{3}{H}} and \overlap{\nuc{3}{H}}{\nuc{4}{He}} overlaps. 
% Also shown in these figures are the Whittaker functions computed for experimental separation energies as well as 
Also shown in these figures are WS fits of GFMC overlaps. Although the absolute overlap signs are rather arbitrary since they depend on the detailed construction of our wave functions, the order of vector couplings in Eqs.~\eqref{eq overlap antisym} and \eqref{eq valence part}, and the isospin Clebsch-Gordan coefficient in \myeqref{eq overlap antisym}, relative signs of different overlaps for a given parent-core combination may be important for interference effects. In our sign convention, the $s$- and $d$-wave overlaps in $A=3$ nuclei have opposite signs consistent with experimentally deduced negative ratios of their ANCs shown in \tabref{tab ANCs s-shell}. Note the slight dips in $s$-wave overlaps near the origin that are consistent with a depletion of one-body densities of $s$-shell nuclei at short range~\cite{PW01} due to a repulsive potential core.

\bfig[t]
   \includegraphics[width = 0.48\textwidth]
                  {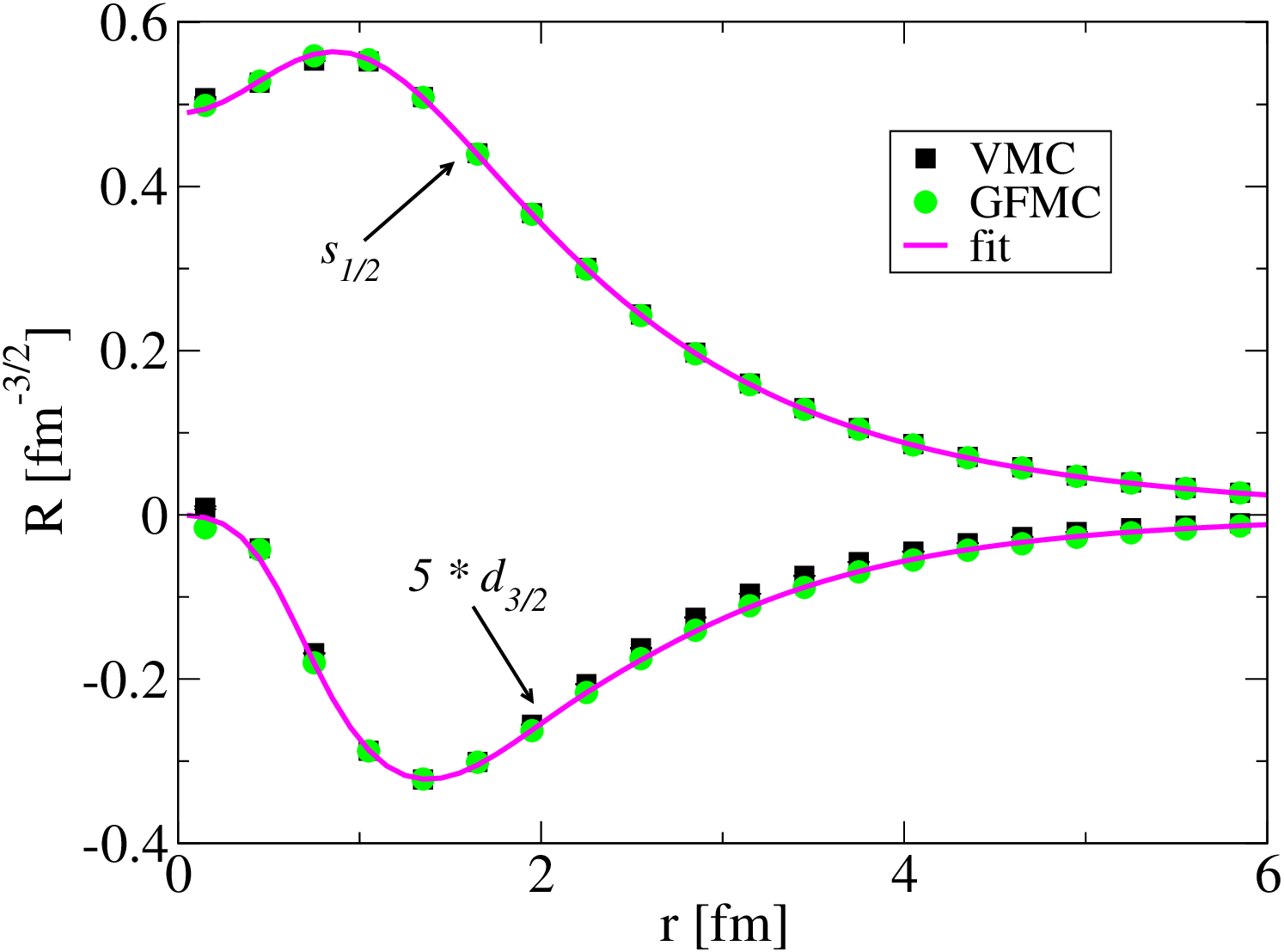}
   \hspace{0.02\textwidth}
   \includegraphics[width = 0.48\textwidth]
                   {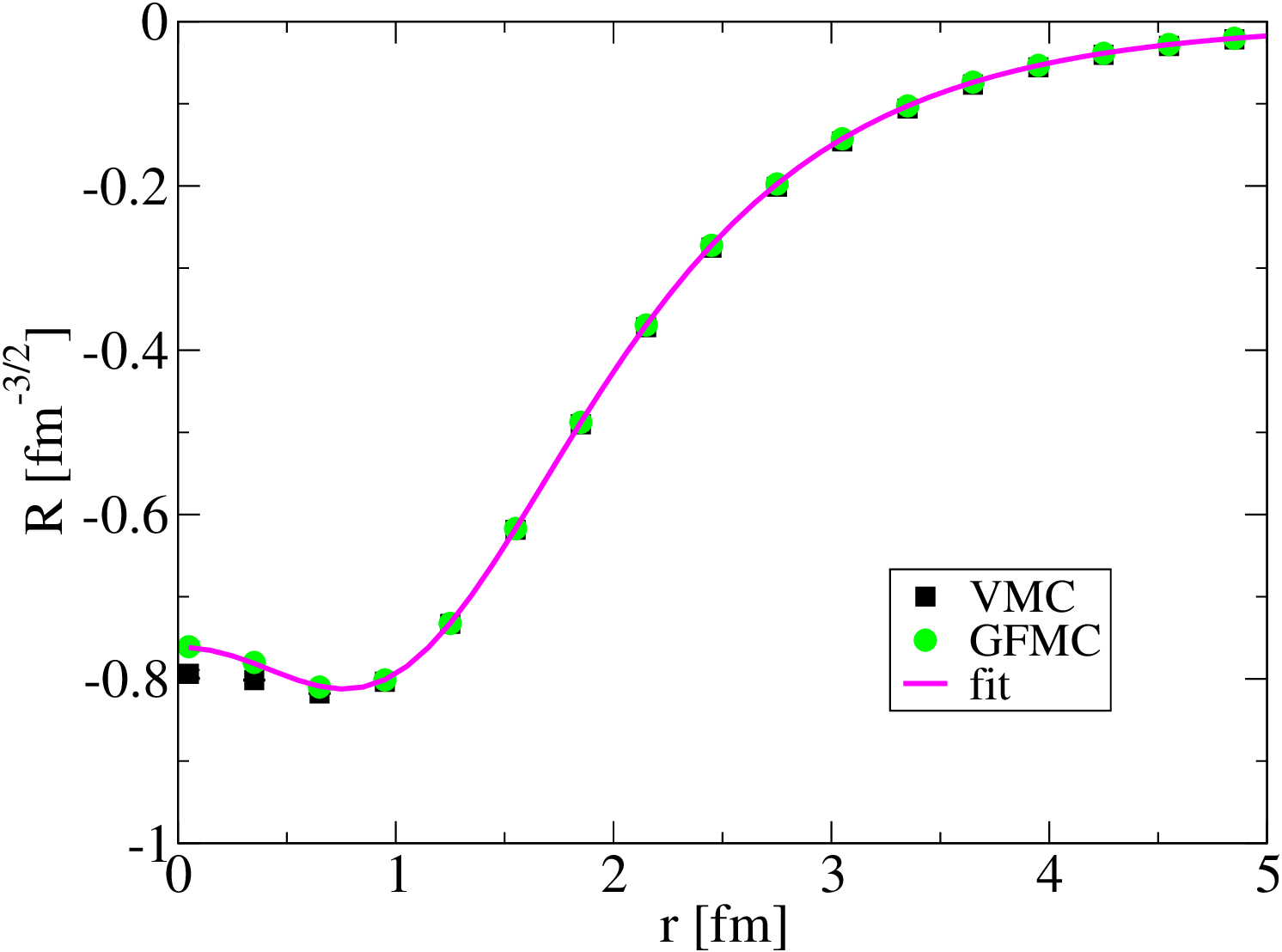}
   \caption{(Color online) VMC and GFMC $s_{1/2}$ and $d_{3/2}$ 
            \overlap{\nuc{2}{H}}{\nuc{3}{H}} (left) and $s_{1/2}$ 
            \overlap{\nuc{3}{H}}{\nuc{4}{He}} (right) overlaps obtained for the 
            AV18+IL7 potential. Only statistical errors are shown. Also shown 
            are WS fits of GFMC overlaps.}
   \label{fig h2n and h3p AV18+IL7}
\efig

VMC and GFMC SFs between $A=6$ and 7 $p$-shell nuclei bound by AV18+IL7 are listed in \tabref{tab SFs p-shell} along with shell-model predictions and experimentally deduced values. The Cohen and Kurath~\cite{CK67} shell-model values (SM) in \tabref{tab SFs p-shell} were corrected for center-of-mass motion effects by a factor $A/(A-1)$~\cite{DF74,HT03} to make them comparable with our results, which, as mentioned in \secref{sec overlaps theory}, are free of such spurious contaminations. In addition, a square of the isospin coupling coefficient from \myeqref{eq overlap antisym} was factored into SM values in \tabref{tab SFs p-shell}. Because it is not clear whether experimental SFs, often compared by their authors with original shell-model values of~\cite{CK67}, were corrected backward for center-of-mass effects or not, we quote them as they appear in the literature. 

\begin{table}
\caption{SFs for overlaps between $A=6$ and 7 $p$-shell nuclei for the 
         AV18+IL7 potential in angular momentum channels $l_j$. The VMC and GFMC          statistical errors are insignificant compared with the estimated 
         systematic uncertainties of 2--3\% or less for GFMC as discussed in 
         \secref{sec inner workings}. SM denotes corrected shell-model values 
         (see text). Experimentally deduced values are also shown.}
\label{tab SFs p-shell}
\begin{ruledtabular}
\begin{tabular}{ccccccccc}
   \multicolumn{2}{c}{parent}    &
   \multicolumn{2}{c}{core}      &
   $l_j$                         &
   \multicolumn{4}{c}{$S$}
   \\ \cline{1-2} \cline{3-4} \cline{6-9}
   $^{A}Z(J^\pi,T)$         &
   Type                     &
   $^{A-1}Z(J^\pi,T)$       &
   Type                     &
                            &
   \multicolumn{1}{c}{SM}   & 
   \multicolumn{1}{c}{VMC}  & 
   \multicolumn{1}{c}{GFMC} & 
   \multicolumn{1}{c}{exp.}
\\ \hline
   \nuc{7}{He}$(\frac{3}{2}^-,\frac{3}{2})$                   &
   I                                                          &
   \nuc{6}{He}$(0^+,1)$                                       &
   I                                                          &
   $p_{3/2}$                                                  &
   0.690                                                      &
   0.532                                                      &
   0.565                                                      &
   0.37(7)~\cite{WR05}
\\ \hline
   \multirow{15}{*}{\nuc{7}{Li}$(\frac{3}{2}^-,\frac{1}{2})$} &
   I                                                          &
   \multirow{2}{*}{\nuc{6}{He}$(0^+,1)$}                      &
   \multirow{2}{*}{I}                                         &
   \multirow{2}{*}{$p_{3/2}$}                                 &
   \multirow{2}{*}{0.691}                                     &
   0.442                                                      &
   0.406                                                      &
   \multirow{2}{*}{0.44(6)\footnotemark[1]~\cite{WS08}, 0.42(4)~\cite{LW99}}
   \\
                                                              &
   II                                                         &
                                                              &
                                                              &
                                                              &
                                                              &
   0.365                                                      &
   0.409                                                      &

   \\ \cline{2-9}
                                                              &
   \multirow{3}{*}{I}                                         &
   \multirow{3}{*}{\nuc{6}{He}$(2^+,1)$}                      &
   \multirow{3}{*}{I}                                         &
   $p_{1/2}$                                                  &
   0.253                                                      &
   0.128                                                      &
                                                              &
   
   \\
                                                              &
                                                              &
                                                              &
                                                              &
   $p_{3/2}$                                                  &
   0.212                                                      &
   0.146                                                      &
                                                              &

   \\   
                                                              &
                                                              &
                                                              &
                                                              &
   sum                                                        &
   0.466                                                      &
   0.274                                                      &
                                                              &
   0.16(2)~\cite{LW99}
   \\ \cline{2-9}
                                                              &
   \multirow{3}{*}{I}                                         &
   \multirow{6}{*}{\nuc{6}{Li}$(1^+,0)$}                      &
   \multirow{3}{*}{I}                                         &
   $p_{1/2}$                                                  &
   0.338                                                      &
   0.229                                                      &
   0.230                                                      &

   \\
                                                              &
                                                              &
                                                              &
                                                              &
   $p_{3/2}$                                                  &
   0.503                                                      &
   0.480                                                      &
   0.438                                                      &

   \\
                                                               &
                                                               &
                                                               &
                                                               &
   sum                                                         &
   0.841                                                       &
   0.709                                                       &
   0.668                                                       &
   0.74(11)\footnotemark[1]~\cite{WS08}, 0.73(5)~\cite{JZ10}
   \\ \cline{2-2} \cline{4-9}
                                                              &
   \multirow{3}{*}{II}                                        &
                                                              &
   \multirow{3}{*}{II}                                        &
   $p_{1/2}$                                                  &
   0.338                                                      &
   0.211                                                      &
   0.229                                                      &
   
   \\
                                                              &
                                                              &
                                                              &
                                                              &
   $p_{3/2}$                                                  &
   0.503                                                      &
   0.401                                                      &
   0.428                                                      &
   
   \\
                                                               &
                                                               &
                                                               &
                                                               &
   sum                                                         &
   0.841                                                       &
   0.612                                                       &
   0.657                                                       &
   0.74(11)\footnotemark[1]~\cite{WS08}, 0.73(5)~\cite{JZ10}
   \\ \cline{2-9}
                                                              &
   I                                                          &
   \multirow{2}{*}{\nuc{6}{Li}$(3^+,0)$}                      &
   \multirow{2}{*}{I}                                         &
   \multirow{2}{*}{$p_{3/2}$}                                 &
   \multirow{2}{*}{0.646}                                     &
   0.500                                                      &
   0.435                                                      &
   \multirow{2}{*}{0.72(14)~\cite{LM69}, 0.58(13)~\cite{To69}}
   \\
                                                              &
   II                                                         &
                                                              &
                                                              &
                                                              &
                                                              &
   0.436                                                      &
   0.447                                                      &

   \\ \cline{2-9}
                                                              &
   I                                                          &
   \multirow{2}{*}{\nuc{6}{Li}$(0^+,1)$}                      &
   \multirow{2}{*}{I}                                         &
   \multirow{2}{*}{$p_{3/2}$}                                 &
   \multirow{2}{*}{0.345}                                     &
   0.221                                                      &
   0.203                                                      &
   \multirow{2}{*}{0.19(3)\footnotemark[1]~\cite{WS08}}
   \\
                                                              &
   II                                                         &
                                                              &
                                                              &
                                                              &
                                                              &
   0.183                                                      &
   0.204                                                      &
   
   \\ \hline
   \multirow{3}{*}{\nuc{7}{Li}$(\frac{1}{2}^-,\frac{1}{2})$}  &
   \multirow{3}{*}{I}                                         &
   \multirow{3}{*}{\nuc{6}{Li}$(1^+,0)$}                      &
   \multirow{3}{*}{I}                                         &
   $p_{1/2}$                                                  &
   0.045                                                      &
   0.069                                                      &
   0.060                                                      &

   \\
                                                              &
                                                              &
                                                              &
                                                              &
   $p_{3/2}$                                                  &
   0.997                                                      &
   0.854                                                      &
   0.759                                                      &

   \\
                                                              &
                                                              &
                                                              &
                                                              &
   sum                                                        &
   1.042                                                      &
   0.923                                                      &
   0.819                                                      &
   1.15~\cite{SM67}, 0.90(9)~\cite{JZ10}
   \\ \hline
   \multirow{5}{*}{\nuc{7}{Be}$(\frac{3}{2}^-,\frac{1}{2})$}  &
   \multirow{5}{*}{I}                                         &
   \multirow{3}{*}{\nuc{6}{Li}$(1^+,0)$}                      &
   \multirow{3}{*}{I}                                         &
   $p_{1/2}$                                                  &
   0.338                                                      &
   0.229                                                      &
   0.225                                                      &
   
   \\
                                                              &
                                                              &
                                                              &
                                                              &
   $p_{3/2}$                                                  &
   0.503                                                      &
   0.480                                                      &
   0.438                                                      &
   
   \\
                                                              &
                                                              &
                                                              &
                                                              &
   sum                                                        &
   0.841                                                      &
   0.709                                                      &
   0.663                                                      &
   
   \\ \cline{3-9}
                                                              &
                                                              &
   \nuc{6}{Li}$(3^+,0)$                                       &
   I                                                          &
   $p_{3/2}$                                                  &
   0.646                                                      &
   0.500                                                      &
   0.457                                                      &
   
   \\ \cline{3-9}
                                                              &
                                                              &
   \nuc{6}{Li}$(0^+,1)$                                       &
   I                                                          &
   $p_{3/2}$                                                  &
   0.345                                                      &
   0.221                                                      &
   0.210                                                      &
   
   \\
\end{tabular}
\end{ruledtabular}
\footnotetext[1]
             {Values obtained from 
             $(\sigma_{\mathrm{exp}}/\sigma_{\mathrm{DWBA}})\times 0.32$, see 
             text.}
\end{table}

In the $p$-shell, the structure improvement provided by GFMC over VMC is known to be more significant compared with the $s$-shell~\cite{PW01}. This is reflected by SFs in \tabref{tab SFs p-shell} that change by as much as 15\% between VMC and GFMC. The results involving the ground states of \nuc{6}{Li} and \nuc{7}{Li} for which both Type~I and Type~II VMC wave functions were used demonstrate that, by closing the gap between 
% Type~I and Type~II 
VMC SFs as big as $\sim$20\% down to less than 3\%, the GFMC method is rather insensitive to starting trial functions.

In \tabref{tab SFs p-shell}, the experimental SFs
% for the \overlap{\nuc{6}{He}$(0^+)$}{\nuc{7}{Li}$(3/2^-)$} and \overlap{\nuc{6}{He}$(2^+)$}{\nuc{7}{Li}$(3/2^-)$} SFs 
from~\cite{LW99} are based on electron scattering data, the other experimental values were deduced from hadronic
% transfer or knockout 
reactions. The experimental SFs from~\cite{WS08} listed in the table are only relative; the authors of that work concluded that their data analysis cannot be trusted to provide an absolute determination of cross sections, and only relative SFs for several reactions were provided upon a renormalization by a somewhat arbitrary factor of 0.32, making them comparable to theoretical VMC values available at the time. In \cite{LW99}, SFs deduced from electron-induced proton knockout on \nuc{7}{Li} leading to the two lowest states of \nuc{6}{He} 
were found in perfect agreement with VMC values available at the time; in \tabref{tab SFs p-shell}, the GFMC SFs involving \nuc{6}{He}$(0^+)$ still agree perfectly with experiment, but the agreement for the reaction leading to \nuc{6}{He}$(2^+)$ has been spoiled by an error in the VMC code discovered after the original work \cite{LW99} had been published. The \overlap{\nuc{6}{He}($2^+$)}{\nuc{7}{Li}$(3/2^-)$} overlap involves an unbound state of \nuc{6}{He} for which GFMC does not find a stable energy; instead, the method produces the core nucleus with an ever-increasing radius by breaking it gradually into \nuc{4}{He} and two neutrons. Consequently, the GFMC SFs for this overlap steadily decrease and as such are absent in \tabref{tab SFs p-shell}. Given the experimental and systematic GFMC uncertainties and bearing in mind all the issues related to the meaning of spectroscopic overlaps and factors mentioned in \secref{sec introduction}, we conclude that the GFMC results in \tabref{tab SFs p-shell} seem to support newer determinations of (relative) SFs.

We do not find a significant difference between SFs of mirror nuclei \nuc{7}{Li} and \nuc{7}{Be}; also, SFs between the ground state of \nuc{7}{Li} and the $J^\pi,T=0^+,1$ isobaric analogs of \nuc{6}{He} and \nuc{6}{Li} differ just by  a factor of 2 due to the difference in isospin coupling coefficients in \myeqref{eq overlap antisym}.
% The experimental information on SFs is rather limited. 
Our SFs for overlaps between $A=6$ and 7 nuclei suggest a substantial quenching of shell-model values by as much as 40\% (except for the weak $p_{1/2}$ channel of the \overlap{\nuc{6}{Li}$(1^+)$}{\nuc{7}{Li}$(1/2^-)$} overlap).

As examples of $p$-shell overlaps, \figref{fig li6n Type I p0.5 and p1.5} shows the $p_{1/2}$ and $p_{3/2}$ \overlap{\nuc{6}{Li}$(1^+)$}{\nuc{7}{Li}$(3/2^-)$}
overlaps obtained for the AV18+IL7 interaction with GFMC wave functions originating from Type~I VMC functions (see \figref{fig he6p IL7 overlaps} for another example). Also shown in the figure are
% are the Whittaker functions computed for experimental separation energies as well as 
WS fits of GFMC overlaps. The figure illustrates that the overlaps may change from VMC to GFMC even when their SFs remain the same ($p_{1/2}$ channel); on the other hand, sometimes the change in SFs is more due to a renormalization than a shape change ($p_{3/2}$ channel). To better appreciate where the changes in SFs come from, we superimpose in \figref{fig li6n Type I p0.5 and p1.5} the ``density" functions $(R \times r)^2$ whose integral in \myeqref{eq SF} gives the SFs.

\bfig[t]
   \includegraphics[width = 0.48\textwidth]
                   {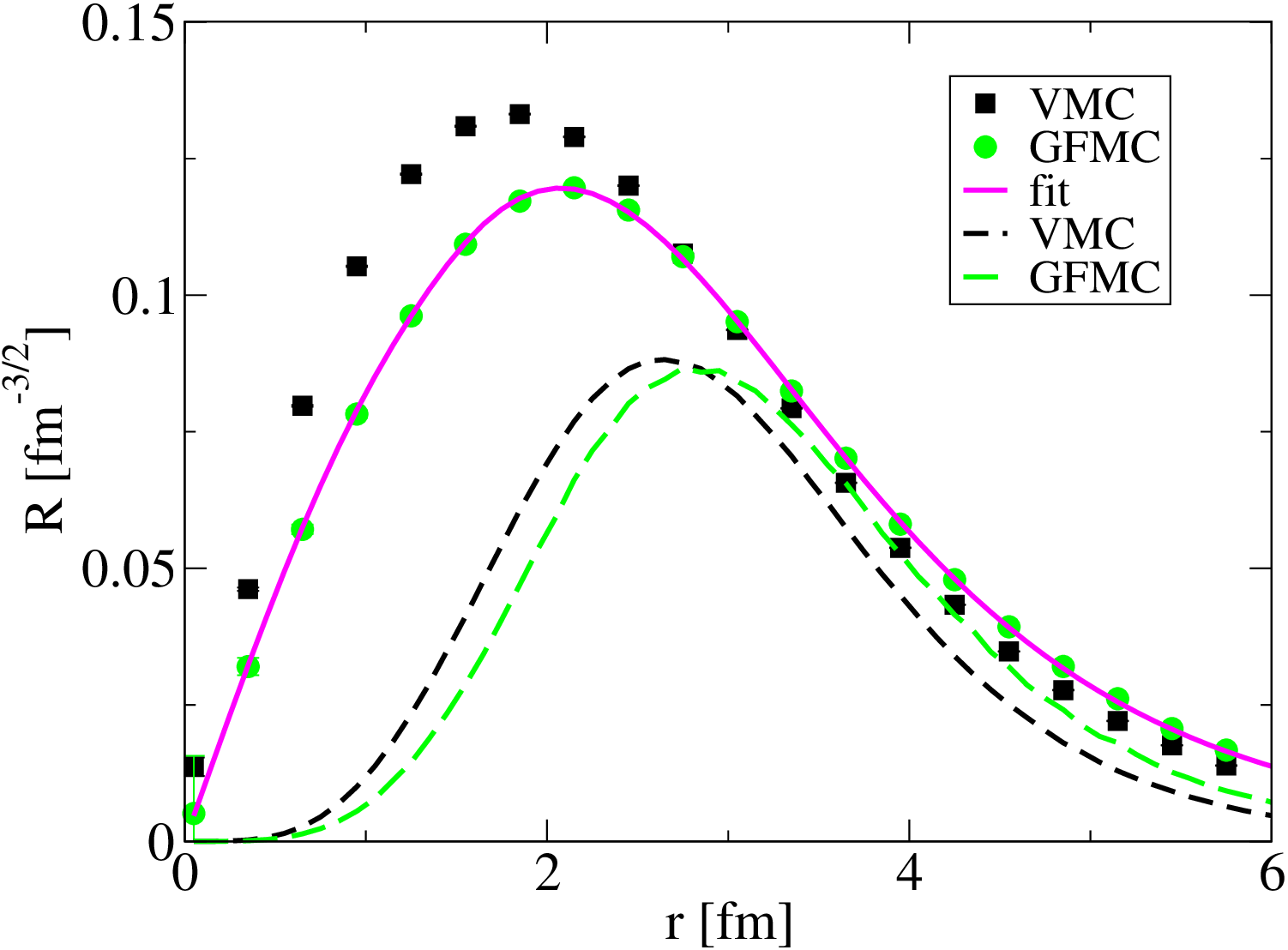}
   \hspace{0.02\textwidth}
   \includegraphics[width = 0.48\textwidth]
                   {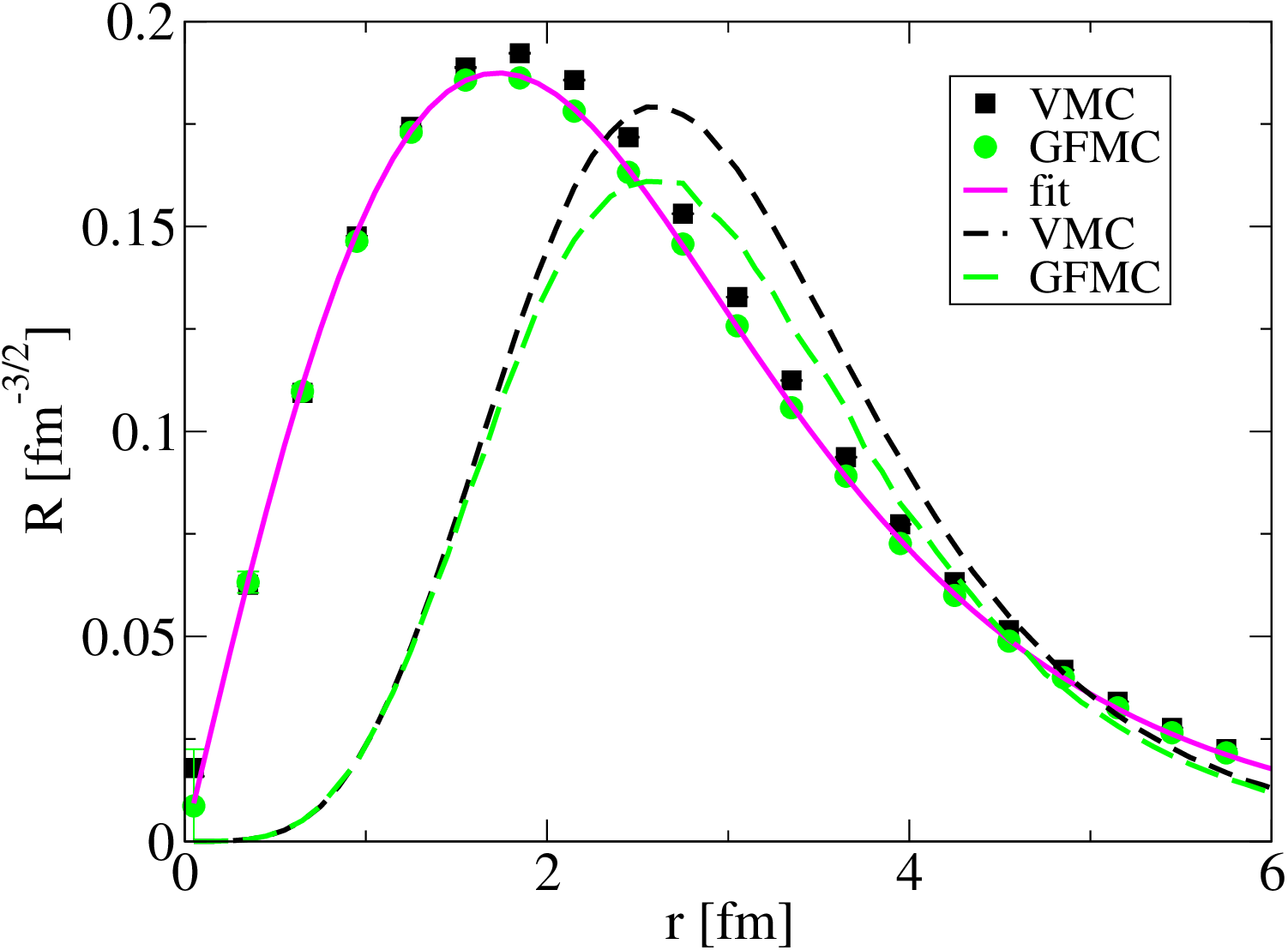}
   \caption{(Color online) VMC and GFMC $p_{1/2}$ (left) and $p_{3/2}$ (right)
            \overlap{\nuc{6}{Li}$(1^+)$}{\nuc{7}{Li}$(3/2^-)$} overlaps obtained
            for the AV18+IL7 potential. Only statistical errors are shown. Also 
            shown are WS fits of GFMC overlaps. Superimposed as dashed lines are
            the ``density" functions $(R\times r)^2$.}
   \label{fig li6n Type I p0.5 and p1.5}
\efig

The \overlap{\nuc{6}{He}}{\nuc{7}{He}} overlap in \tabref{tab SFs p-shell} is particularly challenging for both theory and experiment because it involves a parent nucleus that is particle unbound by about 450~keV relative to the core. In our calculations, \nuc{7}{He} is treated as a bound state and the GFMC propagation yields a stable energy and separation energy shown in \tabref{tab energies} and \tabref{tab separation energies p-shell}. Despite the bound-state approximation to \nuc{7}{He}, GFMC substantially improves the overlap tail compared with VMC, as can be seen in \figref{fig he6n} where the desired asymptotic form is represented by the scattering Coulomb function plotted at $90^\circ$ phase shift.

\bfig[t]
   \includegraphics[width = 0.48\textwidth]
                   {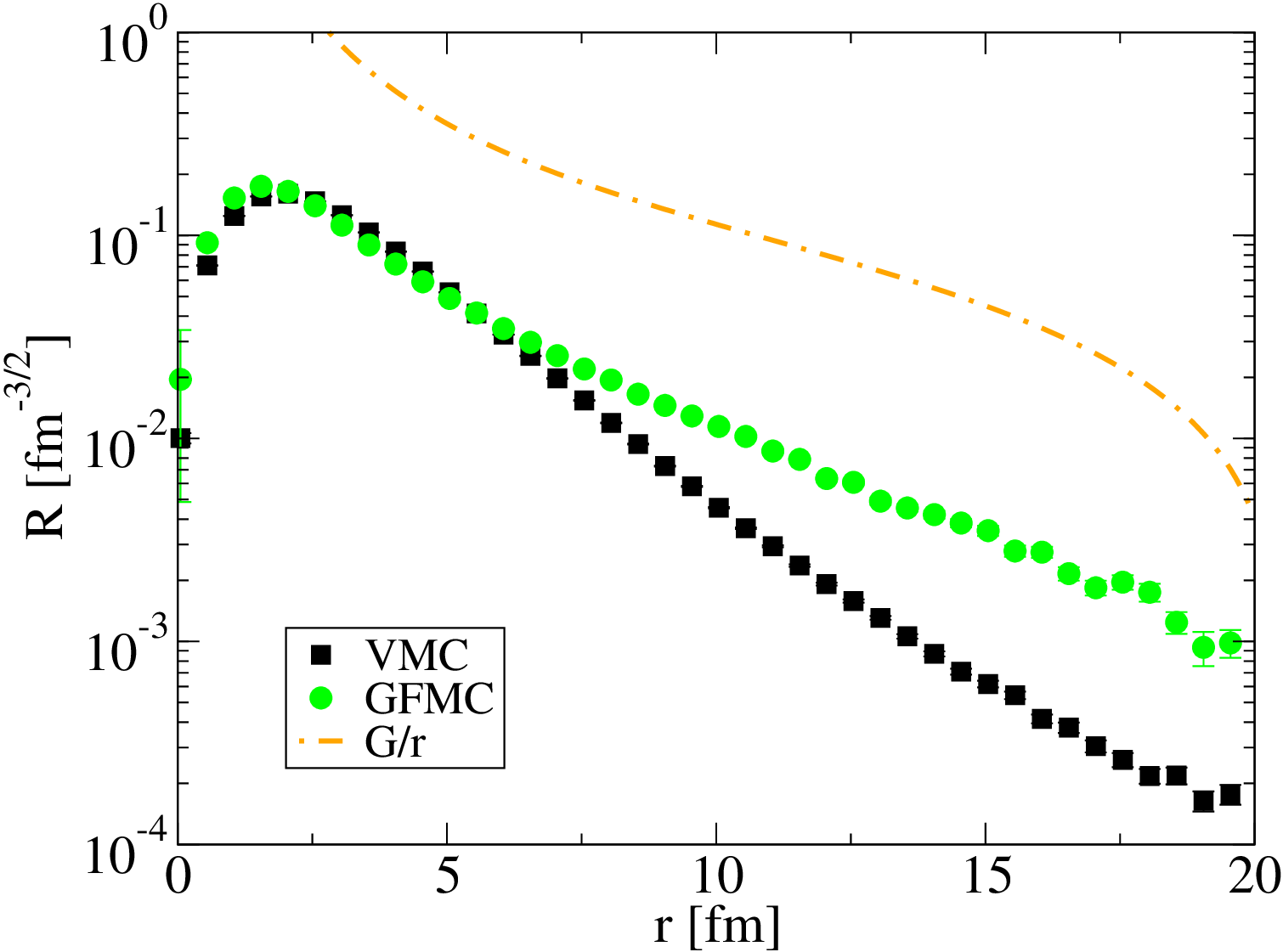}
   \caption{(Color online) VMC and GFMC $p_{3/2}$ 
            \overlap{\nuc{6}{He}$(0^+)$}{\nuc{7}{He}$(3/2^-)$} overlaps obtained
            for the AV18+IL7 potential. Only statistical errors are shown. Also 
            shown is the irregular Coulomb function $G(r)/r$ for the 
            experimental separation energy representing the scattering 
            asymptotics at $90^\circ$ phase shift.}
   \label{fig he6n}
\efig

ANCs, extracted from GFMC overlaps by a fitting procedure outlined in \secref{sec inner workings}, are listed in Tables~\ref{tab ANCs s-shell} and \ref{tab ANCs p-shell} for $s$- and $p$-shell nuclei along with experimentally derived numbers and those of some other realistic methods. Systematic errors on our ANCs were estimated in \secref{sec inner workings} to be $\lesssim$5\%. As was mentioned earlier in this section, the absolute signs of GFMC overlaps and ANCs are not meaningful, but relative signs in different overlap channels for a given parent-core combination matter.

\begin{table}[b]
\caption{ANCs for GFMC overlaps between $s$-shell nuclei for a given $2N$+$3N$ 
         Hamiltonian $H$ and angular momentum channel $l_j$. Systematic 
         uncertainties on GFMC values are 5\% or less as discussed in 
         \secref{sec inner workings}. Also shown are the results of the HH 
         and VMC-IM methods, and experimentally deduced values. The errors on 
         VMC-IM are only statistical. For $A=3$ nuclei, ratios of $d$- to 
         $s$-wave ANCs are also shown.}
\label{tab ANCs s-shell}
\scriptsize
\begin{ruledtabular}
\begin{tabular}{ccccD{.}{.}{2.4}D{.}{.}{2.7}D{.}{.}{2.4}c}
   parent                                       &
   core                                         &
   $l_j$                                        &
   $H$                                          &
   \multicolumn{4}{c}{$C$~$[\text{fm}^{-1/2}]$ }
\\ \cline{1-1} \cline{2-2} \cline{5-8}
   $^{A}Z(J^\pi,T)$           &
   $^{A-1}Z(J^\pi,T)$         &
                              &
                              &
   \multicolumn{1}{c}{HH}     & 
   \multicolumn{1}{c}{VMC-IM} &
   \multicolumn{1}{c}{GFMC}   & 
   \multicolumn{1}{c}{exp.}
\\ \hline
   \multirow{6}{*}{\nuc{3}{H}$(\frac{1}{2}^+,\frac{1}{2})$}  &
   \multirow{6}{*}{\nuc{2}{H}$(1^+,0)$}                      &
   \multirow{2}{*}{$s_{1/2}$}                                &
   AV18+IL7                                                  &
                                                             &
                                                             &
   2.14                                                      &
   \multirow{2}{*}{2.11(3)~\cite{GF79},%
                   2.07(2)~\cite{Ti10},%
                   1.87(14)~\cite{LM78}}
\\
                                                             &
                                                             &
                                                             &
   AV18+UIX                                                  &
   2.15                                                      &
   2.13(1)                                                   &
   2.14                                                      &
   
\\ \cline{3-8}
                                                             &
                                                             &
   \multirow{2}{*}{$d_{3/2}$}                                &
   AV18+IL7                                                  &
                                                             &
                                                             &
   -0.0848                                                   &
   
\\
                                                             &
                                                             &
                                                             &
   AV18+UIX                                                  &
   -0.0925                                                   &
   -0.0979(9)                                                &
   -0.0842                                                   &
   
\\ \cline{3-8}
                                                             &
                                                             &
   \multirow{2}{*}{$C_d/C_s$}                                &
   AV18+IL7                                                  &
                                                             &
                                                             &
   -0.0396                                                   &
   \multirow{2}{*}{-0.0418(15)~\cite{PK10}}
\\                      
                                                             &
                                                             &
                                                             &
   AV18+UIX                                                  &
   -0.0430                                                   &
   -0.0460(5)                                                &
   -0.0393                                                   &

\\ \hline
   \multirow{6}{*}{\nuc{3}{He}$(\frac{1}{2}^+,\frac{1}{2})$} &
   \multirow{6}{*}{\nuc{2}{H}$(1^+,0)$}                      &
   \multirow{2}{*}{$s_{1/2}$}                                &
   AV18+IL7                                                  &
                                                             &
                                                             &
   2.10                                                      &
   \multirow{2}{*}{2.10(16)~\cite{LM78}, 1.76(11)~\cite{BC85}}
\\
                                                             &
                                                             &
                                                             &
   AV18+UIX                                                  & 
   2.16                                                      &
   2.14(1)                                                   &
   2.10                                                      &
   
\\  \cline{3-8}
                                                             &
                                                             &
   \multirow{2}{*}{$d_{3/2}$}                                &
   AV18+IL7                                                  & 
                                                             &
                                                             &
   -0.0762                                                   &
   
\\
                                                             &
                                                             &
                                                             &
   AV18+UIX                                                  & 
   -0.0865                                                   &
   -0.0927(10)                                               &
   -0.0794                                                   &
   
\\ \cline{3-8}
                                                             &
                                                             &
   \multirow{2}{*}{$C_d/C_s$}                                &
   AV18+IL7                                                  &
                                                             &
                                                             &
   -0.0363                                                   &
   \multirow{2}{*}{-0.0389(42)~\cite{PK10}}
\\                      
                                                             &
                                                             &
                                                             &
   AV18+UIX                                                  &
   -0.0400                                                   &
   -0.0432(5)                                                &
   -0.0378                                                   &
   
\\ \hline
   \multirow{4}{*}{\nuc{4}{He}$(0^+,0)$}                     &
   \multirow{2}{*}{\nuc{3}{H}$(\frac{1}{2}^+,\frac{1}{2})$}  &
   \multirow{2}{*}{$s_{1/2}$}                                &
   AV18+IL7                                                  & 
                                                             &
                                                             &
   -6.45                                                     &
   \multirow{2}{*}{7.36(19)~\cite{LM78},%
                   6.70(50)~\cite{LM78}, 5.44(15)~\cite{BB77}}
\\
                                                             &
                                                             &
                                                             &
   AV18+UIX                                                  &
   -6.47                                                     &
   -6.55(2)                                                  &
   -6.49                                                     &
   
\\ \cline{2-8}
                                                             &
   \multirow{2}{*}{\nuc{3}{He}$(\frac{1}{2}^+,\frac{1}{2})$} &
   \multirow{2}{*}{$s_{1/2}$}                                &
   AV18+IL7                                                  & 
                                                             &
                                                             &
   6.45                                                      &
   \multirow{2}{*}{6.77(51)~\cite{LM78}, 6.52(49)~\cite{LM78}}
\\
                                                             &
                                                             &
                                                             &
   AV18+UIX                                                  & 
   6.36                                                      &
   6.42(2)                                                   &
   6.49                                                      &
   
\end{tabular}
\end{ruledtabular}
\end{table}

In the $s$-shell, GFMC and HH overlaps obtained for the AV18+UIX force agree well (see \figref{fig h2n and h3p AV18+UIX}), and so do ANCs in \tabref{tab ANCs s-shell} except for the weak $d$-waves in $A=3$ nuclei (and consequently for the ratios of $d$- to $s$-wave ANCs) for which the two methods differ for unknown reasons. The HH ANCs assume the actual separation energies for a given Hamiltonian, which are close to the experimental values, and so the ANCs would not change dramatically if experimental separation energies were used instead. In~\cite{NW11}, an integral method (IM) was used to compute ANCs in $A \leq 9$ nuclei from AV18+UIX VMC wave functions assuming experimental core-valence separation energies; in \tabref{tab ANCs s-shell}, these ANCs are labeled VMC-IM. For the dominant $s$-wave channels, VMC-IM results are in a good agreement with ours. On the other hand, the difference between GFMC and VMC-IM ANCs for the weaker $d$-waves in $A=3$ nuclei, larger than 15\%, is outside of the systematic error bars set on GFMC ANCs; this discrepancy is most likely due to VMC wave functions employed by VMC-IM that, as shown in \tabref{tab SFs s-shell} in terms of SFs, underestimate the strength of $d$-waves in $A=3$ nuclei. Our ANCs are in a fair agreement with some experimental values but differ from others. Experimentally, particular emphasis was put on the ratio of $d$- to $s$-wave ANCs in \nuc{3}{H} and \nuc{3}{He} most precisely inferred from tensor analysing powers; the experimental values listed in \tabref{tab ANCs s-shell} agree well with ours, including the relative negative phase.

\begin{table}
\caption{ANCs for GFMC overlaps between $A=6$ and 7 $p$-shell nuclei for the
         AV18+IL7 potential in angular momentum channels $l_j$. Systematic 
         uncertainties on GFMC values are 5\% or less as discussed in 
         \secref{sec inner workings}. Experimentally deduced values are also 
         shown.}
\label{tab ANCs p-shell}
\begin{ruledtabular}
\begin{tabular}{cccccR-[.]{1}{2}c}
   \multicolumn{2}{c}{parent}                   &
   \multicolumn{2}{c}{core}                     &
   $l_j$                                        &
   \multicolumn{2}{c}{$C$~$[\text{fm}^{-1/2}]$ }
\\ \cline{1-2} \cline{3-4} \cline{6-7}
   $^{A}Z(J^\pi,T)$         &
   Type                     &
   $^{A-1}Z(J^\pi,T)$       &
   Type                     &
                            &
   \multicolumn{1}{c}{GFMC} & 
   \multicolumn{1}{c}{exp.}
\\ \hline
   \multirow{12}{*}{\nuc{7}{Li}$(\frac{3}{2}^-,\frac{1}{2})$} &
   I                                                          &
   \multirow{2}{*}{\nuc{6}{He}$(0^+,1)$}                      &
   \multirow{2}{*}{I}                                         &
   \multirow{2}{*}{$p_{3/2}$}                                 &
   3.52                                                       &
   \multirow{2}{*}{2.48~\cite{BK91}}
\\
                                                              &
   II                                                         &
                                                              &
                                                              &
                                                              &
   3.65                                                       &
   
\\ \cline{2-7}
                                                              &
   \multirow{3}{*}{I}                                         &
   \multirow{6}{*}{\nuc{6}{Li}$(1^+,0)$}                      &
   \multirow{3}{*}{I}                                         &
   $p_{1/2}$                                                  &
   1.73                                                       &

\\
                                                              &
                                                              &
                                                              &
                                                              &
   $p_{3/2}$                                                  &
   2.29                                                       &

\\
                                                               &
                                                               &
                                                               &
                                                               &
   $\sqrt{\sum C^2}$                                           &
   2.87                                                        &
%= some other values are:
%= 1.83(6), 2.52(6)
%= from Goncharov et al., Czech. J. Phys. 37 (1987) 168
   1.26-2.82~\cite{GM95}
\\ \cline{2-2} \cline{4-7}
                                                              &
   \multirow{3}{*}{II}                                        &
                                                              &
   \multirow{3}{*}{II}                                        &
   $p_{1/2}$                                                  &
   1.85                                                       &
   
\\
                                                              &
                                                              &
                                                              &
                                                              &
   $p_{3/2}$                                                  &
   2.20                                                       &
   
\\
                                                               &
                                                               &
                                                               &
                                                               &
   $\sqrt{\sum C^2}$                                           &
   2.87                                                        &
   1.26-2.82~\cite{GM95}
\\ \cline{2-7}
                                                              &
   I                                                          &
   \multirow{2}{*}{\nuc{6}{Li}$(3^+,0)$}                      &
   \multirow{2}{*}{I}                                         &
   \multirow{2}{*}{$p_{3/2}$}                                 &
   3.50                                                       &
   \multirow{2}{*}{2.06-3.00~\cite{GM95}}
\\
                                                              &
   II                                                         &
                                                              &
                                                              &
                                                              &
   3.63                                                       &
   
\\ \cline{2-7}
                                                              &
   I                                                          &
   \multirow{2}{*}{\nuc{6}{Li}$(0^+,1)$}                      &
   \multirow{2}{*}{I}                                         &
   \multirow{2}{*}{$p_{3/2}$}                                 &
   -2.39                                                      &
   \multirow{2}{*}{1.71-2.62~\cite{GM95}}
\\
                                                              &
   II                                                         &
                                                              &
                                                              &
                                                              &
   -2.46                                                      &
   
\\ \hline
   \multirow{3}{*}{\nuc{7}{Li}$(\frac{1}{2}^-,\frac{1}{2})$}  &
   \multirow{3}{*}{I}                                         &
   \multirow{3}{*}{\nuc{6}{Li}$(1^+,0)$}                      &
   \multirow{3}{*}{I}                                         &
   $p_{1/2}$                                                  &
   -0.573                                                     &

\\
                                                              &
                                                              &
                                                              &
                                                              &
   $p_{3/2}$                                                  &
   -2.85                                                      &

\\
                                                              &
                                                              &
                                                              &
                                                              &
   $\sqrt{\sum C^2}$                                          &
   2.91                                                       &

\\ \hline
   \multirow{5}{*}{\nuc{7}{Be}$(\frac{3}{2}^-,\frac{1}{2})$}  &
   \multirow{5}{*}{I}                                         &
   \multirow{3}{*}{\nuc{6}{Li}$(1^+,0)$}                      &
   \multirow{3}{*}{I}                                         &
   $p_{1/2}$                                                  &
   -1.70                                                      &
   
\\
                                                              &
                                                              &
                                                              &
                                                              &
   $p_{3/2}$                                                  &
   -2.20                                                      &
   
\\
                                                              &
                                                              &
                                                              &
                                                              &
   $\sqrt{\sum C^2}$                                          &
   2.78                                                       &
   
\\ \cline{3-7}
                                                              &
                                                              &
   \nuc{6}{Li}$(3^+,0)$                                       &
   I                                                          &
   $p_{3/2}$                                                  &
   -3.49                                                      &
   
\\ \cline{3-7}
                                                              &
                                                              &
   \nuc{6}{Li}$(0^+,1)$                                       &
   I                                                          &
   $p_{3/2}$                                                  &
   -2.58                                                      &
   
\end{tabular}
\end{ruledtabular}
\end{table}

As shown in \tabref{tab ANCs p-shell}, ANCs for overlaps between $A=6$ and 7 nuclei are constrained rather poorly experimentally. In general, ANCs in the lower $p$-shell are determined mostly from hadronic processes, such as transfer or knockout reactions; see~\cite{Ti10, NW11} and references therein. In \tabref{tab ANCs p-shell}, we show the full range of experimental ANCs from~\cite{GM95} for overlaps between the ground state of \nuc{7}{Li} and the lowest states in \nuc{6}{Li}. Overall, given the experimental uncertainties, it is hard to compare GFMC to experiment. 
% Also, we are not aware of any other realistic calculations of $p$-shell ANCs for AV18+IL7; 
ANCs calculated within VMC-IM for AV18+UIX are in a broad agreement with our AV18+IL7 results, though with some notable differences especially for the $p_{3/2}$ \overlap{\nuc{6}{Li}($1^+$)}{\nuc{7}{Li}$(3/2^-)$} overlap.  Based on simple analytical arguments, the ratio of \overlap{\nuc{6}{Li}($1^+$)}{\nuc{7}{Be}$(3/2^-)$} and \overlap{\nuc{6}{Li}($1^+$)}{\nuc{7}{Li}$(3/2^-)$} ANCs involving mirror nuclei \nuc{7}{Be} and \nuc{7}{Li} was predicted to be 1.02 for both $p_{1/2}$ and $p_{3/2}$ channels~\cite{TJ03}. We obtain 0.95(7) for the $p_{1/2}$ channel and 0.98(7) for the $p_{3/2}$ channel assuming 5\% errors on GFMC ANCs, using averaged Type~I and Type~II \overlap{\nuc{6}{Li}($1^+$)}{\nuc{7}{Li}($3/2^-$)} ANCs from \tabref{tab ANCs p-shell}, and disregarding the signs of involved ANCs.

Finally, we present in \tabref{tab fits} potential parameters appearing in \myeqref{eq fitting potential} that provide good fits to GFMC overlaps. These fits were used to extract ANCs. They also allow our results to be easily used in DWBA or CCBA reaction calculations~\cite{MP78, Th88}. Besides setting the core-valence separation energies to their experimental values, no other constraints were imposed in the fitting procedure. In the $s$-shell, we found it necessary to introduce a short-range repulsive Gaussian potential term to reproduce dips in GFMC overlaps near the origin as seen in \figref{fig h2n and h3p AV18+IL7}; in the $p$-shell, such a term is not necessary. Good fits to $d$-wave overlaps in $A=3$ nuclei seem to require very strong repulsive potential cores and negative radii of the central WS part. Also, for $d$-waves the first few radial bins were omitted from the fit because they are rather uncertain.
%. as can be seen in Figs.~\ref{fig h2n and h3p AV18+UIX} and \ref{fig h2n and h3p AV18+IL7}. 
In order to reproduce the results presented in this work, the WS fits 
% in \tabref{tab fits} 
need to be normalized to SFs from Tables~\ref{tab SFs s-shell} and \ref{tab SFs p-shell}. Our GFMC overlaps are available from~\cite{www-overlaps}.
% At this point it is too early to draw any conclusions about systematics of fitting overlap parameters in the lower $p$-shell, a task that may be possible after completion of the second part of this work for $A>7$ nuclei.

\begin{table}
\caption{Fitting parameters from \myeqref{eq fitting potential} for GFMC 
         overlaps obtained for the AV18+IL7 potential in angular momentum 
         channels $l_j$. All $p$-shell overlaps originated from Type~I VMC wave 
         functions. The precise values of $V_{WS}$ are adjusted to reproduce the
         experimental separation energies $B_{exp}$.}
\label{tab fits}
%\resizebox{20cm}{!}{
\begin{ruledtabular}
\begin{tabular}{ccc
                R-[.][.]{4}{2}R-[.][.]{1}{2}R[.][.]{1}{2}
                R[.][.]{1}{2}R[.][.]{1}{2}
                R[.][.]{1}{2}R[.][.]{1}{2}R[.][.]{1}{2}
                D{.}{.}{2.4}}
   parent                        &
   core                          &
   $l_j$                         &
   \multicolumn{1}{c}{$V_{WS}$}  &
   \multicolumn{1}{c}{$R_{WS}$}  &
   \multicolumn{1}{c}{$a_{WS}$}  &
   \multicolumn{1}{c}{$\beta$}   &
   \multicolumn{1}{c}{$\rho$}    &
   \multicolumn{1}{c}{$V_{so}$}  &
   \multicolumn{1}{c}{$R_{so}$}  &
   \multicolumn{1}{c}{$a_{so}$}  &
   \multicolumn{1}{c}{$B_{exp}$}
\\ \cline{1-1} \cline{2-2}
   $^{A}Z(J^\pi,T)$           &
   $^{A-1}Z(J^\pi,T)$         &
                              &
   \multicolumn{1}{c}{[MeV]}  &
   \multicolumn{1}{c}{[fm]}   &
   \multicolumn{1}{c}{[fm]}   &
                              &
   \multicolumn{1}{c}{[fm]}   &
   \multicolumn{1}{c}{[MeV]}  &
   \multicolumn{1}{c}{[fm]}   &
   \multicolumn{1}{c}{[fm]}   &
   \multicolumn{1}{c}{[MeV]}
\\ \hline 
   \multirow{2}{*}{\nuc{3}{H}$(\frac{1}{2}^+,\frac{1}{2})$}    &
   \multirow{2}{*}{\nuc{2}{H}$(1^+,0)$}                        &
   $s_{1/2}$                                                   &
   -172.88  & 0.55814 & 0.69247 & 1.1537 & 0.6405 &
   \multicolumn{1}{c}{} & \multicolumn{1}{c}{} & \multicolumn{1}{c}{} & 6.2572
\\
                                                               &
                                                               &
   $d_{3/2}$                                                   &
   -2732.9 & -1.1506 & 0.9078 & 1.1637 & 0.40701 &
    2.9535 &  2.1452 & 0.14  & 6.2572
\\ \hline
   \multirow{2}{*}{\nuc{3}{He}$(\frac{1}{2}^+,\frac{1}{2})$}   &
   \multirow{2}{*}{\nuc{2}{H}$(1^+,0)$}                        &
   $s_{1/2}$                                                   &
   -179.94 & 0.54031 & 0.6817 & 1.1294 & 0.64371 &
   \multicolumn{1}{c}{} & \multicolumn{1}{c}{} & \multicolumn{1}{c}{} & 5.4934
\\
                                                               &
                                                               &
   $d_{3/2}$                                                   &
   -8155.1 & -2.191 & 0.90647  & 1.0296 & 0.34857 &
   1.4728  & 2.0738 & 0.059638 & 5.4934
\\ \hline
   \multirow{2}{*}{\nuc{4}{He}$(0^+,0)$}                       &
   \nuc{3}{H}$(\frac{1}{2}^+,\frac{1}{2})$                     &
   $s_{1/2}$                                                   &
   -202.21 & 0.92669 & 0.65737 & 0.87391 & 0.81011 &
   \multicolumn{1}{c}{} & \multicolumn{1}{c}{} & \multicolumn{1}{c}{}& 19.814
\\ \cline{2-12}
                                                               &
   \nuc{3}{He}$(\frac{1}{2}^+,\frac{1}{2})$                    &
   $s_{1/2}$                                                   &
   -200.93 & 0.88285 & 0.68596 & 0.86556 & 0.78838 &
   \multicolumn{1}{c}{} & \multicolumn{1}{c}{} & \multicolumn{1}{c}{} & 20.578
\\ \hline
   \multirow{5}{*}{\nuc{7}{Li}$(\frac{3}{2}^-,\frac{1}{2})$}   &
   \nuc{6}{He}$(0^+,1)$                                        &
   $p_{3/2}$                                                   &
   -58.934 & 2.6801 & 0.92913  & \multicolumn{1}{c}{} & \multicolumn{1}{c}{} &
    1.2597 & 0.9095 & 0.065291 & 9.9758
\\ \cline{2-12}
                                                               &
   \multirow{2}{*}{\nuc{6}{Li}$(1^+,0)$}                       &
   $p_{1/2}$                                                   &
   -41.796 & 3.1831 & 0.84704 & \multicolumn{1}{c}{} & \multicolumn{1}{c}{} &
    2.4462 & 1.354 & 0.17327  & 7.25
\\
                                                               &
                                                               &
   $p_{3/2}$                                                   &
   -69.548 & 1.8943 & 1.1724  & \multicolumn{1}{c}{} & \multicolumn{1}{c}{} &
    2.1328 & 2.3571 & 0.21291 & 7.25
\\ \cline{2-12}
                                                               &
   \nuc{6}{Li}$(3^+,0)$                                        &
   $p_{3/2}$                                                   &
   -62.98  & 2.3506 & 1.1802   & \multicolumn{1}{c}{} & \multicolumn{1}{c}{} &
    1.7223 & 0.5542 & 0.088081 & 9.436
\\ \cline{2-12}
                                                               &
   \nuc{6}{Li}$(0^+,1)$                                        &
   $p_{3/2}$                                                   &
   -59.392  & 2.6402  & 0.96616 & \multicolumn{1}{c}{} & \multicolumn{1}{c}{} &
    0.92785 & 0.91339 & 0.21999 & 10.813
\\ \hline
   \multirow{2}{*}{\nuc{7}{Li}$(\frac{1}{2}^-,\frac{1}{2})$}   &
   \multirow{2}{*}{\nuc{6}{Li}$(1^+,0)$}                       &
   $p_{1/2}$                                                   &
   -33.705 & 3.3855 & 0.30649  & \multicolumn{1}{c}{} & \multicolumn{1}{c}{} &
    1.437  & 1.0883 & 0.034505 & 6.772
\\
                                                               &
                                                               &
   $p_{3/2}$                                                   &
   -64.996 & 2.0446  & 1.1513   & \multicolumn{1}{c}{} & \multicolumn{1}{c}{} &
    1.1429 & 0.87519 & 0.034059 & 6.772
\\ \hline
   \multirow{4}{*}{\nuc{7}{Be}$(\frac{3}{2}^-,\frac{1}{2})$}   &
   \multirow{2}{*}{\nuc{6}{Li}$(1^+,0)$}                       &
   $p_{1/2}$                                                   &
   -39.448 & 3.3242 & 0.75701 & \multicolumn{1}{c}{} & \multicolumn{1}{c}{} &
    3.4705 & 1.4417 & 0.3862  & 5.6055
\\
                                                               &
                                                               &
   $p_{3/2}$                                                   &
   -72.217 & 1.854  & 1.1092  & \multicolumn{1}{c}{} & \multicolumn{1}{c}{} &
    2.6203 & 2.5255 & 0.31134 & 5.6055
\\ \cline{2-12}
                                                               &
   \nuc{6}{Li}$(3^+,0)$                                        &
   $p_{3/2}$                                                   &
   -59.198 & 2.5166  & 1.0688   & \multicolumn{1}{c}{} & \multicolumn{1}{c}{} &
    1.0782 & 0.79152 & 0.032701 & 7.7915
\\ \cline{2-12}
                                                               &
   \nuc{6}{Li}$(0^+,1)$                                        &
   $p_{3/2}$                                                   &
   -59.489 & 2.6359 & 0.96491  & \multicolumn{1}{c}{} & \multicolumn{1}{c}{} &
    1.0627 & 1.0535 & 0.032995 & 9.1685
\end{tabular}
\end{ruledtabular}
%}  %= resizebox
\end{table}

%================================= conclusions =================================
\section{Summary and conclusions}
\label{sec conclusions}

We have reported Green's function Monte Carlo calculations of one-nucleon spectroscopic overlaps in nuclei with mass numbers $A \leq 7$. The calculations have used wave functions derived from a realistic Hamiltonian that reproduces well the low-lying spectra of light nuclei. The overlaps are extrapolated from mixed estimates between VMC and GFMC wave functions, and they are extended to regions beyond the nuclear surface, where the GFMC method may not be accurate, by means of WS fits to the overlap interior. A goal of this work is to provide a consistent set of spectroscopic overlaps, SFs, and ANCs in light nuclei, obtained from our currently best GFMC wave functions, that can be used as structure input in analyzing existing or future experimental data.

The comparison of SFs and ANCs with experimentally deduced values is obscured, as mentioned in \secref{sec introduction}, by the model-dependent way these quantities are extracted from experimental data and by the issues related to the meaning of spectroscopic overlaps. For many overlaps, it is hard to judge the agreement or disagreement between theory and experiment because of the nonexisting or conflicting experimental values. Our (relative) SFs seem to broadly support more recent values deduced from hadronic processes, and some agree particularly well with values provided by electron-scattering experiments. Our calculations reproduce, within error bars, the experimentally well-deduced ratios of $d$- to $s$-wave ANCs in $A=3$ nuclei.We observe a rather substantial (up to 40\%) quenching of GFMC SFs when compared with the traditional shell-model. The GFMC imrpoves the VMC overlaps, but the corrections to SFs are sufficiently small indicating that the VMC values used to analyze experimental data in~\cite{WR05,WR05a,WS08,KA08,GB11,LW99} were reliable.

The GFMC overlaps presented in this paper are available from~\cite{www-overlaps}. The overlaps for somewhat heavier nuclei up to $A \lesssim 10$ will be the subject of a forthcoming paper.

%=============================== acknowledgments================================
\acknowledgments

We thank D.~Kurath and K.~M.~Nollett for many valuable discussions and M.~Viviani and A.~Kievsky for their comments on $s$-shell overlaps and for providing us with hyperspherical harmonics results. The many-body calculations were performed on the parallel computers of the Laboratory Computing Resource Center and of the Mathematics and Computer Science Division, Argonne National Laboratory. This work is supported by the U. S. Department of Energy, Office of Nuclear Physics, under contract No. DE-AC02-06CH11357 and under SciDAC grant No. DE-FC02-07ER41457.

%==================================== Biblio ===================================
\bibliography{references}

\begin{thebibliography}{64}%
\makeatletter
\providecommand \@ifxundefined [1]{%
 \@ifx{#1\undefined}
}%
\providecommand \@ifnum [1]{%
 \ifnum #1\expandafter \@firstoftwo
 \else \expandafter \@secondoftwo
 \fi
}%
\providecommand \@ifx [1]{%
 \ifx #1\expandafter \@firstoftwo
 \else \expandafter \@secondoftwo
 \fi
}%
\providecommand \natexlab [1]{#1}%
\providecommand \enquote  [1]{``#1''}%
\providecommand \bibnamefont  [1]{#1}%
\providecommand \bibfnamefont [1]{#1}%
\providecommand \citenamefont [1]{#1}%
\providecommand \href@noop [0]{\@secondoftwo}%
\providecommand \href [0]{\begingroup \@sanitize@url \@href}%
\providecommand \@href[1]{\@@startlink{#1}\@@href}%
\providecommand \@@href[1]{\endgroup#1\@@endlink}%
\providecommand \@sanitize@url [0]{\catcode `\\12\catcode `\$12\catcode
  `\&12\catcode `\#12\catcode `\^12\catcode `\_12\catcode `\%12\relax}%
\providecommand \@@startlink[1]{}%
\providecommand \@@endlink[0]{}%
\providecommand \url  [0]{\begingroup\@sanitize@url \@url }%
\providecommand \@url [1]{\endgroup\@href {#1}{\urlprefix }}%
\providecommand \urlprefix  [0]{URL }%
\providecommand \Eprint [0]{\href }%
\providecommand \doibase [0]{http://dx.doi.org/}%
\providecommand \selectlanguage [0]{\@gobble}%
\providecommand \bibinfo  [0]{\@secondoftwo}%
\providecommand \bibfield  [0]{\@secondoftwo}%
\providecommand \translation [1]{[#1]}%
\providecommand \BibitemOpen [0]{}%
\providecommand \bibitemStop [0]{}%
\providecommand \bibitemNoStop [0]{.\EOS\space}%
\providecommand \EOS [0]{\spacefactor3000\relax}%
\providecommand \BibitemShut  [1]{\csname bibitem#1\endcsname}%
\let\auto@bib@innerbib\@empty
%</preamble>
\bibitem [{\citenamefont {Clement}(1973)}]{Cl73}%
  \BibitemOpen
  \bibfield  {author} {\bibinfo {author} {\bibfnamefont {C.~F.}\ \bibnamefont
  {Clement}},\ }\href {\doibase DOI: 10.1016/0375-9474(73)90747-1} {\bibfield
  {journal} {\bibinfo  {journal} {Nucl.\ Phys.\ A}\ }\textbf {\bibinfo {volume}
  {213}},\ \bibinfo {pages} {469 } (\bibinfo {year} {1973})}\BibitemShut
  {NoStop}%
\bibitem [{\citenamefont {Bang}\ \emph {et~al.}(1985)\citenamefont {Bang},
  \citenamefont {Gareev}, \citenamefont {Pinkston},\ and\ \citenamefont
  {Vaagen}}]{BG85}%
  \BibitemOpen
  \bibfield  {author} {\bibinfo {author} {\bibfnamefont {J.~M.}\ \bibnamefont
  {Bang}}, \bibinfo {author} {\bibfnamefont {F.~G.}\ \bibnamefont {Gareev}},
  \bibinfo {author} {\bibfnamefont {W.~T.}\ \bibnamefont {Pinkston}}, \ and\
  \bibinfo {author} {\bibfnamefont {J.~S.}\ \bibnamefont {Vaagen}},\ }\href
  {\doibase DOI: 10.1016/0370-1573(85)90132-2} {\bibfield  {journal} {\bibinfo
  {journal} {Phys.\ Rep.}\ }\textbf {\bibinfo {volume} {125}},\ \bibinfo
  {pages} {253 } (\bibinfo {year} {1985})}\BibitemShut {NoStop}%
\bibitem [{\citenamefont {Macfarlane}\ and\ \citenamefont
  {French}(1960)}]{MF60}%
  \BibitemOpen
  \bibfield  {author} {\bibinfo {author} {\bibfnamefont {M.~H.}\ \bibnamefont
  {Macfarlane}}\ and\ \bibinfo {author} {\bibfnamefont {J.~B.}\ \bibnamefont
  {French}},\ }\href {\doibase 10.1103/RevModPhys.32.567} {\bibfield  {journal}
  {\bibinfo  {journal} {Rev. Mod. Phys.}\ }\textbf {\bibinfo {volume} {32}},\
  \bibinfo {pages} {567} (\bibinfo {year} {1960})}\BibitemShut {NoStop}%
\bibitem [{\citenamefont {Satchler}(1983)}]{Sa83}%
  \BibitemOpen
  \bibfield  {author} {\bibinfo {author} {\bibfnamefont {G.~R.}\ \bibnamefont
  {Satchler}},\ }\href@noop {} {\emph {\bibinfo {title} {Direct nuclear
  reactions}}}\ (\bibinfo  {publisher} {Oxford University Press},\ \bibinfo
  {year} {1983})\BibitemShut {NoStop}%
\bibitem [{\citenamefont {Hansen}\ and\ \citenamefont {Tostevin}(2003)}]{HT03}%
  \BibitemOpen
  \bibfield  {author} {\bibinfo {author} {\bibfnamefont {P.~G.}\ \bibnamefont
  {Hansen}}\ and\ \bibinfo {author} {\bibfnamefont {J.~A.}\ \bibnamefont
  {Tostevin}},\ }\href {\doibase 10.1146/annurev.nucl.53.041002.110406}
  {\bibfield  {journal} {\bibinfo  {journal} {Annu.\ Rev.\ Nucl.\ Part.\ Sci.}\
  }\textbf {\bibinfo {volume} {53}},\ \bibinfo {pages} {219} (\bibinfo {year}
  {2003})}\BibitemShut {NoStop}%
\bibitem [{\citenamefont {Thompson}\ and\ \citenamefont {Nunes}(2009)}]{TN09}%
  \BibitemOpen
  \bibfield  {author} {\bibinfo {author} {\bibfnamefont {I.}~\bibnamefont
  {Thompson}}\ and\ \bibinfo {author} {\bibfnamefont {F.~M.}\ \bibnamefont
  {Nunes}},\ }\href@noop {} {\emph {\bibinfo {title} {Nuclear reactions for
  astrophysics}}}\ (\bibinfo  {publisher} {Cambridge University Press, New
  York},\ \bibinfo {year} {2009})\BibitemShut {NoStop}%
\bibitem [{\citenamefont {Pandharipande}\ \emph {et~al.}(1997)\citenamefont
  {Pandharipande}, \citenamefont {Sick},\ and\ \citenamefont {deWitt
  Huberts}}]{PS97}%
  \BibitemOpen
  \bibfield  {author} {\bibinfo {author} {\bibfnamefont {V.~R.}\ \bibnamefont
  {Pandharipande}}, \bibinfo {author} {\bibfnamefont {I.}~\bibnamefont {Sick}},
  \ and\ \bibinfo {author} {\bibfnamefont {P.~K.~A.}\ \bibnamefont {deWitt
  Huberts}},\ }\href {\doibase 10.1103/RevModPhys.69.981} {\bibfield  {journal}
  {\bibinfo  {journal} {Rev. Mod. Phys.}\ }\textbf {\bibinfo {volume} {69}},\
  \bibinfo {pages} {981} (\bibinfo {year} {1997})}\BibitemShut {NoStop}%
\bibitem [{\citenamefont {Kramer}\ \emph {et~al.}(2001)\citenamefont {Kramer},
  \citenamefont {Blok},\ and\ \citenamefont {Lapikás}}]{KB01}%
  \BibitemOpen
  \bibfield  {author} {\bibinfo {author} {\bibfnamefont {G.~J.}\ \bibnamefont
  {Kramer}}, \bibinfo {author} {\bibfnamefont {H.~P.}\ \bibnamefont {Blok}}, \
  and\ \bibinfo {author} {\bibfnamefont {L.}~\bibnamefont {Lapikás}},\ }\href
  {\doibase DOI: 10.1016/S0375-9474(00)00379-1} {\bibfield  {journal} {\bibinfo
   {journal} {Nucl.\ Phys.\ A}\ }\textbf {\bibinfo {volume} {679}},\ \bibinfo
  {pages} {267 } (\bibinfo {year} {2001})}\BibitemShut {NoStop}%
\bibitem [{\citenamefont {Beck}\ \emph {et~al.}(1987)\citenamefont {Beck},
  \citenamefont {Dickmann},\ and\ \citenamefont {Lovas}}]{BD87}%
  \BibitemOpen
  \bibfield  {author} {\bibinfo {author} {\bibfnamefont {R.}~\bibnamefont
  {Beck}}, \bibinfo {author} {\bibfnamefont {F.}~\bibnamefont {Dickmann}}, \
  and\ \bibinfo {author} {\bibfnamefont {R.~G.}\ \bibnamefont {Lovas}},\ }\href
  {\doibase DOI: 10.1016/0003-4916(87)90091-1} {\bibfield  {journal} {\bibinfo
  {journal} {Ann.\ Phys.}\ }\textbf {\bibinfo {volume} {173}},\ \bibinfo
  {pages} {1 } (\bibinfo {year} {1987})}\BibitemShut {NoStop}%
\bibitem [{\citenamefont {Mukhamedzhanov}\ and\ \citenamefont
  {Kadyrov}(2010)}]{MK10}%
  \BibitemOpen
  \bibfield  {author} {\bibinfo {author} {\bibfnamefont {A.~M.}\ \bibnamefont
  {Mukhamedzhanov}}\ and\ \bibinfo {author} {\bibfnamefont {A.~S.}\
  \bibnamefont {Kadyrov}},\ }\href {\doibase 10.1103/PhysRevC.82.051601}
  {\bibfield  {journal} {\bibinfo  {journal} {Phys. Rev. C}\ }\textbf {\bibinfo
  {volume} {82}},\ \bibinfo {pages} {051601} (\bibinfo {year}
  {2010})}\BibitemShut {NoStop}%
\bibitem [{\citenamefont {Lee}\ \emph {et~al.}(2006)\citenamefont {Lee},
  \citenamefont {Tostevin}, \citenamefont {Brown}, \citenamefont {Delaunay},
  \citenamefont {Lynch}, \citenamefont {Saelim},\ and\ \citenamefont
  {Tsang}}]{LT06}%
  \BibitemOpen
  \bibfield  {author} {\bibinfo {author} {\bibfnamefont {J.}~\bibnamefont
  {Lee}}, \bibinfo {author} {\bibfnamefont {J.~A.}\ \bibnamefont {Tostevin}},
  \bibinfo {author} {\bibfnamefont {B.~A.}\ \bibnamefont {Brown}}, \bibinfo
  {author} {\bibfnamefont {F.}~\bibnamefont {Delaunay}}, \bibinfo {author}
  {\bibfnamefont {W.~G.}\ \bibnamefont {Lynch}}, \bibinfo {author}
  {\bibfnamefont {M.~J.}\ \bibnamefont {Saelim}}, \ and\ \bibinfo {author}
  {\bibfnamefont {M.~B.}\ \bibnamefont {Tsang}},\ }\href {\doibase
  10.1103/PhysRevC.73.044608} {\bibfield  {journal} {\bibinfo  {journal} {Phys.
  Rev. C}\ }\textbf {\bibinfo {volume} {73}},\ \bibinfo {pages} {044608}
  (\bibinfo {year} {2006})}\BibitemShut {NoStop}%
\bibitem [{\citenamefont {Bisconti}\ \emph {et~al.}(2007)\citenamefont
  {Bisconti}, \citenamefont {de~Saavedra},\ and\ \citenamefont {Co'}}]{BS07}%
  \BibitemOpen
  \bibfield  {author} {\bibinfo {author} {\bibfnamefont {C.}~\bibnamefont
  {Bisconti}}, \bibinfo {author} {\bibfnamefont {F.~A.}\ \bibnamefont
  {de~Saavedra}}, \ and\ \bibinfo {author} {\bibfnamefont {G.}~\bibnamefont
  {Co'}},\ }\href {\doibase 10.1103/PhysRevC.75.054302} {\bibfield  {journal}
  {\bibinfo  {journal} {Phys. Rev. C}\ }\textbf {\bibinfo {volume} {75}},\
  \bibinfo {pages} {054302} (\bibinfo {year} {2007})}\BibitemShut {NoStop}%
\bibitem [{\citenamefont {Timofeyuk}(2010)}]{Ti10}%
  \BibitemOpen
  \bibfield  {author} {\bibinfo {author} {\bibfnamefont {N.~K.}\ \bibnamefont
  {Timofeyuk}},\ }\href {\doibase 10.1103/PhysRevC.81.064306} {\bibfield
  {journal} {\bibinfo  {journal} {Phys. Rev. C}\ }\textbf {\bibinfo {volume}
  {81}},\ \bibinfo {pages} {064306} (\bibinfo {year} {2010})}\BibitemShut
  {NoStop}%
\bibitem [{\citenamefont {Kievsky}\ \emph {et~al.}(1997)\citenamefont
  {Kievsky}, \citenamefont {Rosati}, \citenamefont {Viviani}, \citenamefont
  {Brune}, \citenamefont {Karwowski}, \citenamefont {Ludwig},\ and\
  \citenamefont {Wood}}]{KR97}%
  \BibitemOpen
  \bibfield  {author} {\bibinfo {author} {\bibfnamefont {A.}~\bibnamefont
  {Kievsky}}, \bibinfo {author} {\bibfnamefont {S.}~\bibnamefont {Rosati}},
  \bibinfo {author} {\bibfnamefont {M.}~\bibnamefont {Viviani}}, \bibinfo
  {author} {\bibfnamefont {C.~R.}\ \bibnamefont {Brune}}, \bibinfo {author}
  {\bibfnamefont {H.~J.}\ \bibnamefont {Karwowski}}, \bibinfo {author}
  {\bibfnamefont {E.~J.}\ \bibnamefont {Ludwig}}, \ and\ \bibinfo {author}
  {\bibfnamefont {M.~H.}\ \bibnamefont {Wood}},\ }\href {\doibase DOI:
  10.1016/S0370-2693(97)00691-6} {\bibfield  {journal} {\bibinfo  {journal}
  {Phys.\ Lett.\ B}\ }\textbf {\bibinfo {volume} {406}},\ \bibinfo {pages} {292
  } (\bibinfo {year} {1997})}\BibitemShut {NoStop}%
\bibitem [{\citenamefont {Viviani}\ \emph {et~al.}(2005)\citenamefont
  {Viviani}, \citenamefont {Kievsky},\ and\ \citenamefont {Rosati}}]{VK05}%
  \BibitemOpen
  \bibfield  {author} {\bibinfo {author} {\bibfnamefont {M.}~\bibnamefont
  {Viviani}}, \bibinfo {author} {\bibfnamefont {A.}~\bibnamefont {Kievsky}}, \
  and\ \bibinfo {author} {\bibfnamefont {S.}~\bibnamefont {Rosati}},\ }\href
  {\doibase 10.1103/PhysRevC.71.024006} {\bibfield  {journal} {\bibinfo
  {journal} {Phys. Rev. C}\ }\textbf {\bibinfo {volume} {71}},\ \bibinfo
  {pages} {024006} (\bibinfo {year} {2005})}\BibitemShut {NoStop}%
\bibitem [{\citenamefont {Navr\'atil}(2004)}]{Na04}%
  \BibitemOpen
  \bibfield  {author} {\bibinfo {author} {\bibfnamefont {P.}~\bibnamefont
  {Navr\'atil}},\ }\href {\doibase 10.1103/PhysRevC.70.054324} {\bibfield
  {journal} {\bibinfo  {journal} {Phys. Rev. C}\ }\textbf {\bibinfo {volume}
  {70}},\ \bibinfo {pages} {054324} (\bibinfo {year} {2004})}\BibitemShut
  {NoStop}%
\bibitem [{\citenamefont {Navr\'atil}\ \emph {et~al.}(2007)\citenamefont
  {Navr\'atil}, \citenamefont {Bertulani},\ and\ \citenamefont
  {Caurier}}]{NB07}%
  \BibitemOpen
  \bibfield  {author} {\bibinfo {author} {\bibfnamefont {P.}~\bibnamefont
  {Navr\'atil}}, \bibinfo {author} {\bibfnamefont {C.~A.}\ \bibnamefont
  {Bertulani}}, \ and\ \bibinfo {author} {\bibfnamefont {E.}~\bibnamefont
  {Caurier}},\ }\href {\doibase DOI: 10.1016/j.nuclphysa.2006.12.082}
  {\bibfield  {journal} {\bibinfo  {journal} {Nucl.\ Phys.\ A}\ }\textbf
  {\bibinfo {volume} {787}},\ \bibinfo {pages} {539 } (\bibinfo {year}
  {2007})},\ \bibinfo {note} {proceedings of the Ninth International Conference
  on Nucleus-Nucleus Collisions - (NN2006)}\BibitemShut {NoStop}%
\bibitem [{\citenamefont {Jensen}\ \emph {et~al.}(2010)\citenamefont {Jensen},
  \citenamefont {Hagen}, \citenamefont {Papenbrock}, \citenamefont {Dean},\
  and\ \citenamefont {Vaagen}}]{JH10}%
  \BibitemOpen
  \bibfield  {author} {\bibinfo {author} {\bibfnamefont {O.}~\bibnamefont
  {Jensen}}, \bibinfo {author} {\bibfnamefont {G.}~\bibnamefont {Hagen}},
  \bibinfo {author} {\bibfnamefont {T.}~\bibnamefont {Papenbrock}}, \bibinfo
  {author} {\bibfnamefont {D.~J.}\ \bibnamefont {Dean}}, \ and\ \bibinfo
  {author} {\bibfnamefont {J.~S.}\ \bibnamefont {Vaagen}},\ }\href {\doibase
  10.1103/PhysRevC.82.014310} {\bibfield  {journal} {\bibinfo  {journal} {Phys.
  Rev. C}\ }\textbf {\bibinfo {volume} {82}},\ \bibinfo {pages} {014310}
  (\bibinfo {year} {2010})}\BibitemShut {NoStop}%
\bibitem [{www()}]{www-overlaps}%
  \BibitemOpen
  \href@noop {} {}\bibinfo {note}
  {\url{http://www.phy.anl.gov/theory/research/overlap/}}\BibitemShut {NoStop}%
\bibitem [{\citenamefont {Nollett}\ and\ \citenamefont {Wiringa}(2011)}]{NW11}%
  \BibitemOpen
  \bibfield  {author} {\bibinfo {author} {\bibfnamefont {K.~M.}\ \bibnamefont
  {Nollett}}\ and\ \bibinfo {author} {\bibfnamefont {R.~B.}\ \bibnamefont
  {Wiringa}},\ }\href {\doibase 10.1103/PhysRevC.83.041001} {\bibfield
  {journal} {\bibinfo  {journal} {Phys. Rev. C}\ }\textbf {\bibinfo {volume}
  {83}},\ \bibinfo {pages} {041001} (\bibinfo {year} {2011})}\BibitemShut
  {NoStop}%
\bibitem [{\citenamefont {Wuosmaa}\ \emph
  {et~al.}(2005{\natexlab{a}})\citenamefont {Wuosmaa}, \citenamefont {Rehm},
  \citenamefont {Greene}, \citenamefont {Henderson}, \citenamefont {Janssens},
  \citenamefont {Jiang}, \citenamefont {Jisonna}, \citenamefont {Moore},
  \citenamefont {Pardo}, \citenamefont {Paul}, \citenamefont {Peterson},
  \citenamefont {Pieper}, \citenamefont {Savard}, \citenamefont {Schiffer},
  \citenamefont {Segel}, \citenamefont {Sinha}, \citenamefont {Tang},\ and\
  \citenamefont {Wiringa}}]{WR05}%
  \BibitemOpen
  \bibfield  {author} {\bibinfo {author} {\bibfnamefont {A.~H.}\ \bibnamefont
  {Wuosmaa}}, \bibinfo {author} {\bibfnamefont {K.~E.}\ \bibnamefont {Rehm}},
  \bibinfo {author} {\bibfnamefont {J.~P.}\ \bibnamefont {Greene}}, \bibinfo
  {author} {\bibfnamefont {D.~J.}\ \bibnamefont {Henderson}}, \bibinfo {author}
  {\bibfnamefont {R.~V.~F.}\ \bibnamefont {Janssens}}, \bibinfo {author}
  {\bibfnamefont {C.~L.}\ \bibnamefont {Jiang}}, \bibinfo {author}
  {\bibfnamefont {L.}~\bibnamefont {Jisonna}}, \bibinfo {author} {\bibfnamefont
  {E.~F.}\ \bibnamefont {Moore}}, \bibinfo {author} {\bibfnamefont {R.~C.}\
  \bibnamefont {Pardo}}, \bibinfo {author} {\bibfnamefont {M.}~\bibnamefont
  {Paul}}, \bibinfo {author} {\bibfnamefont {D.}~\bibnamefont {Peterson}},
  \bibinfo {author} {\bibfnamefont {S.~C.}\ \bibnamefont {Pieper}}, \bibinfo
  {author} {\bibfnamefont {G.}~\bibnamefont {Savard}}, \bibinfo {author}
  {\bibfnamefont {J.~P.}\ \bibnamefont {Schiffer}}, \bibinfo {author}
  {\bibfnamefont {R.~E.}\ \bibnamefont {Segel}}, \bibinfo {author}
  {\bibfnamefont {S.}~\bibnamefont {Sinha}}, \bibinfo {author} {\bibfnamefont
  {X.}~\bibnamefont {Tang}}, \ and\ \bibinfo {author} {\bibfnamefont {R.~B.}\
  \bibnamefont {Wiringa}},\ }\href {\doibase 10.1103/PhysRevLett.94.082502}
  {\bibfield  {journal} {\bibinfo  {journal} {Phys. Rev. Lett.}\ }\textbf
  {\bibinfo {volume} {94}},\ \bibinfo {pages} {082502} (\bibinfo {year}
  {2005}{\natexlab{a}})}\BibitemShut {NoStop}%
\bibitem [{\citenamefont {Wuosmaa}\ \emph
  {et~al.}(2005{\natexlab{b}})\citenamefont {Wuosmaa}, \citenamefont {Rehm},
  \citenamefont {Greene}, \citenamefont {Henderson}, \citenamefont {Janssens},
  \citenamefont {Jiang}, \citenamefont {Jisonna}, \citenamefont {Moore},
  \citenamefont {Pardo}, \citenamefont {Paul}, \citenamefont {Peterson},
  \citenamefont {Pieper}, \citenamefont {Savard}, \citenamefont {Schiffer},
  \citenamefont {Segel}, \citenamefont {Sinha}, \citenamefont {Tang},\ and\
  \citenamefont {Wiringa}}]{WR05a}%
  \BibitemOpen
  \bibfield  {author} {\bibinfo {author} {\bibfnamefont {A.~H.}\ \bibnamefont
  {Wuosmaa}}, \bibinfo {author} {\bibfnamefont {K.~E.}\ \bibnamefont {Rehm}},
  \bibinfo {author} {\bibfnamefont {J.~P.}\ \bibnamefont {Greene}}, \bibinfo
  {author} {\bibfnamefont {D.~J.}\ \bibnamefont {Henderson}}, \bibinfo {author}
  {\bibfnamefont {R.~V.~F.}\ \bibnamefont {Janssens}}, \bibinfo {author}
  {\bibfnamefont {C.~L.}\ \bibnamefont {Jiang}}, \bibinfo {author}
  {\bibfnamefont {L.}~\bibnamefont {Jisonna}}, \bibinfo {author} {\bibfnamefont
  {E.~F.}\ \bibnamefont {Moore}}, \bibinfo {author} {\bibfnamefont {R.~C.}\
  \bibnamefont {Pardo}}, \bibinfo {author} {\bibfnamefont {M.}~\bibnamefont
  {Paul}}, \bibinfo {author} {\bibfnamefont {D.}~\bibnamefont {Peterson}},
  \bibinfo {author} {\bibfnamefont {S.~C.}\ \bibnamefont {Pieper}}, \bibinfo
  {author} {\bibfnamefont {G.}~\bibnamefont {Savard}}, \bibinfo {author}
  {\bibfnamefont {J.~P.}\ \bibnamefont {Schiffer}}, \bibinfo {author}
  {\bibfnamefont {R.~E.}\ \bibnamefont {Segel}}, \bibinfo {author}
  {\bibfnamefont {S.}~\bibnamefont {Sinha}}, \bibinfo {author} {\bibfnamefont
  {X.}~\bibnamefont {Tang}}, \ and\ \bibinfo {author} {\bibfnamefont {R.~B.}\
  \bibnamefont {Wiringa}},\ }\href {\doibase 10.1103/PhysRevC.72.061301}
  {\bibfield  {journal} {\bibinfo  {journal} {Phys. Rev. C}\ }\textbf {\bibinfo
  {volume} {72}},\ \bibinfo {pages} {061301} (\bibinfo {year}
  {2005}{\natexlab{b}})}\BibitemShut {NoStop}%
\bibitem [{\citenamefont {Wuosmaa}\ \emph {et~al.}(2008)\citenamefont
  {Wuosmaa}, \citenamefont {Schiffer}, \citenamefont {Rehm}, \citenamefont
  {Greene}, \citenamefont {Henderson}, \citenamefont {Janssens}, \citenamefont
  {Jiang}, \citenamefont {Jisonna}, \citenamefont {Lighthall}, \citenamefont
  {Marley}, \citenamefont {Moore}, \citenamefont {Pardo}, \citenamefont
  {Patel}, \citenamefont {Paul}, \citenamefont {Peterson}, \citenamefont
  {Pieper}, \citenamefont {Savard}, \citenamefont {Segel}, \citenamefont
  {Siemssen}, \citenamefont {Tang},\ and\ \citenamefont {Wiringa}}]{WS08}%
  \BibitemOpen
  \bibfield  {author} {\bibinfo {author} {\bibfnamefont {A.~H.}\ \bibnamefont
  {Wuosmaa}}, \bibinfo {author} {\bibfnamefont {J.~P.}\ \bibnamefont
  {Schiffer}}, \bibinfo {author} {\bibfnamefont {K.~E.}\ \bibnamefont {Rehm}},
  \bibinfo {author} {\bibfnamefont {J.~P.}\ \bibnamefont {Greene}}, \bibinfo
  {author} {\bibfnamefont {D.~J.}\ \bibnamefont {Henderson}}, \bibinfo {author}
  {\bibfnamefont {R.~V.~F.}\ \bibnamefont {Janssens}}, \bibinfo {author}
  {\bibfnamefont {C.~L.}\ \bibnamefont {Jiang}}, \bibinfo {author}
  {\bibfnamefont {L.}~\bibnamefont {Jisonna}}, \bibinfo {author} {\bibfnamefont
  {J.~C.}\ \bibnamefont {Lighthall}}, \bibinfo {author} {\bibfnamefont {S.~T.}\
  \bibnamefont {Marley}}, \bibinfo {author} {\bibfnamefont {E.~F.}\
  \bibnamefont {Moore}}, \bibinfo {author} {\bibfnamefont {R.~C.}\ \bibnamefont
  {Pardo}}, \bibinfo {author} {\bibfnamefont {N.}~\bibnamefont {Patel}},
  \bibinfo {author} {\bibfnamefont {M.}~\bibnamefont {Paul}}, \bibinfo {author}
  {\bibfnamefont {D.}~\bibnamefont {Peterson}}, \bibinfo {author}
  {\bibfnamefont {S.~C.}\ \bibnamefont {Pieper}}, \bibinfo {author}
  {\bibfnamefont {G.}~\bibnamefont {Savard}}, \bibinfo {author} {\bibfnamefont
  {R.~E.}\ \bibnamefont {Segel}}, \bibinfo {author} {\bibfnamefont {R.~H.}\
  \bibnamefont {Siemssen}}, \bibinfo {author} {\bibfnamefont {X.~D.}\
  \bibnamefont {Tang}}, \ and\ \bibinfo {author} {\bibfnamefont {R.~B.}\
  \bibnamefont {Wiringa}},\ }\href {\doibase 10.1103/PhysRevC.78.041302}
  {\bibfield  {journal} {\bibinfo  {journal} {Phys. Rev. C}\ }\textbf {\bibinfo
  {volume} {78}},\ \bibinfo {pages} {041302} (\bibinfo {year}
  {2008})}\BibitemShut {NoStop}%
\bibitem [{\citenamefont {Kanungo}\ \emph {et~al.}(2008)\citenamefont
  {Kanungo}, \citenamefont {Andreyev}, \citenamefont {Buchmann}, \citenamefont
  {Davids}, \citenamefont {Hackman}, \citenamefont {Howell}, \citenamefont
  {Khalili}, \citenamefont {Mills}, \citenamefont {Rodal}, \citenamefont
  {Pieper}, \citenamefont {Pearson}, \citenamefont {Ruiz}, \citenamefont
  {Ruprecht}, \citenamefont {Shotter}, \citenamefont {Tanihata}, \citenamefont
  {Vockenhuber}, \citenamefont {Walden},\ and\ \citenamefont {Wiringa}}]{KA08}%
  \BibitemOpen
  \bibfield  {author} {\bibinfo {author} {\bibfnamefont {R.}~\bibnamefont
  {Kanungo}}, \bibinfo {author} {\bibfnamefont {A.~N.}\ \bibnamefont
  {Andreyev}}, \bibinfo {author} {\bibfnamefont {L.}~\bibnamefont {Buchmann}},
  \bibinfo {author} {\bibfnamefont {B.}~\bibnamefont {Davids}}, \bibinfo
  {author} {\bibfnamefont {G.}~\bibnamefont {Hackman}}, \bibinfo {author}
  {\bibfnamefont {D.}~\bibnamefont {Howell}}, \bibinfo {author} {\bibfnamefont
  {P.}~\bibnamefont {Khalili}}, \bibinfo {author} {\bibfnamefont
  {B.}~\bibnamefont {Mills}}, \bibinfo {author} {\bibfnamefont {E.~P.}\
  \bibnamefont {Rodal}}, \bibinfo {author} {\bibfnamefont {S.~C.}\ \bibnamefont
  {Pieper}}, \bibinfo {author} {\bibfnamefont {J.}~\bibnamefont {Pearson}},
  \bibinfo {author} {\bibfnamefont {C.}~\bibnamefont {Ruiz}}, \bibinfo {author}
  {\bibfnamefont {G.}~\bibnamefont {Ruprecht}}, \bibinfo {author}
  {\bibfnamefont {A.}~\bibnamefont {Shotter}}, \bibinfo {author} {\bibfnamefont
  {I.}~\bibnamefont {Tanihata}}, \bibinfo {author} {\bibfnamefont
  {C.}~\bibnamefont {Vockenhuber}}, \bibinfo {author} {\bibfnamefont
  {P.}~\bibnamefont {Walden}}, \ and\ \bibinfo {author} {\bibfnamefont {R.~B.}\
  \bibnamefont {Wiringa}},\ }\href {\doibase DOI:
  10.1016/j.physletb.2007.12.024} {\bibfield  {journal} {\bibinfo  {journal}
  {Phys.\ Lett.\ B}\ }\textbf {\bibinfo {volume} {660}},\ \bibinfo {pages} {26
  } (\bibinfo {year} {2008})}\BibitemShut {NoStop}%
\bibitem [{\citenamefont {Grinyer}\ \emph {et~al.}(2011)\citenamefont
  {Grinyer}, \citenamefont {Bazin}, \citenamefont {Gade}, \citenamefont
  {Tostevin}, \citenamefont {Adrich}, \citenamefont {Bowen}, \citenamefont
  {Brown}, \citenamefont {Campbell}, \citenamefont {Cook}, \citenamefont
  {Glasmacher}, \citenamefont {McDaniel}, \citenamefont {Navr\'atil},
  \citenamefont {Obertelli}, \citenamefont {Quaglioni}, \citenamefont {Siwek},
  \citenamefont {Terry}, \citenamefont {Weisshaar},\ and\ \citenamefont
  {Wiringa}}]{GB11}%
  \BibitemOpen
  \bibfield  {author} {\bibinfo {author} {\bibfnamefont {G.~F.}\ \bibnamefont
  {Grinyer}}, \bibinfo {author} {\bibfnamefont {D.}~\bibnamefont {Bazin}},
  \bibinfo {author} {\bibfnamefont {A.}~\bibnamefont {Gade}}, \bibinfo {author}
  {\bibfnamefont {J.~A.}\ \bibnamefont {Tostevin}}, \bibinfo {author}
  {\bibfnamefont {P.}~\bibnamefont {Adrich}}, \bibinfo {author} {\bibfnamefont
  {M.~D.}\ \bibnamefont {Bowen}}, \bibinfo {author} {\bibfnamefont {B.~A.}\
  \bibnamefont {Brown}}, \bibinfo {author} {\bibfnamefont {C.~M.}\ \bibnamefont
  {Campbell}}, \bibinfo {author} {\bibfnamefont {J.~M.}\ \bibnamefont {Cook}},
  \bibinfo {author} {\bibfnamefont {T.}~\bibnamefont {Glasmacher}}, \bibinfo
  {author} {\bibfnamefont {S.}~\bibnamefont {McDaniel}}, \bibinfo {author}
  {\bibfnamefont {P.}~\bibnamefont {Navr\'atil}}, \bibinfo {author}
  {\bibfnamefont {A.}~\bibnamefont {Obertelli}}, \bibinfo {author}
  {\bibfnamefont {S.}~\bibnamefont {Quaglioni}}, \bibinfo {author}
  {\bibfnamefont {K.}~\bibnamefont {Siwek}}, \bibinfo {author} {\bibfnamefont
  {J.~R.}\ \bibnamefont {Terry}}, \bibinfo {author} {\bibfnamefont
  {D.}~\bibnamefont {Weisshaar}}, \ and\ \bibinfo {author} {\bibfnamefont
  {R.~B.}\ \bibnamefont {Wiringa}},\ }\href {\doibase
  10.1103/PhysRevLett.106.162502} {\bibfield  {journal} {\bibinfo  {journal}
  {Phys. Rev. Lett.}\ }\textbf {\bibinfo {volume} {106}},\ \bibinfo {pages}
  {162502} (\bibinfo {year} {2011})}\BibitemShut {NoStop}%
\bibitem [{\citenamefont {Lapik\'as}\ \emph {et~al.}(1999)\citenamefont
  {Lapik\'as}, \citenamefont {Wesseling},\ and\ \citenamefont
  {Wiringa}}]{LW99}%
  \BibitemOpen
  \bibfield  {author} {\bibinfo {author} {\bibfnamefont {L.}~\bibnamefont
  {Lapik\'as}}, \bibinfo {author} {\bibfnamefont {J.}~\bibnamefont
  {Wesseling}}, \ and\ \bibinfo {author} {\bibfnamefont {R.~B.}\ \bibnamefont
  {Wiringa}},\ }\href {\doibase 10.1103/PhysRevLett.82.4404} {\bibfield
  {journal} {\bibinfo  {journal} {Phys. Rev. Lett.}\ }\textbf {\bibinfo
  {volume} {82}},\ \bibinfo {pages} {4404} (\bibinfo {year}
  {1999})}\BibitemShut {NoStop}%
\bibitem [{\citenamefont {Pieper}\ and\ \citenamefont {Wiringa}(2001)}]{PW01}%
  \BibitemOpen
  \bibfield  {author} {\bibinfo {author} {\bibfnamefont {S.~C.}\ \bibnamefont
  {Pieper}}\ and\ \bibinfo {author} {\bibfnamefont {R.~B.}\ \bibnamefont
  {Wiringa}},\ }\href {\doibase 10.1146/annurev.nucl.51.101701.132506}
  {\bibfield  {journal} {\bibinfo  {journal} {Annu.\ Rev.\ Nucl.\ Part.\ Sci.}\
  }\textbf {\bibinfo {volume} {51}},\ \bibinfo {pages} {53} (\bibinfo {year}
  {2001})}\BibitemShut {NoStop}%
\bibitem [{\citenamefont {Pieper}\ \emph {et~al.}(2002)\citenamefont {Pieper},
  \citenamefont {Varga},\ and\ \citenamefont {Wiringa}}]{PV02}%
  \BibitemOpen
  \bibfield  {author} {\bibinfo {author} {\bibfnamefont {S.~C.}\ \bibnamefont
  {Pieper}}, \bibinfo {author} {\bibfnamefont {K.}~\bibnamefont {Varga}}, \
  and\ \bibinfo {author} {\bibfnamefont {R.~B.}\ \bibnamefont {Wiringa}},\
  }\href {\doibase 10.1103/PhysRevC.66.044310} {\bibfield  {journal} {\bibinfo
  {journal} {Phys. Rev. C}\ }\textbf {\bibinfo {volume} {66}},\ \bibinfo
  {pages} {044310} (\bibinfo {year} {2002})}\BibitemShut {NoStop}%
\bibitem [{\citenamefont {Pieper}\ \emph {et~al.}(2004)\citenamefont {Pieper},
  \citenamefont {Wiringa},\ and\ \citenamefont {Carlson}}]{PW04}%
  \BibitemOpen
  \bibfield  {author} {\bibinfo {author} {\bibfnamefont {S.~C.}\ \bibnamefont
  {Pieper}}, \bibinfo {author} {\bibfnamefont {R.~B.}\ \bibnamefont {Wiringa}},
  \ and\ \bibinfo {author} {\bibfnamefont {J.}~\bibnamefont {Carlson}},\ }\href
  {\doibase 10.1103/PhysRevC.70.054325} {\bibfield  {journal} {\bibinfo
  {journal} {Phys. Rev. C}\ }\textbf {\bibinfo {volume} {70}},\ \bibinfo
  {pages} {054325} (\bibinfo {year} {2004})}\BibitemShut {NoStop}%
\bibitem [{\citenamefont {Pieper}(2005)}]{Pi05}%
  \BibitemOpen
  \bibfield  {author} {\bibinfo {author} {\bibfnamefont {S.~C.}\ \bibnamefont
  {Pieper}},\ }\href {\doibase DOI: 10.1016/j.nuclphysa.2005.02.018} {\bibfield
   {journal} {\bibinfo  {journal} {Nucl.\ Phys.\ A}\ }\textbf {\bibinfo
  {volume} {751}},\ \bibinfo {pages} {516 } (\bibinfo {year} {2005})},\
  \bibinfo {note} {proceedings of the 22nd International Nuclear Physics
  Conference (Part 1)}\BibitemShut {NoStop}%
\bibitem [{\citenamefont {Nollett}\ \emph {et~al.}(2007)\citenamefont
  {Nollett}, \citenamefont {Pieper}, \citenamefont {Wiringa}, \citenamefont
  {Carlson},\ and\ \citenamefont {Hale}}]{NP07}%
  \BibitemOpen
  \bibfield  {author} {\bibinfo {author} {\bibfnamefont {K.~M.}\ \bibnamefont
  {Nollett}}, \bibinfo {author} {\bibfnamefont {S.~C.}\ \bibnamefont {Pieper}},
  \bibinfo {author} {\bibfnamefont {R.~B.}\ \bibnamefont {Wiringa}}, \bibinfo
  {author} {\bibfnamefont {J.}~\bibnamefont {Carlson}}, \ and\ \bibinfo
  {author} {\bibfnamefont {G.~M.}\ \bibnamefont {Hale}},\ }\href {\doibase
  10.1103/PhysRevLett.99.022502} {\bibfield  {journal} {\bibinfo  {journal}
  {Phys. Rev. Lett.}\ }\textbf {\bibinfo {volume} {99}},\ \bibinfo {pages}
  {022502} (\bibinfo {year} {2007})}\BibitemShut {NoStop}%
\bibitem [{\citenamefont {Pervin}\ \emph {et~al.}(2007)\citenamefont {Pervin},
  \citenamefont {Pieper},\ and\ \citenamefont {Wiringa}}]{PP07}%
  \BibitemOpen
  \bibfield  {author} {\bibinfo {author} {\bibfnamefont {M.}~\bibnamefont
  {Pervin}}, \bibinfo {author} {\bibfnamefont {S.~C.}\ \bibnamefont {Pieper}},
  \ and\ \bibinfo {author} {\bibfnamefont {R.~B.}\ \bibnamefont {Wiringa}},\
  }\href {\doibase 10.1103/PhysRevC.76.064319} {\bibfield  {journal} {\bibinfo
  {journal} {Phys. Rev. C}\ }\textbf {\bibinfo {volume} {76}},\ \bibinfo
  {pages} {064319} (\bibinfo {year} {2007})}\BibitemShut {NoStop}%
\bibitem [{\citenamefont {Marcucci}\ \emph {et~al.}(2008)\citenamefont
  {Marcucci}, \citenamefont {Pervin}, \citenamefont {Pieper}, \citenamefont
  {Schiavilla},\ and\ \citenamefont {Wiringa}}]{MP08}%
  \BibitemOpen
  \bibfield  {author} {\bibinfo {author} {\bibfnamefont {L.~E.}\ \bibnamefont
  {Marcucci}}, \bibinfo {author} {\bibfnamefont {M.}~\bibnamefont {Pervin}},
  \bibinfo {author} {\bibfnamefont {S.~C.}\ \bibnamefont {Pieper}}, \bibinfo
  {author} {\bibfnamefont {R.}~\bibnamefont {Schiavilla}}, \ and\ \bibinfo
  {author} {\bibfnamefont {R.~B.}\ \bibnamefont {Wiringa}},\ }\href {\doibase
  10.1103/PhysRevC.78.065501} {\bibfield  {journal} {\bibinfo  {journal} {Phys.
  Rev. C}\ }\textbf {\bibinfo {volume} {78}},\ \bibinfo {pages} {065501}
  (\bibinfo {year} {2008})}\BibitemShut {NoStop}%
\bibitem [{\citenamefont {Wiringa}\ \emph {et~al.}(1995)\citenamefont
  {Wiringa}, \citenamefont {Stoks},\ and\ \citenamefont {Schiavilla}}]{WS95}%
  \BibitemOpen
  \bibfield  {author} {\bibinfo {author} {\bibfnamefont {R.~B.}\ \bibnamefont
  {Wiringa}}, \bibinfo {author} {\bibfnamefont {V.~G.~J.}\ \bibnamefont
  {Stoks}}, \ and\ \bibinfo {author} {\bibfnamefont {R.}~\bibnamefont
  {Schiavilla}},\ }\href {\doibase 10.1103/PhysRevC.51.38} {\bibfield
  {journal} {\bibinfo  {journal} {Phys. Rev. C}\ }\textbf {\bibinfo {volume}
  {51}},\ \bibinfo {pages} {38} (\bibinfo {year} {1995})}\BibitemShut {NoStop}%
\bibitem [{\citenamefont {Pieper}(2008)}]{Pi08}%
  \BibitemOpen
  \bibfield  {author} {\bibinfo {author} {\bibfnamefont {S.~C.}\ \bibnamefont
  {Pieper}},\ }\href {\doibase 10.1063/1.2932280} {\bibfield  {journal}
  {\bibinfo  {journal} {AIP Conference Proceedings}\ }\textbf {\bibinfo
  {volume} {1011}},\ \bibinfo {pages} {143} (\bibinfo {year}
  {2008})}\BibitemShut {NoStop}%
\bibitem [{\citenamefont {Pudliner}\ \emph {et~al.}(1995)\citenamefont
  {Pudliner}, \citenamefont {Pandharipande}, \citenamefont {Carlson},\ and\
  \citenamefont {Wiringa}}]{PP95}%
  \BibitemOpen
  \bibfield  {author} {\bibinfo {author} {\bibfnamefont {B.~S.}\ \bibnamefont
  {Pudliner}}, \bibinfo {author} {\bibfnamefont {V.~R.}\ \bibnamefont
  {Pandharipande}}, \bibinfo {author} {\bibfnamefont {J.}~\bibnamefont
  {Carlson}}, \ and\ \bibinfo {author} {\bibfnamefont {R.~B.}\ \bibnamefont
  {Wiringa}},\ }\href {\doibase 10.1103/PhysRevLett.74.4396} {\bibfield
  {journal} {\bibinfo  {journal} {Phys. Rev. Lett.}\ }\textbf {\bibinfo
  {volume} {74}},\ \bibinfo {pages} {4396} (\bibinfo {year}
  {1995})}\BibitemShut {NoStop}%
\bibitem [{\citenamefont {Metropolis}\ \emph {et~al.}(1953)\citenamefont
  {Metropolis}, \citenamefont {Rosenbluth}, \citenamefont {Rosenbluth},
  \citenamefont {Teller},\ and\ \citenamefont {Teller}}]{MR53}%
  \BibitemOpen
  \bibfield  {author} {\bibinfo {author} {\bibfnamefont {N.}~\bibnamefont
  {Metropolis}}, \bibinfo {author} {\bibfnamefont {A.~W.}\ \bibnamefont
  {Rosenbluth}}, \bibinfo {author} {\bibfnamefont {M.~N.}\ \bibnamefont
  {Rosenbluth}}, \bibinfo {author} {\bibfnamefont {A.~H.}\ \bibnamefont
  {Teller}}, \ and\ \bibinfo {author} {\bibfnamefont {E.}~\bibnamefont
  {Teller}},\ }\href@noop {} {\bibfield  {journal} {\bibinfo  {journal} {J.\
  Chem.\ Phys.}\ }\textbf {\bibinfo {volume} {21}},\ \bibinfo {pages} {1087}
  (\bibinfo {year} {1953})}\BibitemShut {NoStop}%
\bibitem [{\citenamefont {Wiringa}\ \emph {et~al.}(2000)\citenamefont
  {Wiringa}, \citenamefont {Pieper}, \citenamefont {Carlson},\ and\
  \citenamefont {Pandharipande}}]{WP00}%
  \BibitemOpen
  \bibfield  {author} {\bibinfo {author} {\bibfnamefont {R.~B.}\ \bibnamefont
  {Wiringa}}, \bibinfo {author} {\bibfnamefont {S.~C.}\ \bibnamefont {Pieper}},
  \bibinfo {author} {\bibfnamefont {J.}~\bibnamefont {Carlson}}, \ and\
  \bibinfo {author} {\bibfnamefont {V.~R.}\ \bibnamefont {Pandharipande}},\
  }\href {\doibase 10.1103/PhysRevC.62.014001} {\bibfield  {journal} {\bibinfo
  {journal} {Phys. Rev. C}\ }\textbf {\bibinfo {volume} {62}},\ \bibinfo
  {pages} {014001} (\bibinfo {year} {2000})}\BibitemShut {NoStop}%
\bibitem [{\citenamefont {Wiringa}(1991)}]{Wi91}%
  \BibitemOpen
  \bibfield  {author} {\bibinfo {author} {\bibfnamefont {R.~B.}\ \bibnamefont
  {Wiringa}},\ }\href {\doibase 10.1103/PhysRevC.43.1585} {\bibfield  {journal}
  {\bibinfo  {journal} {Phys. Rev. C}\ }\textbf {\bibinfo {volume} {43}},\
  \bibinfo {pages} {1585} (\bibinfo {year} {1991})}\BibitemShut {NoStop}%
\bibitem [{\citenamefont {Pudliner}\ \emph {et~al.}(1997)\citenamefont
  {Pudliner}, \citenamefont {Pandharipande}, \citenamefont {Carlson},
  \citenamefont {Pieper},\ and\ \citenamefont {Wiringa}}]{PP97}%
  \BibitemOpen
  \bibfield  {author} {\bibinfo {author} {\bibfnamefont {B.~S.}\ \bibnamefont
  {Pudliner}}, \bibinfo {author} {\bibfnamefont {V.~R.}\ \bibnamefont
  {Pandharipande}}, \bibinfo {author} {\bibfnamefont {J.}~\bibnamefont
  {Carlson}}, \bibinfo {author} {\bibfnamefont {S.~C.}\ \bibnamefont {Pieper}},
  \ and\ \bibinfo {author} {\bibfnamefont {R.~B.}\ \bibnamefont {Wiringa}},\
  }\href {\doibase 10.1103/PhysRevC.56.1720} {\bibfield  {journal} {\bibinfo
  {journal} {Phys. Rev. C}\ }\textbf {\bibinfo {volume} {56}},\ \bibinfo
  {pages} {1720} (\bibinfo {year} {1997})}\BibitemShut {NoStop}%
\bibitem [{\citenamefont {Abramowitz}\ and\ \citenamefont
  {Stegun}(1972)}]{AS72}%
  \BibitemOpen
  \bibfield  {author} {\bibinfo {author} {\bibfnamefont {M.}~\bibnamefont
  {Abramowitz}}\ and\ \bibinfo {author} {\bibfnamefont {I.~A.}\ \bibnamefont
  {Stegun}},\ }\href@noop {} {\emph {\bibinfo {title} {Handbook of mathematical
  functions}}}\ (\bibinfo  {publisher} {Dover Publications},\ \bibinfo {year}
  {1972})\BibitemShut {NoStop}%
\bibitem [{\citenamefont {Nollett}\ \emph {et~al.}(2001)\citenamefont
  {Nollett}, \citenamefont {Wiringa},\ and\ \citenamefont {Schiavilla}}]{NW01}%
  \BibitemOpen
  \bibfield  {author} {\bibinfo {author} {\bibfnamefont {K.~M.}\ \bibnamefont
  {Nollett}}, \bibinfo {author} {\bibfnamefont {R.~B.}\ \bibnamefont
  {Wiringa}}, \ and\ \bibinfo {author} {\bibfnamefont {R.}~\bibnamefont
  {Schiavilla}},\ }\href {\doibase 10.1103/PhysRevC.63.024003} {\bibfield
  {journal} {\bibinfo  {journal} {Phys. Rev. C}\ }\textbf {\bibinfo {volume}
  {63}},\ \bibinfo {pages} {024003} (\bibinfo {year} {2001})}\BibitemShut
  {NoStop}%
\bibitem [{\citenamefont {Nollett}(2001)}]{No01}%
  \BibitemOpen
  \bibfield  {author} {\bibinfo {author} {\bibfnamefont {K.~M.}\ \bibnamefont
  {Nollett}},\ }\href {\doibase 10.1103/PhysRevC.63.054002} {\bibfield
  {journal} {\bibinfo  {journal} {Phys. Rev. C}\ }\textbf {\bibinfo {volume}
  {63}},\ \bibinfo {pages} {054002} (\bibinfo {year} {2001})}\BibitemShut
  {NoStop}%
\bibitem [{\citenamefont {Cohen}\ and\ \citenamefont {Kurath}(1967)}]{CK67}%
  \BibitemOpen
  \bibfield  {author} {\bibinfo {author} {\bibfnamefont {S.}~\bibnamefont
  {Cohen}}\ and\ \bibinfo {author} {\bibfnamefont {D.}~\bibnamefont {Kurath}},\
  }\href {\doibase DOI: 10.1016/0375-9474(67)90285-0} {\bibfield  {journal}
  {\bibinfo  {journal} {Nucl.\ Phys.\ A}\ }\textbf {\bibinfo {volume} {101}},\
  \bibinfo {pages} {1 } (\bibinfo {year} {1967})}\BibitemShut {NoStop}%
\bibitem [{\citenamefont {Macfarlane}\ and\ \citenamefont
  {Pieper}(1978)}]{MP78}%
  \BibitemOpen
  \bibfield  {author} {\bibinfo {author} {\bibfnamefont {M.~H.}\ \bibnamefont
  {Macfarlane}}\ and\ \bibinfo {author} {\bibfnamefont {S.}~\bibnamefont
  {Pieper}},\ }\href@noop {} {\bibfield  {journal} {\bibinfo  {journal}
  {Argonne National Laboratory Report ANL-76-11, Rev. 1}\ } (\bibinfo {year}
  {1978})}\BibitemShut {NoStop}%
\bibitem [{\citenamefont {Thompson}(1988)}]{Th88}%
  \BibitemOpen
  \bibfield  {author} {\bibinfo {author} {\bibfnamefont {I.~J.}\ \bibnamefont
  {Thompson}},\ }\href@noop {} {\bibfield  {journal} {\bibinfo  {journal}
  {Comput.\ Phys.\ Rep.}\ }\textbf {\bibinfo {volume} {7}},\ \bibinfo {pages}
  {167} (\bibinfo {year} {1988})}\BibitemShut {NoStop}%
\bibitem [{\citenamefont {Kievsky}\ \emph {et~al.}(2008)\citenamefont
  {Kievsky}, \citenamefont {Rosati}, \citenamefont {Viviani}, \citenamefont
  {Marcucci},\ and\ \citenamefont {Girlanda}}]{KR08}%
  \BibitemOpen
  \bibfield  {author} {\bibinfo {author} {\bibfnamefont {A.}~\bibnamefont
  {Kievsky}}, \bibinfo {author} {\bibfnamefont {S.}~\bibnamefont {Rosati}},
  \bibinfo {author} {\bibfnamefont {M.}~\bibnamefont {Viviani}}, \bibinfo
  {author} {\bibfnamefont {L.~E.}\ \bibnamefont {Marcucci}}, \ and\ \bibinfo
  {author} {\bibfnamefont {L.}~\bibnamefont {Girlanda}},\ }\href
  {http://stacks.iop.org/0954-3899/35/i=6/a=063101} {\bibfield  {journal}
  {\bibinfo  {journal} {J.\ Phys.\ G: Nucl.\ Part.\ Phys.}\ }\textbf {\bibinfo
  {volume} {35}},\ \bibinfo {pages} {063101} (\bibinfo {year}
  {2008})}\BibitemShut {NoStop}%
\bibitem [{\citenamefont {Viviani}\ and\ \citenamefont {Kievsky}(2011)}]{VK11}%
  \BibitemOpen
  \bibfield  {author} {\bibinfo {author} {\bibfnamefont {M.}~\bibnamefont
  {Viviani}}\ and\ \bibinfo {author} {\bibfnamefont {A.}~\bibnamefont
  {Kievsky}},\ }\href@noop {} {}\bibinfo {howpublished} {private communication}
  (\bibinfo {year} {2011})\BibitemShut {NoStop}%
\bibitem [{\citenamefont {Forest}\ \emph {et~al.}(1996)\citenamefont {Forest},
  \citenamefont {Pandharipande}, \citenamefont {Pieper}, \citenamefont
  {Wiringa}, \citenamefont {Schiavilla},\ and\ \citenamefont {Arriaga}}]{FP96}%
  \BibitemOpen
  \bibfield  {author} {\bibinfo {author} {\bibfnamefont {J.~L.}\ \bibnamefont
  {Forest}}, \bibinfo {author} {\bibfnamefont {V.~R.}\ \bibnamefont
  {Pandharipande}}, \bibinfo {author} {\bibfnamefont {S.~C.}\ \bibnamefont
  {Pieper}}, \bibinfo {author} {\bibfnamefont {R.~B.}\ \bibnamefont {Wiringa}},
  \bibinfo {author} {\bibfnamefont {R.}~\bibnamefont {Schiavilla}}, \ and\
  \bibinfo {author} {\bibfnamefont {A.}~\bibnamefont {Arriaga}},\ }\href
  {\doibase 10.1103/PhysRevC.54.646} {\bibfield  {journal} {\bibinfo  {journal}
  {Phys. Rev. C}\ }\textbf {\bibinfo {volume} {54}},\ \bibinfo {pages} {646}
  (\bibinfo {year} {1996})}\BibitemShut {NoStop}%
\bibitem [{\citenamefont {Wiringa}(2006)}]{Wi06}%
  \BibitemOpen
  \bibfield  {author} {\bibinfo {author} {\bibfnamefont {R.~B.}\ \bibnamefont
  {Wiringa}},\ }\href {\doibase 10.1103/PhysRevC.73.034317} {\bibfield
  {journal} {\bibinfo  {journal} {Phys. Rev. C}\ }\textbf {\bibinfo {volume}
  {73}},\ \bibinfo {pages} {034317} (\bibinfo {year} {2006})}\BibitemShut
  {NoStop}%
\bibitem [{\citenamefont {Schiavilla}\ \emph {et~al.}(1986)\citenamefont
  {Schiavilla}, \citenamefont {Pandharipande},\ and\ \citenamefont
  {Wiringa}}]{SP86}%
  \BibitemOpen
  \bibfield  {author} {\bibinfo {author} {\bibfnamefont {R.}~\bibnamefont
  {Schiavilla}}, \bibinfo {author} {\bibfnamefont {V.~R.}\ \bibnamefont
  {Pandharipande}}, \ and\ \bibinfo {author} {\bibfnamefont {R.~B.}\
  \bibnamefont {Wiringa}},\ }\href {\doibase DOI: 10.1016/0375-9474(86)90003-5}
  {\bibfield  {journal} {\bibinfo  {journal} {Nucl.\ Phys.\ A}\ }\textbf
  {\bibinfo {volume} {449}},\ \bibinfo {pages} {219 } (\bibinfo {year}
  {1986})}\BibitemShut {NoStop}%
\bibitem [{\citenamefont {Dieperink}\ and\ \citenamefont
  {de~Forest}(1974)}]{DF74}%
  \BibitemOpen
  \bibfield  {author} {\bibinfo {author} {\bibfnamefont {A.~E.~L.}\
  \bibnamefont {Dieperink}}\ and\ \bibinfo {author} {\bibfnamefont
  {T.}~\bibnamefont {de~Forest}},\ }\href {\doibase 10.1103/PhysRevC.10.543}
  {\bibfield  {journal} {\bibinfo  {journal} {Phys. Rev. C}\ }\textbf {\bibinfo
  {volume} {10}},\ \bibinfo {pages} {543} (\bibinfo {year} {1974})}\BibitemShut
  {NoStop}%
\bibitem [{\citenamefont {Jun}\ \emph {et~al.}(2010)\citenamefont {Jun},
  \citenamefont {Zhi-Hong}, \citenamefont {Bing}, \citenamefont {Xi-Xiang},
  \citenamefont {Zhi-Chang}, \citenamefont {Jian-Cheng}, \citenamefont
  {You-Bao}, \citenamefont {Gang}, \citenamefont {Sheng}, \citenamefont
  {Bao-Xiang}, \citenamefont {Sheng-Quan}, \citenamefont {Yun-Ju},
  \citenamefont {Er-Tao}, \citenamefont {Qi-Wen},\ and\ \citenamefont
  {Wei-Ping}}]{JZ10}%
  \BibitemOpen
  \bibfield  {author} {\bibinfo {author} {\bibfnamefont {S.}~\bibnamefont
  {Jun}}, \bibinfo {author} {\bibfnamefont {L.}~\bibnamefont {Zhi-Hong}},
  \bibinfo {author} {\bibfnamefont {G.}~\bibnamefont {Bing}}, \bibinfo {author}
  {\bibfnamefont {B.}~\bibnamefont {Xi-Xiang}}, \bibinfo {author}
  {\bibfnamefont {L.}~\bibnamefont {Zhi-Chang}}, \bibinfo {author}
  {\bibfnamefont {L.}~\bibnamefont {Jian-Cheng}}, \bibinfo {author}
  {\bibfnamefont {W.}~\bibnamefont {You-Bao}}, \bibinfo {author} {\bibfnamefont
  {L.}~\bibnamefont {Gang}}, \bibinfo {author} {\bibfnamefont {Z.}~\bibnamefont
  {Sheng}}, \bibinfo {author} {\bibfnamefont {W.}~\bibnamefont {Bao-Xiang}},
  \bibinfo {author} {\bibfnamefont {Y.}~\bibnamefont {Sheng-Quan}}, \bibinfo
  {author} {\bibfnamefont {L.}~\bibnamefont {Yun-Ju}}, \bibinfo {author}
  {\bibfnamefont {L.}~\bibnamefont {Er-Tao}}, \bibinfo {author} {\bibfnamefont
  {F.}~\bibnamefont {Qi-Wen}}, \ and\ \bibinfo {author} {\bibfnamefont
  {L.}~\bibnamefont {Wei-Ping}},\ }\href
  {http://stacks.iop.org/0256-307X/27/i=5/a=052101} {\bibfield  {journal}
  {\bibinfo  {journal} {Chin.\ Phys.\ Lett.}\ }\textbf {\bibinfo {volume}
  {27}},\ \bibinfo {pages} {052101} (\bibinfo {year} {2010})}\BibitemShut
  {NoStop}%
\bibitem [{\citenamefont {Li}\ and\ \citenamefont {Mark}(1969)}]{LM69}%
  \BibitemOpen
  \bibfield  {author} {\bibinfo {author} {\bibfnamefont {T.~Y.}\ \bibnamefont
  {Li}}\ and\ \bibinfo {author} {\bibfnamefont {S.~K.}\ \bibnamefont {Mark}},\
  }\href {\doibase DOI: 10.1016/0375-9474(69)90895-1} {\bibfield  {journal}
  {\bibinfo  {journal} {Nucl.\ Phys.\ A}\ }\textbf {\bibinfo {volume} {123}},\
  \bibinfo {pages} {147 } (\bibinfo {year} {1969})}\BibitemShut {NoStop}%
\bibitem [{\citenamefont {Towner}(1969)}]{To69}%
  \BibitemOpen
  \bibfield  {author} {\bibinfo {author} {\bibfnamefont {I.~S.}\ \bibnamefont
  {Towner}},\ }\href {\doibase DOI: 10.1016/0375-9474(69)90401-1} {\bibfield
  {journal} {\bibinfo  {journal} {Nucl.\ Phys.\ A}\ }\textbf {\bibinfo {volume}
  {126}},\ \bibinfo {pages} {97 } (\bibinfo {year} {1969})}\BibitemShut
  {NoStop}%
\bibitem [{\citenamefont {Schiffer}\ \emph {et~al.}(1967)\citenamefont
  {Schiffer}, \citenamefont {Morrison}, \citenamefont {Siemssen},\ and\
  \citenamefont {Zeidman}}]{SM67}%
  \BibitemOpen
  \bibfield  {author} {\bibinfo {author} {\bibfnamefont {J.~P.}\ \bibnamefont
  {Schiffer}}, \bibinfo {author} {\bibfnamefont {G.~C.}\ \bibnamefont
  {Morrison}}, \bibinfo {author} {\bibfnamefont {R.~H.}\ \bibnamefont
  {Siemssen}}, \ and\ \bibinfo {author} {\bibfnamefont {B.}~\bibnamefont
  {Zeidman}},\ }\href {\doibase 10.1103/PhysRev.164.1274} {\bibfield  {journal}
  {\bibinfo  {journal} {Phys. Rev.}\ }\textbf {\bibinfo {volume} {164}},\
  \bibinfo {pages} {1274} (\bibinfo {year} {1967})}\BibitemShut {NoStop}%
\bibitem [{\citenamefont {Girard}\ and\ \citenamefont {Fuda}(1979)}]{GF79}%
  \BibitemOpen
  \bibfield  {author} {\bibinfo {author} {\bibfnamefont {B.~A.}\ \bibnamefont
  {Girard}}\ and\ \bibinfo {author} {\bibfnamefont {M.~G.}\ \bibnamefont
  {Fuda}},\ }\href {\doibase 10.1103/PhysRevC.19.583} {\bibfield  {journal}
  {\bibinfo  {journal} {Phys. Rev. C}\ }\textbf {\bibinfo {volume} {19}},\
  \bibinfo {pages} {583} (\bibinfo {year} {1979})}\BibitemShut {NoStop}%
\bibitem [{\citenamefont {Locher}\ and\ \citenamefont {Mizutani}(1978)}]{LM78}%
  \BibitemOpen
  \bibfield  {author} {\bibinfo {author} {\bibfnamefont {M.~P.}\ \bibnamefont
  {Locher}}\ and\ \bibinfo {author} {\bibfnamefont {T.}~\bibnamefont
  {Mizutani}},\ }\href {\doibase DOI: 10.1016/0370-1573(78)90156-4} {\bibfield
  {journal} {\bibinfo  {journal} {Phys.\ Rep.}\ }\textbf {\bibinfo {volume}
  {46}},\ \bibinfo {pages} {43 } (\bibinfo {year} {1978})}\BibitemShut
  {NoStop}%
\bibitem [{\citenamefont {Purcell}\ \emph {et~al.}(2010)\citenamefont
  {Purcell}, \citenamefont {Kelley}, \citenamefont {Kwan}, \citenamefont
  {Sheu},\ and\ \citenamefont {Weller}}]{PK10}%
  \BibitemOpen
  \bibfield  {author} {\bibinfo {author} {\bibfnamefont {J.~E.}\ \bibnamefont
  {Purcell}}, \bibinfo {author} {\bibfnamefont {J.~H.}\ \bibnamefont {Kelley}},
  \bibinfo {author} {\bibfnamefont {E.}~\bibnamefont {Kwan}}, \bibinfo {author}
  {\bibfnamefont {C.~G.}\ \bibnamefont {Sheu}}, \ and\ \bibinfo {author}
  {\bibfnamefont {H.~R.}\ \bibnamefont {Weller}},\ }\href {\doibase DOI:
  10.1016/j.nuclphysa.2010.08.012} {\bibfield  {journal} {\bibinfo  {journal}
  {Nucl.\ Phys.\ A}\ }\textbf {\bibinfo {volume} {848}},\ \bibinfo {pages} {1 }
  (\bibinfo {year} {2010})}\BibitemShut {NoStop}%
\bibitem [{\citenamefont {Blinov}\ \emph {et~al.}(1985)\citenamefont {Blinov},
  \citenamefont {Chuvilo}, \citenamefont {Drobot}, \citenamefont {Ergakov},
  \citenamefont {Grechko}, \citenamefont {Korolev}, \citenamefont {Selektor},
  \citenamefont {Soloviev}, \citenamefont {Shulyachenko}, \citenamefont
  {Turov}, \citenamefont {Vanyushin},\ and\ \citenamefont {Zombkovsky}}]{BC85}%
  \BibitemOpen
  \bibfield  {author} {\bibinfo {author} {\bibfnamefont {A.~V.}\ \bibnamefont
  {Blinov}}, \bibinfo {author} {\bibfnamefont {I.~V.}\ \bibnamefont {Chuvilo}},
  \bibinfo {author} {\bibfnamefont {V.~V.}\ \bibnamefont {Drobot}}, \bibinfo
  {author} {\bibfnamefont {V.~A.}\ \bibnamefont {Ergakov}}, \bibinfo {author}
  {\bibfnamefont {V.~E.}\ \bibnamefont {Grechko}}, \bibinfo {author}
  {\bibfnamefont {Y.~V.}\ \bibnamefont {Korolev}}, \bibinfo {author}
  {\bibfnamefont {Y.~M.}\ \bibnamefont {Selektor}}, \bibinfo {author}
  {\bibfnamefont {V.~V.}\ \bibnamefont {Soloviev}}, \bibinfo {author}
  {\bibfnamefont {V.~N.}\ \bibnamefont {Shulyachenko}}, \bibinfo {author}
  {\bibfnamefont {V.~F.}\ \bibnamefont {Turov}}, \bibinfo {author}
  {\bibfnamefont {I.~A.}\ \bibnamefont {Vanyushin}}, \ and\ \bibinfo {author}
  {\bibfnamefont {S.~M.}\ \bibnamefont {Zombkovsky}},\ }\href
  {http://stacks.iop.org/0305-4616/11/i=5/a=009} {\bibfield  {journal}
  {\bibinfo  {journal} {J.\ Phys.\ G: Nucl.\ Phys.}\ }\textbf {\bibinfo
  {volume} {11}},\ \bibinfo {pages} {623} (\bibinfo {year} {1985})}\BibitemShut
  {NoStop}%
\bibitem [{\citenamefont {Blokhintsev}\ \emph {et~al.}(1977)\citenamefont
  {Blokhintsev}, \citenamefont {Borbely},\ and\ \citenamefont
  {Dolinskii}}]{BB77}%
  \BibitemOpen
  \bibfield  {author} {\bibinfo {author} {\bibfnamefont {L.~D.}\ \bibnamefont
  {Blokhintsev}}, \bibinfo {author} {\bibfnamefont {I.}~\bibnamefont
  {Borbely}}, \ and\ \bibinfo {author} {\bibfnamefont {E.~I.}\ \bibnamefont
  {Dolinskii}},\ }\href@noop {} {\bibfield  {journal} {\bibinfo  {journal}
  {Sov. J. Part. Nucl.}\ }\textbf {\bibinfo {volume} {8}},\ \bibinfo {pages}
  {485} (\bibinfo {year} {1977})},\ \bibinfo {note} {{Fiz. Elem. Chastits At.
  Yadra} \textbf{8}, 1189 (1977)}\BibitemShut {NoStop}%
\bibitem [{\citenamefont {Bekbaev}\ \emph {et~al.}(1991)\citenamefont
  {Bekbaev}, \citenamefont {Kim}, \citenamefont {Mukhamedzhanov},\ and\
  \citenamefont {Timofeyuk}}]{BK91}%
  \BibitemOpen
  \bibfield  {author} {\bibinfo {author} {\bibfnamefont {S.~M.}\ \bibnamefont
  {Bekbaev}}, \bibinfo {author} {\bibfnamefont {G.}~\bibnamefont {Kim}},
  \bibinfo {author} {\bibfnamefont {A.~M.}\ \bibnamefont {Mukhamedzhanov}}, \
  and\ \bibinfo {author} {\bibfnamefont {N.~K.}\ \bibnamefont {Timofeyuk}},\
  }\href@noop {} {\bibfield  {journal} {\bibinfo  {journal} {Sov. J. Nucl.
  Phys.}\ }\textbf {\bibinfo {volume} {54}},\ \bibinfo {pages} {232} (\bibinfo
  {year} {1991})}\BibitemShut {NoStop}%
\bibitem [{\citenamefont {{Gulamov}}\ \emph {et~al.}(1995)\citenamefont
  {{Gulamov}}, \citenamefont {{Mukhamedzhanov}},\ and\ \citenamefont
  {{Nie}}}]{GM95}%
  \BibitemOpen
  \bibfield  {author} {\bibinfo {author} {\bibfnamefont {I.~R.}\ \bibnamefont
  {{Gulamov}}}, \bibinfo {author} {\bibfnamefont {A.~M.}\ \bibnamefont
  {{Mukhamedzhanov}}}, \ and\ \bibinfo {author} {\bibfnamefont {G.~K.}\
  \bibnamefont {{Nie}}},\ }\href@noop {} {\bibfield  {journal} {\bibinfo
  {journal} {Phys. At. Nucl.}\ }\textbf {\bibinfo {volume} {58}},\ \bibinfo
  {pages} {1689} (\bibinfo {year} {1995})},\ \bibinfo {note} {{Yad.} Fiz.
  \textbf{58}, 1789 (1995)}\BibitemShut {NoStop}%
\bibitem [{\citenamefont {Timofeyuk}\ \emph {et~al.}(2003)\citenamefont
  {Timofeyuk}, \citenamefont {Johnson},\ and\ \citenamefont
  {Mukhamedzhanov}}]{TJ03}%
  \BibitemOpen
  \bibfield  {author} {\bibinfo {author} {\bibfnamefont {N.~K.}\ \bibnamefont
  {Timofeyuk}}, \bibinfo {author} {\bibfnamefont {R.~C.}\ \bibnamefont
  {Johnson}}, \ and\ \bibinfo {author} {\bibfnamefont {A.~M.}\ \bibnamefont
  {Mukhamedzhanov}},\ }\href {\doibase 10.1103/PhysRevLett.91.232501}
  {\bibfield  {journal} {\bibinfo  {journal} {Phys. Rev. Lett.}\ }\textbf
  {\bibinfo {volume} {91}},\ \bibinfo {pages} {232501} (\bibinfo {year}
  {2003})}\BibitemShut {NoStop}%
\end{thebibliography}%

\end{document}